\documentclass[aps,prd,reprint,superscriptaddress,nofootinbib, longbibliography]{revtex4-2}

\usepackage{amssymb}
\usepackage{cuted}
\usepackage{siunitx}
\usepackage{mathtools}
\usepackage{amsmath}
\usepackage[utf8]{inputenc}
\usepackage{graphicx}
\usepackage{cancel}
\usepackage{bbm}
\usepackage{color}
\usepackage{comment}
\usepackage{lineno}
\usepackage{dcolumn}   % needed for some tables
\usepackage{bm}        % for math
\usepackage{amssymb}   % for math
\usepackage{amsmath,amssymb}
\usepackage{verbatim}
\usepackage[T1]{fontenc}
\usepackage{lipsum}
\usepackage{subcaption}
\usepackage{caption}
\usepackage{float}
\usepackage[newcommands]{ragged2e}
\usepackage{graphicx,caption}

\usepackage[dvipsnames]{xcolor}
\usepackage[colorlinks = true,
            linkcolor  = RoyalBlue,
            urlcolor   = RoyalBlue,
            citecolor  = RoyalBlue]{hyperref}
            
\usepackage{amsmath}
\usepackage{graphicx}
\usepackage{color}

\newcommand{\hp}{\hat{p}}
\newcommand{\tp}{\tilde{p}}
\newcommand{\ud}{\mathrm{d}}
\newcommand{\feq}{f_\text{eq}}
\DeclareMathOperator\arctanh{arctanh}

\newcommand{\bv}{\Bar{v}}
\newcommand{\bu}{\Bar{u}}
\newcommand{\bgamma}{\Bar{\gamma}}
\newcommand{\bpi}{\Bar{\pi}}
\newcommand{\bnu}{\Bar{\nu}}
\newcommand{\balpha}{\Bar{\alpha}}
\newcommand{\bbeta}{\Bar{\beta}}
\newcommand{\map}{|\mathbf{p}|}

\newcommand{\aref}[1]{\hyperref[#1]{appendix~\ref*{#1}}}

\begin{document}
\title{Cooper-Frye spectra of hadrons with viscous corrections including feed down from resonance decays}

\author{A. Kirchner}
\email[]{kirchner@thphys.uni-heidelberg.de }
\affiliation{Institut für Theoretische Physik der Universität Heidelberg, 69120 Heidelberg, Germany}

\author{E. Grossi}
\email[]{eduardo.grossi@unifi.it }
\affiliation{Dipartimento di Fisica, Universit\`a di Firenze and INFN Sezione di Firenze, via G. Sansone 1,
50019 Sesto Fiorentino, Italy}

\author{S. Floerchinger}
\email[]{stefan.floerchinger@uni-jena.de }
\affiliation{Friedrich-Schiller-Universit\"at Jena, 07743 Jena, Germany}

\date{\today}

\begin{abstract}
A method to calculate hadron momentum spectra after feed down from resonance decays in the context of ultra-relativistic heavy ion collisions described by relativistic fluid dynamics is presented. The conceptual setup uses the Cooper-Frye freeze-out integration together with an integral operator describing resonance decays. We provide explicit expressions for the integration over the freeze-out surface for a smooth and symmetric background solution, as well as for linearized perturbations around it. A major advantage of our method is that many integrals can be precomputed independently of a concrete hydrodynamic simulation. Additionally, we examine the influence of adding heavier resonances to the decay chain on the spectrum of pions and show how to include a phase with partial chemical equilibrium in order to separate the chemical from the kinetic freeze-out.
\end{abstract}

\maketitle

\section{Introduction}
One of the forefronts of modern physics are heavy ion collisions and the closely related study of quantum chromodynamics and its thermodynamic phases, especially the quark-gluon plasma (QGP) \cite{Busza_2018,ALICE:2010suc,STAR:2005gfr,PHENIX:2004vcz}. The two main research facilities examining the QGP are the Large Hadron Collider (LHC) at CERN in Geneva and the Relativistic Heavy Ion Collider (RHIC) at Brookhaven National Laboratory, Long Island. A big success in the study of heavy ion collisions was the description of the QGP as a relativistic fluid \cite{Heinz_2013}, which allows for a description in terms of macroscopic quantities, such as temperature $T(x)$ and fluid velocity $u^\mu(x)$. During the evolution of a heavy ion collision, the fireball expands and cools down, which also changes the effective degrees of freedom from quarks and gluons to hadrons. In the thermodynamic sense, this change in degrees of freedom is believed to be a cross-over and not a phase transition \cite{ding2015thermodynamics}. A fluid dynamic description has the advantage of being universal enough to describe the crossover regime when a realistic thermodynamic equation of state and transport properties are employed. However, to compare with experimental data one needs a transition from fluid fields to hadrons and their momentum distribution.

The first attempts to describe the freeze-out were made by the likes of Hagedorn \cite{Hagedorn:1965st,Hagedorn:1967tlw}, Landau \cite{Landau:1953gs} and Milekhin \cite{milekhin1959hydrodynamic}. The standard description which conserves energy and momentum, as well as other conserved quantum numbers, but does not yet include feed down from resonance decays, was given by Cooper and Frye \cite{PhysRevD.10.186}. Current state-of-the-art simulations are still based on Cooper and Frye's ansatz but include resonance decays and a hadronic rescattering phase, usually based on Monte-Carlo generators \cite{SMASH:2016zqf,Bleicher_1999,Nijs_2021,Bass_2017}. However, this method is numerically costly, because all intermediate states of a decay chain appear during the simulation.

The following calculations are based on a scheme presented in ref. \cite{Mazeliauskas_2019}, which allows us to obtain the final particle spectra after decays directly from the initial thermal distributions on the freeze-out surface without the need to calculate in-between resonances. Applications of the results presented in the following concern the extraction of phenomenological QGP properties from experimental data \cite{Devetak_2020} and related investigations we plan to present in the future. For this purpose we will show how to apply the freeze-out scheme to a hydrodynamic simulation based on a background-fluctuation splitting ansatz. Obtaining the spectra, including feed down from resonance decays, directly from the freeze-out surface, allows for a more efficient calculation of the observables and a more precise parameter extraction.

Moreover, one can separate the chemical from the kinetic freeze-out, with a phase of so-called partial chemical equilibrium (PCE) between them. This freeze-out prescription is more universal than descriptions based on hadronic scatterings and hadronic transport because the latter need detailed information on scattering cross sections, which can be challenging to measure experimentally.

This paper is structured as follows: In \autoref{sec_Cooper_Frye} we discuss the general properties of the Cooper-Frye freeze-out prescription, also addressing the issue of possible negative contributions to the freeze-out integrals. In \autoref{sec_FastResoIdea} we give a brief overview of how to incorporate feed down from resonance decays into the Cooper-Frye prescription using a method originally presented in ref. \cite{Mazeliauskas_2019}. The application of the formalism to mode-by-mode fluid dynamics, together with our conventions for coordinates and fluid fields, are presented in \autoref{sec_fieldConventions}, \autoref{sec_modeByMode} and \aref{sec_transformed_fields}, respectively. Subsequently, we also present a way to obtain fully analytical expressions for the integrations over azimuthal angle and rapidities on the freeze-out surface in the case of thermal spectra without feed down from resonances. In the final sections, we examine the influence of different resonance lists specifically for pions in the low transverse momentum region (\autoref{sec_resonanceListComp}) and give a brief overview of how to include a phase with partial chemical equilibrium (\autoref{sec_PCE}). Some conclusions are drawn in \autoref{sec_conclusion}. In \aref{appendix_tensor_decomposition}, we give the complete tensor decomposition for symmetric tensors with up to four Lorentz indices. In \aref{sec_analytic_res_kernel_bg} and \aref{sec_analytic_res_kernel_pert} we present the expressions for the freeze-out kernels for the background and the perturbation part, together with the initialization and transformation rules of the distribution functions in \aref{appendix_spectra_trafo_rules_distrs}. In the final part of the appendix, we give the expressions for the analytic thermal kernels on the freeze-out surface for the background (\aref{sec_thermal_bg_ker}) and the perturbation part (\aref{sec_thermal_pert_ker}) together with a way of expressing the appearing integrations over the azimuthal and rapidity angle in terms of Bessel functions (\aref{sec_integrations}).

\section{Cooper-Frye freeze-out} \label{sec_Cooper_Frye}
In this section we revisit the freeze-out prescription of Cooper and Frye \cite{PhysRevD.10.186}. We discuss how particle spectra determined as integrals over different hypersurfaces are related, why negative contributions can arise on the constant temperature freeze-out hypersurface, and how one obtains nevertheless physical results for the asymptotic distribution of particles at very large times.

In any collider experiment the particles are registered once they reach the detector, far away in space and time from the collision vertex. The interaction with the detector can determine the particles' momenta and sometimes their energy. 
It is only possible to measure the so-called out-states, which are stable states that result from the collision after a very long time evolution with respect to the intrinsic duration of the interaction. At that point any interaction has ceased and the out-states are free particle states (with respect to short range interactions) that can be labeled with momentum $\mathbf{p}$, energy $E_p$, and a species index identifying the particular particle quantum numbers. 

The measurement results are typically a collection of momenta $p_i$ for each particle species that reach the detector. The particles measured have a lifetime big enough to fly from the vertex to the detector without further decay.

The distribution function of particles produced can then be constructed from the collection of momenta and energies for each particle species,
\begin{equation}
   \left\langle E_p\frac{\ud N_i}{\ud^3 p} \right\rangle,
\end{equation}
that is, the number of particles of the type $i$ in a given momentum shell $\ud^3 p$ weighted with the energy to make it a relativistic covariant, averaged over the collection of events considered. 
Precise spatial information on where the particles were produced can not be accessed experimentally.

The particles are already so far apart during the propagation from the collision zone to the detector that their time evolution can be assumed to be
non-interacting. Therefore, it is possible to introduce the single-particle distribution function $f(x,p)$ as the distribution of particles in phase space $\ud^3x \ud^3p$, at least as an useful approximate concept\footnote{The distribution function should be seen as defining the single-particle Wigner distribution in a quantum field theory \cite{calzetta_hu_2008}.}.

The final covariant momentum distribution can then be computed as 
\begin{equation}
    \left\langle E_p\frac{\ud N_i}{\ud^3 p} \right\rangle = - \int_{\Sigma} \ud \Sigma_{\mu} p^{\mu} f(x,p), \label{eq:spectra_def}
\end{equation}
with a Cauchy hypersurface $\Sigma$ positioned at a given time much larger than any other typical time scale of the interaction. 
The latter is a three-dimensional hypersurface with future oriented  time-like normal vector $\ud \Sigma^{\mu}$, i.e., 
$g_{\mu\nu} \Sigma^\mu \Sigma^\nu <0 $ and $\ud \Sigma^0>0$ (we use the metric signature $(-,+,+,+)$) and can be thought of as one instant in time. Observe that for a Cauchy hypersurface $- \ud \Sigma_\mu p^\mu \geq 0$, as a consequence of $p^0>0$. For $f(x,p)\geq 0$ one has then only positive contributions to the integral in \autoref{eq:spectra_def}. The precise choice of the hypersurface is arbitrary if there is no interaction between the particles and, consequently, the distribution function obeys the collisionless Boltzmann equation \footnote{We neglect here the effect of resonance decays which will be treated separately later on.},

\begin{equation}\label{eq:boltzmann}
    p^{\mu}\nabla_{\mu} f(x,p)= 0,
\end{equation}
where $p^{\mu}$ is the on-shell momentum of the particle
\begin{equation}
    p^{\mu}= (E_p, \mathbf{p}). 
\end{equation}
This can be seen as an evolution equation for the distribution function $f(x,p)$,
\begin{align}
    p^\mu \partial_\mu f(x,p)=0.
\end{align}
The currents $J^{\mu}_p= - p^{\mu}f(x,p)$ are conserved for each on-shell momentum $p$,
\begin{equation}
    \nabla_{\mu} J^{\mu}_p=0. \label{eq:conservation laws}
\end{equation}
As a consequence of \autoref{eq:conservation laws}, for fixed momentum $p^\mu$ the distribution function does not change, $f(x,p)=f(y,p)$, for spacetime points related by a particle trajectroy $x^\mu-y^\mu= \tau p^\mu/m$. It is therefore ensured that $f(x,p)$ has physical properties - such as $f(x,p)\geq 0$ and $f(x,p)=0$ except when $p^0>0$ and $p^2+m^2=0$ - in all region that are connected to a point where these conditions are obeyed, by a free particle trajectory.

We now discuss how the conservation laws in \autoref{eq:conservation laws} and Gauss's theorem allow to relate the integrations over two different Cauchy surfaces with equivalent results. This enables a choice more convenient for calculating and modeling the freeze-out's complicated dynamics.

\begin{figure}
	\centering
	\includegraphics[width=.45\textwidth]{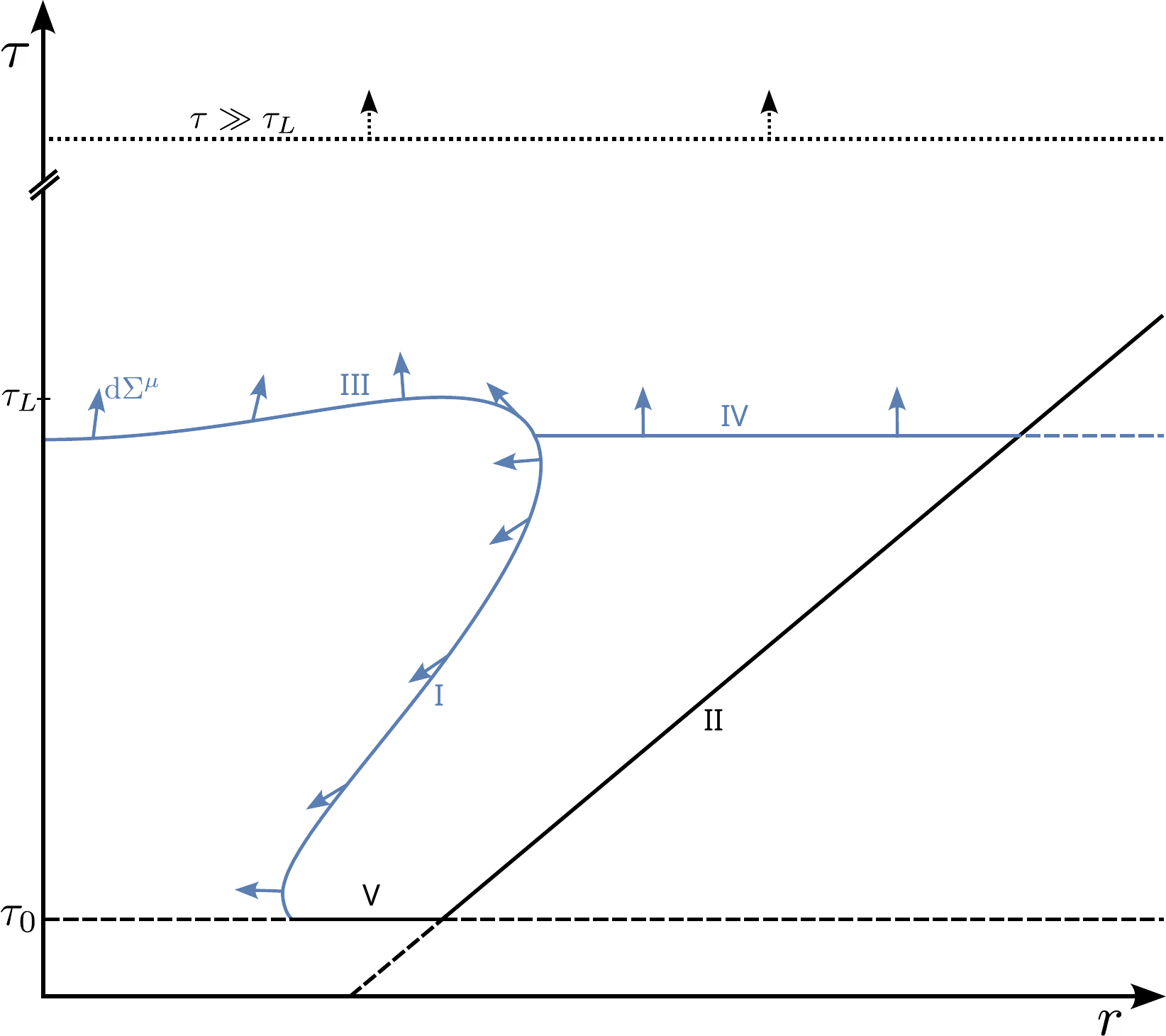}
	\caption{Illustrative cartoon of a freeze-out surface. The curves I and III define the constant temperature hypersurface. Curve IV is one possible extension of III to form a Cauchy surface. The black line II is the outward-going part of the lightcone originating from the outmost point of the fireball at its creation time $\tau_0$. Curve V indicates the corona of the fireball at its creation time. The lifetime of the fireball is indicated by $\tau_L$. The arrows indicate the direction of the normal vectors $n^\mu \propto \ud\Sigma^\mu$.}
	\label{fig:Freezeout}
\end{figure} 
Let us consider a simplified version of a freeze-out, depicted in \autoref{fig:Freezeout}. The fireball lies in the region enclosed by curves I and III; the free streaming phase is assumed to be everywhere outside these curves. The curve V indicates the region of a possible corona at the creation of the fireball where the particle density might be non-vanishing but is assumed to be very small. The black line II is the light cone originating from the outermost point of the fireball and its corona at the creation time $\tau_0$ of the fireball. Since most of the energy of the collision is deposited inside the fireball, we assume that the contributions from the corona to the final hadron spectra are negligible. The space-time region on the right of this light cone line is causally disconnected from the evolution of the fireball. Therefore we assume the distribution function to be zero on the light cone and in the region outside. We will show the equivalence of the integration in the asymptotic regime close to the detector at some large time $\tau \gg \tau_L$, with $\tau_L$ being the lifetime of the fireball, and on the freeze-out surface (I+III), in two steps:
First, we consider a space-time volume $\Omega$ enclosed by the time axis, the hypersurface at constant large time $\tau\gg \tau_L$, the light cone, and the hypersurface III+IV. Inside this space-time volume, the hadrons are free-streaming. Therefore we can write 
\begin{align}
    \int_\Omega \nabla_\mu J^\mu_p=0.
\end{align}
Applying Gauss's theorem and using that $f(x,p)=0$ on the time axis and the light cone, we can write
\begin{align}
   \int_{\Sigma(\tau \gg \tau_L)}\ud \Sigma_{\mu} J^{\mu}_p - \int_{\Sigma_{\text{IV}} \cup\Sigma_{\text{III}}} \ud \Sigma_{\mu} J^{\mu}_p=0. \label{eq_first_step}
\end{align}
Note that in both integrals over Cauchy surfaces we take the normal vector to be future oriented. Making use of Gauss's theorem a second time for the space-time volume enclosed by I, II, and IV, lets us write
\begin{align}
    \int_{\Sigma(\tau \gg \tau_L)}\ud \Sigma_{\mu} J^{\mu}_p =\int_{\Sigma_\text{I}}\ud \Sigma_{\mu} J^{\mu}_p +\int_{\Sigma_{\text{III}}}\ud \Sigma_{\mu} J^{\mu}_p.
\end{align}
Here the normal vector $n^\mu \propto \ud \Sigma^\mu$ is still future oriented in region III and oriented to the inside (as illustrated in \autoref{fig:Freezeout}), in region I.

We want to emphasize two important points at this step of the calculation: Considering the first in-between step given in \autoref{eq_first_step}, we want to stress that the integrand of the second integration is strictly positive. This originates again from the integration being over a Cauchy hypersurface, which corresponds to an interpretation of integrating a particle density to obtain the particle number. A second important point to make here is that this interpretation breaks down when considering the second step
\begin{align}
    E_p\frac{\ud N}{\ud^3 p} = \int_{\Sigma_{\text{III}}}\ud \Sigma_{\mu} J^{\mu}_p+\int_{\Sigma_{\text{I}}}\ud \Sigma_{\mu} J^{\mu}_p,\label{eq_cf_with_minus}
\end{align}
since the hypersurface, I is no longer part of a Cauchy surface. Vividly the interpretation breaks down because surface I is no longer one instant of time, which means that the integrand cannot be seen as density anymore. Note that in \autoref{eq_cf_with_minus} we took the orientation of the hypersurface I to point inwards. This follows the standard prescription for regions of spacetime and leads to a normal vector that is continuous at the intersection point between regions I and III, where the normal vector is light-like. Moreover, one can combine the surfaces $\Sigma_{\text{III}}$ and $\Sigma_{\text{I}}$ so that
\begin{align}
    E_p\frac{\ud N}{\ud^3 p} &= \int_{T=const.}\ud \Sigma_{\mu} J^{\mu}_p.\label{eq:Copper-Frye-Standart-form}
\end{align}

The same reasoning can be applied to the two-particle correlation function, and with little additional effort for any n-particle correlation. For simplicity, we will only demonstrate this for the two-particle correlation function. 
The two-particle correlation function that can be measured experimentally is defined as 
\begin{equation}
    C_2(p,k)=E_p E_k\left\langle  \frac{d N }{d^3 p} \frac{d N }{d^3 k}  \right\rangle.
\end{equation}
This correlation function is double differential, depending on the two on-shell momenta of the particles, $p$ and $k$.  

In the most straightforward and reasonable approximation, the two-point function is the average of the product of the fluxes of the distribution functions across the Cauchy surface at some large time in the future, similar to the one-point function, 
\begin{equation}
    C(p,k)= \int_{\Sigma(\tau)} d \Sigma_{\mu} 
    \int_{\Sigma(\tau)} d \Sigma_{ \nu} p^{\mu}p^{\nu} \langle f(x,p) f(y,k) \rangle .
\end{equation}
For most of the evolution, the distribution functions undergo a free-streaming phase, i.e., the phase space distributions are conserved (\autoref{eq:boltzmann}).
The integration domain at infinity can be deformed, like in the case of the one-particle distribution function for the entire region where we have free-streaming dynamics. Assuming again that after the freeze-out, collisions are practically negligible, the two-particle correlation function is determined by the flux across the constant temperature hypersurface
\begin{equation}
    C(p,k)= \int_{T=\text{const}} d \Sigma_{\mu} 
    \int_{T=\text{const}} d \Sigma_{ \nu} p^{\mu}p^{\nu} \langle f(x,p) f(y,k) \rangle .
\end{equation}
The correlation function is determined from the correlation at the freeze-out, and the subsequent free-streaming dynamic does not alter the two-particle spectra.

In summary, we have confirmed the freeze-out prescription of Cooper and Frye as it is used in practice with a freeze-out surface that is not necessarily a Cauchy surface. In that case there can be positive and negative contributions to the integral, but the equivalence to a Cauchy surface integral at a later time ensures that the resulting particle spectrum has physical properties and the assumption of free streaming is valid. This follows from the free streaming dynamics as long as $f(x,p)$ has physical properties on the freeze-out surface.

\section{Particle production with resonance decays} \label{sec_FastResoIdea}
This section will present how to incorporate resonance decays in the calculation of hadron spectra with the Cooper-Frye freeze-out, roughly following ref. \cite{Mazeliauskas_2019}. Using the fact that we want to describe a multitude of particle decays, we can express the final spectrum of a decay product $b$ as
\begin{equation}
    E_p \frac{\ud N_b}{\ud^3 p} = \int \frac{\ud^3 q}{(2\pi)^3 2 E_q} D^a_{b|c}(p,q) E_q \frac{\ud N_a}{\ud^3 q}
\end{equation}
where we introduce the linear decay map $D^a_b(p,q)$ which gives the Lorentz invariant probability of particle $a$ having momentum $q$ decaying into particle $b$ with momentum $p$. Assuming for simplicity an isotropic two-body decay $a \to b +c$, one finds after evaluating the phase space integration for particle $c$
\begin{equation}
    D^a_{b|c} (p_\mu q^\mu) =B \frac{4\pi^2 m_a}{p^a_{b|c}} \delta(p_\mu q^\mu + m_a E^a_{b|c}).
\end{equation}
Here $B$ is the branching ratio of the process and we define $E^a_{b|c}=\sqrt{m_b^2+(p^a_{b|c})^2}$ with
\begin{equation}
    p^a_{b|c} = \frac{1}{2m_a} \sqrt{((m_a+m_b)^2-m_c^2)((m_a-m_b)^2-m_c^2)}.
\end{equation}
Since the decay operator is linear, treating more complicated decays, such as decay chains or decays into multiple final states, is easy: Three- and higher body decays, such as $a\to b +c+d $, can be included by treating particles $c$ and $d$ as one fictitious particle $e$ with mass $m_e^2=-(p_c+p_d)^2$. A decay cascade like $a \to b + c \to c + d + e $ can be treated by applying multiple decay operators
\begin{equation}
    D^a_{e|f} (p_\mu q^\mu) = \int \frac{\ud^3 k}{(2\pi)^3 2 E_k} D^a_{b|c}(p_\nu k^\nu) D^b_{d|e}(k_\nu q^\nu),
\end{equation}
where we again introduced the fictitious particle $f= c+d$. Using the notation introduced above for the Cooper-Frye freeze-out, the vector distribution function for a decay particle $b$ is given by
\begin{equation}
    g_b^\mu(p,\Phi) = \sum_a \frac{\nu_a}{\nu_b} \int \frac{\ud^3 q}{(2\pi)^3 2E_q} D^a_{b|c}(p_\nu q^\nu) f_a(-u_\nu q^\nu,\Phi) q^\mu. \label{gmu_integration}
\end{equation}
Because $g^\mu_b$ only depends on fluid fields $\Phi$ on the freeze-out surface (we denote by $\Phi$ the collection of temperature $T$, chemical potentials $\mu_i$, fluid velocity $u^\rho$ and non-equilibrium fields like shear stress, bulk viscous pressure and so on), it can be calculated independently from the fluid simulation and stored once for each particle. The final particle spectrum is then obtained by integrating over the freeze-out hyper-surface as in \autoref{eq:Copper-Frye-Standart-form} without having to compute intermediate particle spectra. In the following, we describe how to perform the integration in \autoref{gmu_integration} over the decay operator for a primary resonance in the case of ideal hydrodynamics (we will show later how to deal with corrections stemming from viscous terms).

First, $g^\mu_a = f_a p^\mu$ gives the vector distribution function for a primary resonance. In this case, the distribution function for particle $a$ will either be a Bose- or a Fermi-distribution $f_a=f_a(E_p,T,\mu_i)=(e^{(E_p-\sum_i Q_{a,i}\mu_i)/T} \pm 1)^{-1}$ which only depends on Lorentz scalars like the temperature $T$, chemical potentials $\mu_i$ and the particle energy $E_p=-u_\mu p^\mu$ in the frame co-moving with the fluid. Note that the chemical potentials can be conjugated to any conserved charge $Q_{a,i}$ (over which the index $i$ is running), such as baryon numbers, electrical charge or the conserved particle number appearing in a partial chemical equilibrium, described in \autoref{sec_PCE}. Using the Lorentz-invariance of the decay, we can write the vector distribution function of the decay products as a unique sum of two scalar functions
\begin{equation}
    g_b^\mu=f^\text{eq}_{1,b}(E_p) (p^\mu - E_p u^\mu) + f^\text{eq}_{0,b}(E_p) E_p u^\mu.
\end{equation}
In this case, $p^\mu$ and $E_p u^\mu$ are the only available Lorentz vectors. We also use two scalar functions $f^\text{eq}_{s}$, with $s=1,2$, which only depend on the aforementioned Lorentz scalars. Under closer examination we find that the two Lorentz vectors are two irreducible representations of the rotation group $SO(3)$ in the fluid rest frame, where $\hp^\mu=p^\mu - E_p u^\mu =(0,\mathbf{p})$ transforms as a vector and $\tp^\mu=E_p u^\mu =(E_p,\mathbf{0})$ transforms as a scalar. The linearity of the decay operator ensures that these irreducible representations of $SO(3)$ do not mix during the decay process. Using this allows us to simplify the integration \autoref{gmu_integration}. By only considering one irreducible term and contracting with $p_\mu$, we can obtain the following iterative relations for scalar distribution functions corresponding to each irreducible representation
\begin{align}
\begin{split}
    f_{0,b}^\text{eq} &=B \frac{\nu_a}{\nu_b} \frac{m_a^2}{m_b^2} \frac{1}{2} \int_{-1}^1 \ud w \; f_{0,a}^\text{eq}(E(w)) \frac{E(w)}{E_p},\\
    f_{1,b}^\text{eq} &=B \frac{\nu_a}{\nu_b} \frac{m_a^2}{m_b^2} \frac{1}{2} \int_{-1}^1 \ud w \; f_{1,a}^\text{eq}(E(w)) \frac{Q(w)}{|\mathbf{p}|}. 
   \end{split} \label{eq_transformation_rules}
\end{align}
Here we introduced the abbreviations
\begin{align}
    E(w) &= \frac{m_a E^a_{b|c} E_p}{m_b^2} - w \frac{m_a p^a_{b|c} |\mathbf{p}|}{m_b^2}, \\
    Q(w) &= \frac{m_a E^a_{b|c} |\mathbf{p}|}{m_b^2} - w \frac{m_a p^a_{b|c} E_p}{m_b^2},
\end{align}
together with the substitution $q^z=w \frac{m_a  p^a_{b|c}}{m_b}$ for the momentum in $z$-direction of $q$. Moreover, $w$ can be interpreted as the fraction of particle $a$'s momentum in the direction of the fluid velocity in the rest frame of particle $b$.

\begin{figure}
	%\centering
	\includegraphics[width=.45\textwidth]{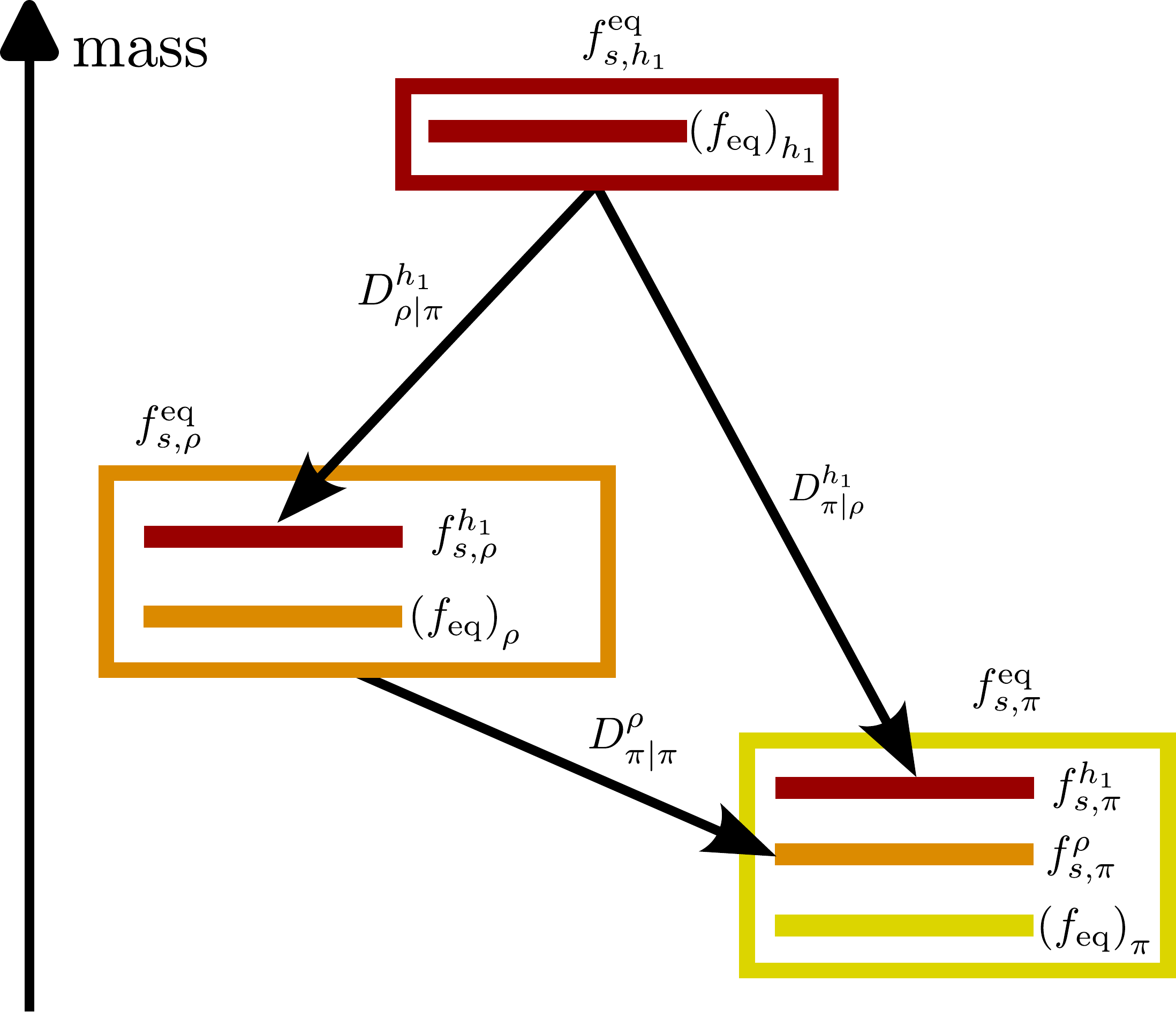}
 \caption{Schematic of the decay of the $h_1$ resonance. The occupation number of $h_1$ is initialized with its thermal distribution (red line) and contributes to the $\rho$ and $\pi$ distribution functions via the corresponding decay operators. The full $\rho$ distribution function (orange box) is then given by the sum of the thermally initialized spectra (orange line), and the contributions from the $h_1$ decays (red line). The full pion distribution is given similarly by the sum of the thermal distribution plus all contributions from resonance decays.}
	\label{fig:FastResoScheme}
\end{figure}

Now we can apply this iterative rule to show how to calculate a chain of resonances for a practical example. We will consider the decay chain $h_1 \to (\rho, \; \pi) \to \pi$, for this illustration neglecting all decays feeding down into these particles (see \autoref{fig:FastResoScheme}). In this case, we will initialize the distribution functions of the $h_1$ as described before for a primary resonance, thermally produced on the freeze-out hypersurface
\begin{equation}
    f_{s,h_1}^\text{eq}= (\feq)_{h_1}.
\end{equation}
Since $h_1$ is a meson, $\feq$ will be a Bose-Einstein distribution. Applying the transformation rules established in \autoref{eq_transformation_rules}, we find the feed down to the spectra of $\pi$ and $\rho$. Since the $\rho$ itself is not a stable particle, we have to apply the transformation again to the full distribution function $f_{s,\rho}^\text{eq}$ to account for the decay $\rho \to \pi \pi$. This allows us to write the final distribution functions as
\begin{align}
    f_{s,\rho}^\text{eq} &= (\feq)_\rho + f_{s,\rho}^{h_1}, \\
    f_{s,\pi}^\text{eq} &= (\feq)_\pi + f_{s,\pi}^{h_1}+ f_{s,\pi}^{\rho}.
\end{align}
$f_{s,\pi}^{\rho}$ represents the contribution from the decay $\rho \to \pi \pi$, irrespective of the primary origin of the $\rho$. In practice, the easiest and most efficient way of including large amounts of resonances is to start from the heaviest resonance, sum thermal and resonance contributions to its decay products, and then go down in mass considering the contributions for the lower mass states repetitively.

So far, we have mostly been discussing the theoretical concepts of including resonance decays in calculating the final hadron spectra during a heavy-ion collision. For a practical simulation, one has to apply these concepts to fluid dynamic calculations, to which we will turn now, starting with some conventions and field definitions.

\section{Conventions and field definitions} \label{sec_fieldConventions}
The following two sections demonstrate how to apply the previously described freeze-out routine to a hydrodynamic simulation, including viscous corrections. We will work with a background-fluctuation splitting ansatz for the hydrodynamic fields, which will be explained in more detail after introducing the coordinate system.

When describing a multitude of high-energy heavy-ion collisions, invariance under longitudinal boosts and azimuthal rotations are good symmetries. A convenient choice of coordinates are Bjorken coordinates, where the line element reads
\begin{equation}
    \ud s^2 = -\ud \tau^2 + \ud r ^2 + r^2 \ud \phi^2 +\tau^2 \ud \eta^2.
\end{equation}
We will develop our equations for a setting with a background-fluctuation splitting. The combined fluid fields $\Phi \left(\tau,r,\phi,\eta\right)$ describing a single heavy-ion collision depend on all four coordinates. However, our ansatz is based on the description of event ensembles of heavy-ion collisions, which display more symmetries than a single event. To be more specific, when choosing the described ensemble to be within one centrality class, and restrict one to observables measured at mid-rapidity, we find an approximate boost invariance of the measured spectra. Additionally, the description of an ensemble allows us to employ a statistical symmetry in azimuthal angle. Therefore we perform the decomposition $\Phi\left(\tau,r,\phi,\eta\right)=\Bar{\Phi}\left(\tau,r\right)+ \delta \Phi \left(\tau,r,\phi,\eta\right)$. The background configuration $\Bar{\Phi}$ displays the aforementioned symmetries and only depends on time and radius. 

In this approach of a symmetric background, with fluctuations encoded in the perturbations, the freeze-out hypersurface on which the hadronization happens could also be split into a background configuration and a correction from fluctuations. However, as discussed in ref. \cite{Floerchinger_2014} it is possible, and in fact easier, to take the freeze-out surface to be defined by the background fields only. Instead of having to work with a fluctuating geometry only the fluid fields (include the temperature) are fluctuating in this prescription. The background freeze-out hypersurface is given by a one dimensional $\tau$-$r$-curve which can be parametrized by $\left( \tau(c), r (c) \right)$, with $c \in [0,1]$ without loss of generality. From this, the hypersurface element entails being 
\begin{equation}
    \ud \Sigma_\mu =r(c) \tau(c) \left( -\frac{\partial r}{\partial c}, \frac{\partial\tau}{\partial c} ,0,0\right) \ud c \ud \phi \ud \eta.\label{eq_surface-element}
\end{equation}
For the particle momentum, we use a parametrization in terms of transverse mass $m_T=\sqrt{E^2-p_3^2}$ and transverse momentum $p_T=\sqrt{p_1^2+p_2^2}$, as well as the momentum-space rapidity $\eta_P = \arctanh(p_3/E)$ and the momentum azimuthal angle $\phi_P=\arctan(p_2/p_1)$, in cartesian lab coordinates
%\begin{equation}
    %p_\mu = \left( -m_T \cosh(\eta_P), p_T \cos (\phi_P), p_T \sin (\phi_P), m_T \sinh (\eta_P) \right).
%\end{equation}
\begin{equation}
    p^\mu = \left( m_T \cosh(\eta_P), p_T \cos (\phi_P), p_T \sin (\phi_P), m_T \sinh (\eta_P) \right).
\end{equation}
After transforming into Bjorken coordinates, the momentum emerges as
\begin{align}
	p_\mu  &= \left(- m_T  \cosh(\eta-\eta_P), p_T \cos(\phi-\phi_P) ,\right.\nonumber\\
 &\left.-r p_T \sin(\phi-\phi_P),-\tau m_T \sinh(\eta-\eta_P) \right).
\end{align}
%with the coordinate transformation given by $\Lambda^\nu_\mu=\partial x^\nu/ \partial x'^\mu$.\\
For parametrizing the fluid fields, we will use the tetrad formalism \cite{weinberg1972gravitation}. This formalism defines the tetrad field by choosing a local orthogonal frame. The tetrad field $v_\mu^a(x)$ defines a transformation from the general coordinate frame with metric $g_{\mu\nu}(x)$ into the Minkowski metric $\eta_{ab}=diag(-1,1,1,1)$, such that
\begin{equation}
    g_{\mu\nu}(x) = v_\mu^a(x) v_\nu^b(x) \eta_{ab}.
\end{equation}
We also introduce the inverse tetrad $v^\mu_a(x)$ such that
\begin{equation}
    v_\mu^a(x) v^\nu_a(x) = \delta_\mu^\nu \quad \quad v_\mu^a (x) v^\mu_b (x) = \delta^a_b.
\end{equation}
In our case, choosing this local frame to be aligned with the (background) fluid velocity is particularly convenient. With this choice, the tetrad and its inverse are given by
\begin{align}
    v^\mu_a(x)&=\begin{pmatrix}
        \bgamma & \bv \bgamma & 0 & 0 \\
        \bv \bgamma & \bgamma &0 & 0 \\
        0 & 0 & 1/r & 0 \\
        0 & 0 & 0 & 1/\tau
    \end{pmatrix} ,
    \end{align}
    and
    \begin{align}
        v_\mu^a(x)&=\begin{pmatrix}
        \bgamma & -\bv \bgamma & 0 & 0 \\
        -\bv \bgamma & \bgamma &0 & 0 \\
        0 & 0 & r & 0 \\
        0 & 0 & 0 & \tau
    \end{pmatrix},
\end{align}
where $\bv$ is the spatial background velocity in the radial direction (here and in the following, we suppress its argument $c$) and $\bgamma=\frac{1}{\sqrt{1-\bv^2}}$. In the orthogonal frame, $\Bar{u}^a=(1,0,0,0)$ gives the background fluid velocity. Using the normalization of the fluid velocity $u_\mu u^\mu =-1$, we can parametrize the perturbations around the background fluid velocity as $\delta u^a=(0,v_1,v_2,v_3)$. Using the orthogonality $\pi^{\mu\nu}u_\mu=0$ (up to linear order in perturbations), symmetry $\pi^{\mu\nu}=\pi^{\nu\mu}$ and the tracelessness $\pi^\mu_\mu=0$ of the shear stress tensor together with the boost and rotational invariance of its background part, we can parametrize the shear stress tensor $\pi^{ab}=\Bar{\pi}^{ab}+\delta \pi^{ab}$ as
\begin{align}
    \Bar{\pi}^{ab}&= \begin{pmatrix}
        0 & 0 & 0 & 0 \\
        0 & -\Bar{\pi}^t & 0 & 0 \\
        0 & 0 & \Bar{\pi}^{22} & 0 \\
        0 & 0 & 0 & \Bar{\pi}^{33}
    \end{pmatrix},
    \end{align}
    and
    \begin{align}
   \delta \pi^{ab}&= \begin{pmatrix}
         0 & \delta v_1\Bar{\pi}^t & - \delta v_2 \Bar{\pi}^{22} & - \delta v_3 \Bar{\pi}^{33} \\
        \delta  v_1 \Bar{\pi}^t & -\delta \pi^t & \delta \pi^{12} & \delta \pi^{13} \\
         -\delta v_2 \Bar{\pi}^{22} & \delta \pi^{12} & \delta \pi^{22} & \delta \pi^{23} \\
         -\delta v_3 \Bar{\pi}^{33} & \delta \pi^{13} &\delta  \pi^{23} & \delta \pi^{33}
    \end{pmatrix},
\end{align}
where we use the abbreviations $\Bar{\pi}^t=\Bar{\pi}^{22}+\Bar{\pi}^{33}$ and $\pi^t=\pi^{22}+\pi^{33}$. Similarly the diffusion current $\nu_i^a=\Bar{\nu}_i^a+\delta \nu_i^a$ can be parametrized using its orthogonality to the fluid velocity $\nu^\mu u_\mu =0$ (up to linear order in perturbations) via
\begin{align}
    \Bar{\nu}_i^a &= (0,\bnu_i,0,0),
    \end{align}
    and
    \begin{align}
    \delta \nu_i^a &= (-\delta v_1 \bnu_i , \delta \nu^i_1,\delta \nu^i_2,\delta \nu^i_3),
\end{align}
where the index $i$ again runs over the conserved charges. In the following we will restrict ourselves to taking only the baryon number conservation into account, which leaves us with seven independent background fields $\Bar{\Phi} = \left( \bbeta, \bv, \Bar{\pi}^{22},\Bar{\pi}^{33}, \bpi_B,\balpha,\bnu \right)$ and fourteen independent perturbation fields $\delta \Phi=\left(\delta \beta, \delta v_1, \delta v_2, \delta v_3,\delta  \pi^{12},\delta \pi^{13},\delta \pi^{22},\delta \pi^{23},\delta \pi^{33} ,\right.$ $\left.\delta \pi_B,\delta \alpha,\delta \nu_1, \delta \nu_2,\delta \nu_3 \right)$. Here we have introduced $\beta=1/T$ and $\alpha=\mu/T$ for convenience. The expressions for the fields transformed back to the Bjorken coordinate frame are given in \aref{sec_transformed_fields}. After defining the hydrodynamic variables and equations, we can now examine the application of the resonance decay prescription to the result of the hydrodynamic simulation.

\section{Application to viscous and mode by mode hydrodynamics} \label{sec_modeByMode}
For the application to a hydrodynamic simulation, we need to consider the distribution function $f(E_p,\beta,\alpha)$ again. The description above was based on an ideal hydrodynamic approximation, neglecting viscous corrections. However, it is well established that ideal hydrodynamics are insufficient for a realistic description of experimental data \cite{Romatschke_2007,Schenke_2011}. A better description is obtained by introducing out-of-equilibrium fields accounting for the viscous effects of the quark-gluon plasma. This will introduce additional dependencies in the vector distribution function, on the dissipative fields which we take to be the shear stress $\pi^{\mu\nu}$, bulk viscous pressure $\pi_B$ and baryon number diffusion current $\nu^\mu$. One could generalize this to additional propagated conserved currents, for example electric charge currents, and would have several diffusion currents. We will assume that the fluid is sufficiently close to thermal equilibrium at freeze-out to write the full distribution function as the equilibrium part plus deviations 
\begin{equation}
	f(\Phi)= \feq(\beta,v,\alpha) + \Delta f_\text{shear} + \Delta f_\text{bulk} + \Delta f_\text{diff},
\end{equation} 
with $\Delta f_\text{shear}$, $\Delta f_\text{bulk}$, $\Delta f_\text{diff}$ being linear in $\pi^{\mu\nu}$, $\pi_B$ and $\nu^\mu$, respectively. Including higher-order corrections or additional terms is straightforward with the method discussed below. At this point, it is important to stress that $\Delta f_i$ corrections to the distribution function are fundamentally different from the $\delta \Phi$, which means that to obtain a consistent description, one also needs to apply the background-fluctuation splitting to the $\Delta f_i$. Conceptually the $\Delta f_i$ are corrections appearing due to dissipative effects. Regardless of the dissipative fields' magnitude, they can still be split into a background with perturbations around them. This will result in $\Delta f_i$ terms proportional to $\delta \Phi$, which is of course different from $(\delta \Phi)^2$.
For the explicit form of these out-of-equilibrium corrections, we will be using the standard forms used in almost all state-of-the-art hydrodynamic simulations \cite{Denicol_2018,Paquet_2016}
\begin{align}
	\Delta f_\text{shear}&=\feq(1 \pm \feq) \frac{p_\mu p_\nu \pi^{\mu\nu}}{2 (e+p) T^2},\\
	\Delta f_\text{bulk}&=\feq (1 \pm \feq) \left[ \frac{E_p}{T}(\frac{1}{3}-c_s^2)-\frac{1}{3} \frac{m^2}{T E_p} \right] \frac{\tau_B \pi_B}{\zeta},\\
	\Delta f_\text{diff}&=\feq(1\pm\feq) \left[ \frac{n_B}{e+p} - \frac{Q_B}{E_P}\right] \frac{\nu^\mu p_\mu}{\kappa}. %- \frac{\feq}{p_{HQ}}p_\mu \nu^\mu.
\end{align} 
Here $m$ is the mass of the primary resonance, $c_s^2$ is the speed of sound, $\tau_b/\zeta$ is the ratio of the bulk relaxation time and bulk viscosity and $n_B$ and $Q_B$ are the baryon density and baryon charge respectively, together with the baryon diffusion coefficient $\kappa$ (including additional conserved charges with their respective currents can easily be done). The addition of these corrections to the distribution function also translates into the vector distribution function, which we now write as
\begin{equation}
	g^\mu = g^\mu_\text{eq} + g^{\mu\nu\rho}_\text{shear} \pi_{\nu\rho} + g^{\mu}_\text{bulk} \pi_B + g^{\mu\rho}_\text{diff} \nu_\rho.
\end{equation}
The concrete form of the correction to the vector distribution function can be obtained by comparing it with its definition, e.g., $ g^{\mu\nu\rho}_\text{shear} \pi_{\nu\rho} = \Delta f_\text{shear} p^\mu$. For the calculation of the explicit expressions, we employ the same scheme as before, using the decay operator's properties and decomposing the appearing terms under $SO(3)$. Multiple particle momenta appear when considering the correction arising from the shear stress. This leads to a term proportional to $p^\mu p^\nu p^\rho$ and even to $p^\mu p^\nu p^\rho p^\sigma$ when considering the linearisation of the shear stress correction. The decompositions of these higher order terms can also be obtained by substituting $p^\mu =\tp^\mu + \hp^\mu$. In these higher order cases, some terms, such as $\hp^\mu \hp^\nu$ arise, which are not directly irreducible. This can be mended by removing the trace from these terms (symmetric and traceless tensors are irreducible representations of $SO(3)$). The full tensor decompositions for $p^\mu$, $p^\mu p^\nu$, $p^\mu p^\nu p^\rho$ and $p^\mu p^\nu p^\rho p^\sigma$ are given in \aref{appendix_tensor_decomposition}. With these decompositions at hand, we can now come back to the inclusion of additional terms, which now is only a matter of the tensor structure: 

When comparing contributions from the equilibrium and the bulk part, we notice that they both can be written as $j p^\mu$ where the factor $j$ contains the distribution functions and numerical prefactors, such as the $\tau_B/\zeta$, etc., of the bulk correction term. Since they share the same tensor structure $p^\mu$, both contributions share the same tensor decomposition and, therefore, will have the same kernel expressions, except for the distribution functions. The same, of course, also holds for terms that are proportional to a different tensor in particle momenta (exceptions can arise when considering fields with different symmetries, such as $\pi^{\mu\nu}$, where $\bpi^{\mu\nu}\bu_\mu=0$, but $\delta \pi^{\mu\nu}\bu_\mu\neq 0$). The expressions for the applied tensor decompositions and the definitions for the distribution functions and transformation rules are given in \aref{appendix_spectra_trafo_rules_distrs} for the aforementioned corrections and their linearisations.

The final hadron spectrum can then be written as
\begin{equation}
	\frac{\ud N}{2\pi p_T \ud p_T \ud \eta_P} = \frac{\ud \Bar{N}}{2\pi p_T \ud p_T \ud \eta_P} +\frac{\delta \ud N}{ p_T \ud p_T \ud \phi_P \ud \eta_P}
\end{equation}
where we again perform a split into a symmetric background and a fluctuation part \cite{Floerchinger_2019,Floerchinger:2013rya}. Formally both terms are integrations over the full freeze-out surface. However, we can precompute the azimuthal and rapidity integrations, which leaves us with only the integration in the $\tau-r$-plane to be concretely done. This lets us write for the background
\begin{align}
	&\frac{\ud \Bar{N}}{2\pi p_T \ud p_T \ud \eta_P} = \frac{\nu}{(2\pi)^3} \int_0^1 \ud c  \tau (c) r (c) \nonumber \\
	&\left( \frac{\partial r}{\partial c} \left[ K_1^\text{eq} + \frac{\bpi^{22}}{2(e+p)T^2}K_1^\text{shear} +\frac{\bpi^{33}}{2(e+p)T^2}K_3^\text{shear}\right. \right.  \nonumber\\
 & \left. \left. +\bpi_B K_1^\text{bulk} +\bgamma \bnu K_1^\text{diff} \right]\right. \nonumber\\
 &\left. - \frac{\partial \tau}{\partial c} \left[  K_2^\text{eq} + \frac{\bpi^{22}}{2(e+p)T^2}K_2^\text{shear} +\frac{\bpi^{33}}{2(e+p)T^2}K_4^\text{shear}\right. \right. \nonumber\\
 &\left. \left. +\bpi_B K_2^\text{bulk} + \bgamma \bnu K_2^\text{diff}  \right] \right).
\end{align}
For the perturbation fields, we introduce a decomposition into azimuthal and longitudinal Fourier modes
\begin{equation}
    \delta \Phi = \sum_{m=-\infty}^\infty \int_{-\infty}^{\infty} \frac{\ud k}{2\pi} \; \delta \Phi_{m,k} e^{im\phi+ik\eta}.
\end{equation}
The $\phi \to \phi +2 \pi$ symmetry ensures that the azimuthal quantum number $m$ can only appear in integer values $m \in \mathbb{Z}$. Omitting the dependence of the perturbation fields on the quantum numbers, we can write the perturbation spectrum as
\begin{align}
    \frac{\delta \ud N}{ p_T \ud p_T \ud \phi_P \ud \eta_P} &\propto \frac{\partial f_\text{eq}}{\partial \Phi_i} \delta \Phi_i+\frac{\partial \Delta f_\text{shear}}{\partial \Phi_i} \delta \Phi_i\nonumber\\
    &+\frac{\partial \Delta f_\text{bulk}}{\partial \Phi_i} \delta \Phi_i+\frac{\partial \Delta f_\text{diff}}{\partial \Phi_i} \delta \Phi_i, \label{eq:shorthand}
\end{align}
where we have introduced a shorthand notation for the appearing kernel expressions. The individual terms are then expressed in terms of kernels $K_i^j$. The full expression for the perturbation spectrum is given in \aref{sec_analytic_res_kernel_pert}. Note that the kernels $K_i^j$ only depend on the fluid velocity $\bv$ and the transverse particle momentum $p_T$ and can be precomputed and stored. The explicit expressions for the kernels are given in \aref{sec_analytic_res_kernel_bg} for the background and in \aref{sec_analytic_res_kernel_pert} for the perturbations. In the case of omitting resonance decays for the calculation of hadron spectra, we are able to give closed, analytical expressions for the angular and rapidity integrations on the freeze-out surface, presented in the next chapter.

\section{Analytic expression for thermal spectra} \label{sec_thermal_spec}

This section will present analytic expressions for the hadron spectrum when omitting resonance decays. This is useful for consistency checks as well as for particles (such as the $\phi$ meson) that are not produced through resonance decays. We are roughly following ref. \cite{Floerchinger_2014}.

In the case of purely thermal spectra, the above description still holds with the simplification that the decay operator does not need to be applied. This implies that all distribution functions are the same $f_s=f$, eliminating many appearing terms. The hadron spectrum with viscous corrections is then given by
\begin{widetext}
\begin{align}
    \frac{\ud N}{p_T \ud p_T \ud \phi_P \ud \eta_P} &= \frac{\nu}{(2\pi)^3}\int \ud c \ud \phi \ud \eta \tau(c) r(c)\;\left( \frac{\partial r}{\partial c} m_T \cosh(\eta_P-\eta) - \frac{\partial \tau}{\partial c} p_T \cos(\phi_P-\phi) \right) \sum_{j=0}^\infty \left[ 1+ (1+j)\frac{p_\mu p^\nu \pi^{\mu\nu}}{2(e+p)T^2} \right.\nonumber\\
    &\left.+ (1+j)\left(\beta E_p (\frac{1}{3}-c_s^2) - \frac{1}{3} \frac{\beta m^2}{E_p} \right) \frac{\tau_B \pi_B}{\zeta} +(1+j)\left(\frac{n_b}{e+p}-\frac{Q_B}{E_p}\right) \frac{p_\mu \nu^\mu}{\kappa}\right] (\pm 1)^j e^{(1+j)(\beta p_\mu u^\mu + \alpha)},
\end{align}
\end{widetext}
where we already expanded the distribution function $f$ in terms of Boltzmann weights $\exp \left[ \beta p_\nu u^\nu +\alpha \right]$ for a later step in the calculation. Applying the background fluctuation splitting, we can write the background spectrum as
\begin{align}
    \frac{\ud \Bar{N}}{2\pi p_T \ud p_T  \ud \eta_P} &= \frac{\nu}{2\pi^2}\int \ud c \tau(c) r(c) \sum_{j=0}^\infty (\pm 1)^j e^{(1+j)\alpha}\nonumber\\
    &\times\left[ \frac{\partial r}{\partial c} m_T \Bar{\Phi}_i \Bar{Y}^a_i - \frac{\partial \tau}{\partial c} p_T \Bar{\Phi}_i \Bar{Y}^b_i \right],
\end{align}
and the perturbation spectrum as
\begin{align}
    \frac{\ud \Bar{N}}{ p_T \ud p_T \ud \phi_P \ud \eta_P} &= \frac{\nu}{2\pi^2}\int \ud c \tau(c) r(c) \sum_{j=0}^\infty (\pm 1)^j e^{(1+j)\alpha}\nonumber\\
    &\times(1+j)\left[ \frac{\partial r}{\partial c} m_T \delta\Phi_i Y^a_i - \frac{\partial \tau}{\partial c} p_T \delta\Phi_i Y^b_i \right],
\end{align}
with an implied summation over the index $i$. We again employ a similar kernel structure as before, with $\Bar{Y}$ being the background kernels and $Y$ being the perturbation kernels. In this case, most of the kernels are given by analytic expressions in terms of Bessel functions. The explicit expressions are given in \aref{sec_thermal_bg_ker} and \aref{sec_thermal_pert_ker}. The calculation of these expressions is straightforward after the realization that the $\phi$ and $\eta$ integrations factorize for each term, resulting in an integration of a polynomial in trigonometric or hyperbolic functions times the exponential Boltzmann factor. How to express these integrations in terms of Bessel functions is shown in \aref{sec_integrations}. Note that the contributions from the bulk and diffusion correction terms are an exception, which cannot be fully expressed in Bessel functions due to the appearing $1/E_p$. This gives rise to a contribution proportional to $1/(p_T \bgamma \cos(\phi_P-\phi)-m_T \bgamma \bv \cosh(\eta_P-\eta))$ which cannot be expressed in terms of Bessel functions. However, the bulk and diffusion kernels can still be expressed in terms of Bessel functions using a series expansion of the $1/(E_p/m)$ term, which we will now demonstrate for the bulk correction:

Starting from the problematic part $\beta m^2/(3E_p)$ of the bulk correction term, we employ a Taylor expansion of $1/x$ with $x=E_p/m$. This expansion is valid for $|x-1|<1$, which, at midrapidity, can be rewritten as $p_T < \sqrt{3}m$. Together with the binomial theorem, we can express this term in polynomials of cosine and hyperbolic cosine
\begin{align}
    \frac{\beta m^2}{3 E_p} &=\frac{\beta m}{3} \sum_{n=0}^\infty \sum_{k=0}^n \sum_{l=0}^{n-l} (-1)^{n+k+l} \frac{n!}{k!l!(n-k-l)!}\nonumber\\
    &\times (m_T/m \bgamma \bv \cosh(\eta_P-\eta))^{n-k-l}\nonumber\\
    &\times (p_T/m \bgamma \cos(\phi_P-\phi))^l.
\end{align}
This way of calculating the spectra has the benefit that it can be included directly into the simulation without the need for outside input, as in the case with resonances above. This of course can only be done when neglecting resonance decays, which, for most particles, is not feasible for comparing with real data, as we discuss in the next section.

\section{Comparison of different resonance lists} \label{sec_resonanceListComp}
When considering the spectra after resonance decays, a crucial ingredient is the set of resonances considered. Over time, more resonances get discovered, and more and more information about their decay, such as branching ratios, is collected. A classification of the information content and quality of resonance data is given by the number of stars in the particle data group (PDG) booklet \cite{10.1093/ptep/ptaa104}. A resonance with four stars is very well known in their grading system, whereas a resonance with only one star is not known well. In the following, we compare two different PDG lists, one from the listing from 2005, in the following referred to as PDG2005, and one from 2016, referred to as PDG2016. In both cases, we take particles with two to four stars into account. We include also the lesser-known resonances with only one star in the list PDG2016+. To see if so far undetected resonances could improve current fits, we also add a list that includes the PDG2016+ list together with resonances predicted based on the quark model. This list will be referred to as QM2016+. All lists are taken from refs. \cite{Alba:2017hhe,Alba:2017mqu,Alba:2020jir}.

Using these resonance lists we compare the pion spectra obtained using the hydro framework Fluid$u$M \cite{Floerchinger_2019} in the $0-1\%$ centrality class. To have a consistent comparison, we adjust the normalization for the initial profiles taken from $\textsc{T}_\textsc{R}\textsc{ENTo}$ \cite{Moreland:2014oya}, such that all the resonance lists give the same integrated multiplicity.

\begin{figure}[t]
    \centering
    \includegraphics[width=.48\textwidth]{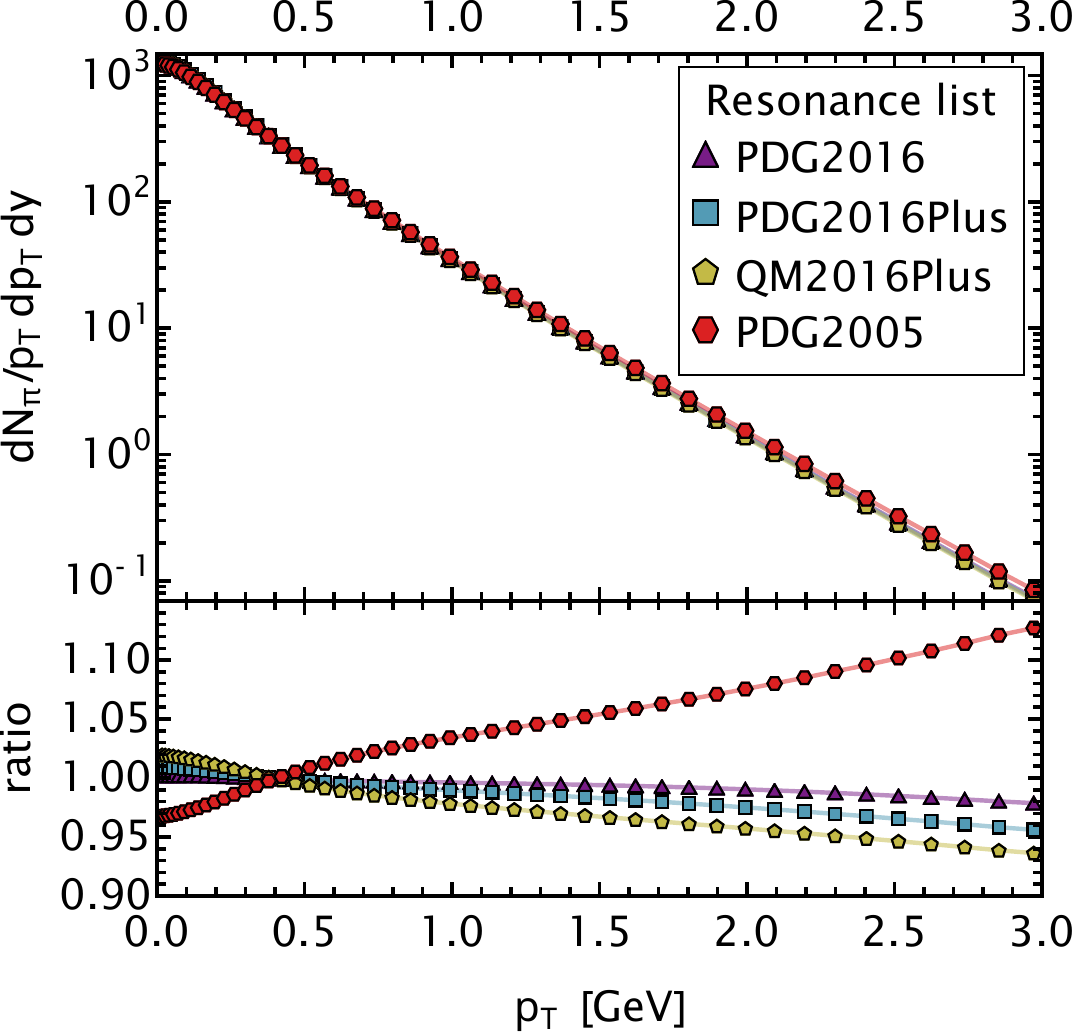}
    \caption{Effect of different resonance lists on the final pion spectrum after decays. The ratio shown is taken between the individual result of each decay list to the average of all four results.}
    \label{fig:PionResSpectra}
\end{figure}

\begin{figure}
    \centering
    \includegraphics[width=.48 \textwidth]{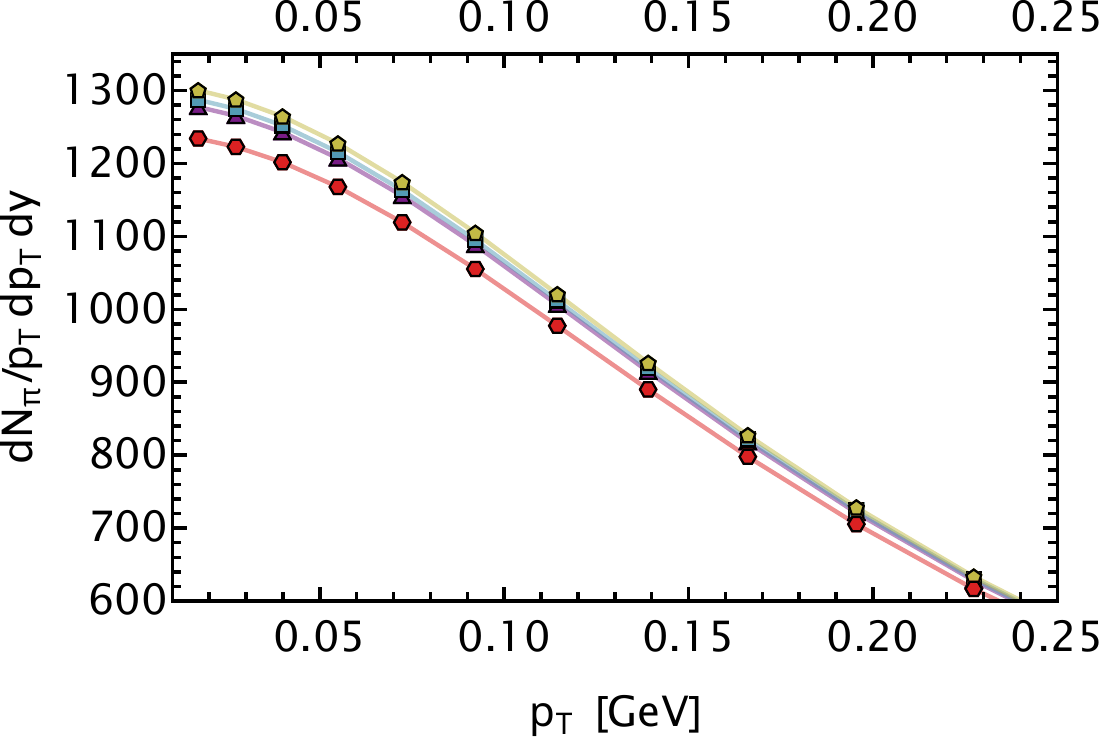}
    \caption{Zoom into the low momentum part of the pion spectra for different resonance lists. Including resonances with higher mass (yellow and blue markers) increases the pion yield at low momenta.}
    \label{fig:pion_zoom}
\end{figure}

We notice that including more resonances increases the number of pions at low momentum, as seen in \autoref{fig:PionResSpectra} and \autoref{fig:pion_zoom}. Including all measured PDG states and the predictions from the quark model results in an increase of $2 \%$ for the lowest momentum pions compared to the PDG2016 list. Comparing with fit results in the same centrality class ($0-5\% $), the ratio of data over model approaches values of around $1.4$ for the low momentum pions \cite{Devetak_2020,Nijs_2021}. With the current state of the PDG list and predictions from the quark model, we can conclude that heavy, so far poorly or unknown resonances will only have a minor effect in solving the low-$p_T$ pion puzzle.% Another possibility of modifying the hadron spectra is the inclusion of new physics into the description as a new stage in evolution. In the next section, we will investigate this by introducing a new stage, the so-called phase of partial chemical equilibrium.

\section{Partial chemical equilibrium} \label{sec_PCE}
In the above-presented work, we implicitly assumed a sudden freeze-out from a fluid to free streaming hadronic resonances which subsequently decay. This sudden transition is the kinetic freeze-out. We want to stress that the setup is general enough to allow for a scenario where chemical equilibrium is lost already earlier. This generalization is the subject of the present section.

From recent fits to data, it is clear that a two-stage freeze-out, first a chemical, then a kinetic one, plays an important role \cite{Nijs_2021}. One of the most commonly used ways of accounting for these two freeze-outs is the introduction of a hadronic rescattering phase \cite{SMASH:2016zqf,Bleicher_1999}. Another approach is to extend the fluid dynamic description to include both freeze-outs, employing a phase with a partial chemical equilibrium (PCE) \cite{BEBIE199295,Huovinen_2008} between the two freeze-outs. This extension of the hydrodynamic phase has the conceptual benefit of being more universal than kinetic theory since it does not rely on scattering cross-sections and is computationally less costly than a full transport simulation. Similarly to the instantaneous freeze-out from above, we assume that the chemical composition is frozen out at a fixed temperature $T_{chem}$; but we do not assume that the kinetic distributions also freeze out at this temperature. This is equivalent to the assumption that the density of the fireball is for temperatures below the chemical freeze-out, but above the kinetic freeze-out temperature, low enough such that inelastic collisions lose efficiency to change the chemical composition, but elastic collisions are still effective in maintaining kinetic equilibrium. The loss of chemical equilibrium entails building chemical potentials associated with conserved charges. Due to the possibility of particle decays, the conserved charges are not directly the particle numbers of each particle species but rather effectively conserved particle numbers corresponding to linear superpositions. The conserved effective particle numbers are given by $\Bar{N}_i=N_i + \sum_j b_j^i N_j$, where $N_i$ is the actual particle number of species $i$ and $b_j^i$ is the number of particles of type $i$ created in the decay $i\to j$, including branching ratios. The summation here goes over all particles and resonances $j$ with a lifetime smaller than the characteristic lifetime of the system, which we assume to be $10 fm/c$. The particles with a lifetime larger than the characteristic lifetime of the fireball are $\pi ,\, K , \, \eta ,\, \omega ,\, p ,\, n ,\, \eta' ,\, \phi ,\, \Lambda ,\,\Sigma ,\, \Xi ,\, \Lambda(1520) ,\, \Xi(1530) ,\, \Omega$ together with their corresponding antiparticles. The chemical potential for any resonance can be obtained via $\mu_j = \sum_i b_j^i \mu_i$ where we sum over all decay products $i$ of a resonance $j$.

For the following description of the concrete calculation, we will stick to the leading order of the hydrodynamic gradient expansion. The following paragraph will clarify this statement. 

Using energy-momentum conservation together with the Gibbs-Duhem relation, one finds as evolution equation for the entropy density current
\begin{align}
    \partial_\mu (s u^\mu) = -\alpha \partial_\mu (\Bar{n} u^\mu), \label{eq:entropy_current}
\end{align}
with $\alpha$ being the corresponding chemical potential to the conserved particle number $\bar{n}$. Note that this equation has been derived using the ideal form of the energy-momentum tensor, which we referred to as the leading order of the hydrodynamic gradient expansion before. Including higher-order terms of the expansion, i.e., viscous corrections, would result in additional terms on the left-hand side proportional to the viscous fields and their derivatives. In this ideal approximation, the equations of motion governing the evolution of the chemical potential can then be computed independently from the history of the collision. If we were to include viscous corrections to this stage of the collision, the relation given below would be modified by integrations over the full history of the viscous fields during the collision. In the ideal approximation the chemical potential depends only on the two freeze-out temperatures $T_{chem}$ and $T_{kin}$. It can be incorporated easily into the kernel calculation for the spectra described above. Since the left-hand side of \autoref{eq:entropy_current} is zero by construction, we can obtain the chemical potential $\mu$ at a given temperature $T<T_{chem}$ via the implicit relation $\Bar{n}_i/s(T_{chem},0)=\Bar{n}_i/s(T,\mu)$.

After further expansion, when the system is at $T_{kin}$, the efficiency of the elastic collisions has dropped so far that also the kinetic distributions freeze out. The further evolution of the system is now governed by free streaming, and the decay of resonances.
\section{Conclusion} \label{sec_conclusion}
We presented and discussed different aspects of the Cooper-Frye freeze-out prescription. We started with a discussion of the relation between the actual freeze-out surface employed in practical calculations, such as at fixed temperature, to integrals over Cauchy surfaces and to the asymptotic regime at very late times where particles are detected experimentally. We have shown in \autoref{sec_Cooper_Frye}, that the usual prescription indeed leads to consistent results, despite the fact that the freeze-out integral can have negative contributions if it is not a Cauchy hypersurface. The argument is based on an infinite set of conserved currents for free streaming dynamics. 

Afterwards we presented a method to calculate the final hadron spectra after feed down from resonance decays directly from integrals over the initial freeze-out surface (\autoref{sec_FastResoIdea}). In \autoref{sec_fieldConventions} and \autoref{sec_modeByMode}, we showed how this method can be applied to a setting using relativistic fluid dynamics and a mode expansion scheme. Here the major benefits of this method become apparent: Assuming a symmetric freeze-out hypersurface, the numerically costly integrations over azimuthal and rapidity angles can be precomputed since they only depend on the temperature with which the freeze-out hypersurface is defined, as well as chemical potentials, the fluid velocity and particle momentum.

Moreover, we also presented closed expressions that can be used to calculate the spectra of thermally produced particles, before resonance decays with azimuthal and rapidity angle integrations already performed in terms of Bessel functions in \autoref{sec_thermal_spec}. Subsequently, we demonstrated in \autoref{sec_resonanceListComp} that the inclusion of heavy resonances, predicted by the quark model, increases the yield of pions at low momenta slightly but not enough to explain the pion excess at low momenta when comparing theory predictions to experimental data. Finally, we discussed how to include a phase with partial chemical equilibrium that separates the chemical from the kinetic freeze-out. The enhancement of the chemical potentials in this regime is driven by the conservation of effective particle numbers with respect to resonance decays.

The corrections $\Delta f$ to phase space distribution functions due to dissipative terms in the fluid used in this work have been obtained in the standard linear scheme. In future work, it would be interesting to address how the distribution function changes at non-linear order in dissipative corrections, and to determine the effects on the particle spectra. Another point of interest for future works is to directly calculate the two-point function, including resonance decays with the above-presented methods. The previously described results are, of course, sufficient for the calculation of the two-point function assuming linearized fluid dynamics. Additionally, one further topic for research is the extension of the partial chemical equilibrium to the next order in gradients in fluid velocity.

%One of the central questions occurring to the authors during this work was how different (conserved) currents interact within the medium. A more precise formulation of this question can be given by asking how the different $\Delta f$ correction terms change when including more currents and the interaction between the currents. 

\section*{Acknowledgment}
The authors wish to thank A. Mazeliauskas, F. Capellino and K. Reygers for useful discussions. 
This work is part of and supported by the DFG Collaborative Research Centre SFB 1225 (ISOQUANT).

\appendix 
\section{Hydrodynamic fields in the general Bjorken coordinate frame} \label{sec_transformed_fields}
In this section we give the expressions for the fluid velocity, shear stress tensor and diffusion current transformed back to Bjorken coordinates. The transformations from the orthogonal frame to Bjorken coordinates are given by
\begin{align}
    u^\mu &= v^\mu_a u^a,\\
    \nu^\mu &= v^\mu_a \nu^a,\\
    \pi^{\mu\rho} &= v^\mu_a v^\rho_b \pi^{ab}.
\end{align}
The full expressions for the background fluid velocity and diffusion current together with their perturbations read as
\begin{align}
    u^\mu &= (\bgamma , \bv\bgamma,0,0)+(\bv\bgamma \delta v_1, \bgamma \delta v_1, \delta v_2/r, \delta v_3/\tau) \\
    \nu^\mu &= (\bv\bgamma \bnu, \bgamma\bnu,0,0)\nonumber\\
    &+(\bgamma(\bv\delta \nu_1-\bnu \delta v_1),\bgamma(\delta \nu_1-\bv\bnu \delta v_1),\delta \nu_2/r,\delta \nu_3/\tau)
\end{align}
Note that the diffusion current now also depends on the background fluid velocity. Similarly the shear stress tensor also gains contributions from the background fluid velocity through the transformation and is given by
    \begin{widetext}
\begin{align}
    \pi^{\mu\nu} &= \begin{pmatrix}
        -\bv^2 \bgamma^2 \Bar{\pi}^t & -\bv \bgamma^2 \Bar{\pi}^t & 0 & 0 \\
        -\bv \bgamma^2 \Bar{\pi}^t & -\bgamma^2 \Bar{\pi}^t & 0 & 0 \\
        0 & 0 & \Bar{\pi}^{22}/r^2 & 0 \\
        0 & 0 & 0 & \Bar{\pi}^{33}/\tau^2
    \end{pmatrix} \\
    &+ \begin{pmatrix}
        -\bv\bgamma^2 (\bv \delta \pi^t - 2 \delta v_1 \Bar{\pi}^t) & -\bgamma (\bv \delta \pi^t - \delta v_1 (1+\bv^2)\Bar{\pi}^t) & \bgamma(\bv\delta \pi^{12} -\delta v_2 \Bar{\pi}^{22})/r & \bgamma(\bv \delta \pi^{13}-\delta v_3 \Bar{\pi}^{33})/\tau  \\
        -\bgamma (\bv \delta \pi^t - \delta v_1 (1+\bv^2)\Bar{\pi}^t) & -\bgamma^2(\delta \pi^t - 2 \delta v_1 \bv \Bar{\pi}^t ) & \bgamma(\delta \pi^{12} -\delta v_2 \bv \Bar{\pi}^{22})/r & \bgamma ( \delta \pi^{13} -\bv \delta v_3 \Bar{\pi}^{33})/\tau \\
        \bgamma(\bv\delta \pi^{12} -\delta v_2 \Bar{\pi}^{22})/r & \bgamma( \delta \pi^{12}-\delta v_2 \bv \Bar{\pi}^{22})/\tau & \delta \pi^{22}/r^2 &  \delta \pi^{23}/r\tau \\
         \bgamma(\bv \delta \pi^{13}-\delta v_3 \Bar{\pi}^{33})/\tau &  \bgamma( \delta \pi^{13}-\delta v_3 \bv \Bar{\pi}^{33})/\tau & \delta \pi^{23}/r\tau & \delta \pi^{33}/\tau^2
    \end{pmatrix}.
\end{align}
\end{widetext}
Take note of that this transformation is not strictly needed. Another possibility would be to set up the fluid dynamic simulation in the orthogonal frame where the components of the tetrads are promoted to independent fluid fields with their own evolution equations.

\section{Tensor decomposition} \label{appendix_tensor_decomposition}
In the following we present the tensor decompositions needed for the calculation of the kernels presented in \autoref{sec_modeByMode}, for tensors of up to four Lorentz indices, i.e. $p^\mu$, $p^\mu p^\nu$, $p^\mu p^\nu p^\rho$ and $p^\mu p^\nu p^\rho p^\sigma $. Since our tensors are fully symmetric in Lorentz indices, we will only give the symmetric terms of the decomposition. 

\subsection{Decomposition of $p^\mu $}
The decomposition of $p^\mu$ is given by one scalar 
\begin{align}
    \tp^\mu = E_p u^\mu
\end{align}
and one vector
\begin{align}
    \hp^\mu = p^\mu -E_p u^\mu .
\end{align}
\subsection{Decomposition of $p^\mu p^\nu $}
The decomposition of $p^\mu p^\nu$ is given by two scalars
\begin{align}
    &1.) \tp^\mu \tp^\nu \\
    &2.) \frac{1}{3}|\mathbf{p}|^2 \Delta^{\mu\nu},
\end{align}
two vectors
\begin{align}
    &1.)\hp^\mu \tp^\nu\\
    &2.)\tp^\mu \hp^\nu,
\end{align}
and one traceless, symmetric tensor
\begin{align}
    \hp^\mu \hp^\nu -\frac{1}{3} |\mathbf{p}|^2 \Delta^{\mu\nu}.
\end{align}
\subsection{Decomposition of $p^\mu p^\nu p^\rho $}
The decomposition of $p^\mu p^\nu p^\rho $ is given by four scalars
\begin{align}
&1.) \; \tp^\mu \tp^\nu \tp^\rho \\
&2.) \;\frac{1}{3}|\mathbf{p}|^2 \tp^\mu \Delta^{\nu \rho} \text{ + 2 permutations},
\end{align}
six vectors
\begin{align}
	&1.) \;\tp^\mu \tp^\nu \hp^\rho \text{ + 2 permutations}\\
	&2.) \;\frac{1}{5}|\mathbf{p}|^2 \Delta^{\mu \nu} \hp^\rho \text{ + 2 permutations},
\end{align}
three symmetric and traceless two tensors
\begin{align}
	&1.)\; \tp^\mu (\hp^\nu \hp^\rho -\frac{1}{3}|\mathbf{p}|^2\Delta^{\nu \rho}) \text{ + 2 permutations},
\end{align}
and one symmetric and traceless three tensor
\begin{align}
	\hp^\mu \hp^\nu \hp^\rho -\frac{1}{5} |\mathbf{p}|^2 (\hp^\mu \Delta^{\nu\rho} +\hp^\nu \Delta^{\mu \rho}+ \hp^\rho \Delta^{\mu\nu}).
\end{align}
\subsection{Decomposition of $p^\mu p^\nu p^\rho p^\sigma$}
The decomposition of $p^\mu p^\nu p^\rho p^\sigma$ is given by eight scalars
\begin{align}
	&1.)\; \tp^\mu \tp^\nu \tp^\rho \tp^\sigma\\
	&2.)\; \frac{1}{3}|\mathbf{p}|^2\Delta^{\mu \nu} \tp^\rho \tp^\sigma\text{ + 5 permutations} \\
	&3.)\;\frac{1}{15} |\mathbf{p}|^4(\Delta^{\mu\nu}\Delta^{\rho\sigma}+\Delta^{\mu\rho}\Delta^{\nu\sigma}+\Delta^{\mu\sigma}\Delta^{\nu\rho}),
\end{align}
eight vectors
\begin{align}
	&1.)\; \hp^\mu\tp^\nu\tp^\rho\tp^\sigma\text{ + 3 permutations} \\ 
	&2.)\; \frac{1}{5}|\mathbf{p}|^2 (\Delta^{\mu\nu}\hp^\rho + \Delta^{\mu\rho}\hp^\nu +\Delta^{\nu\rho} \hp^\rho)\tp^\sigma \\
 &\text{ + 3 permutations},
\end{align}
twelve symmetric and traceless two tensors
\begin{align}
	&1.)\; (\hp^\mu \hp^\nu -\frac{1}{3} |\mathbf{p}|^2 \Delta^{\mu\nu})\tp^\rho \tp^\sigma\text{ + 5 permutations}\\
	&2.)\;\frac{1}{7} (\hp^\mu \hp^\nu -\frac{1}{3} |\mathbf{p}|^2 \Delta^{\mu\nu})\Delta^{\rho\sigma}\text{ + 5 permutations},
\end{align}
four symmetric and traceless three tensors
\begin{align}
	&1.)\; (\hp^\mu \hp^\nu \hp^\rho - \frac{1}{5}|\mathbf{p}|^2(\Delta^{\mu\nu}\hp^\rho+\Delta^{\mu\rho}p^\nu+\Delta^{\nu \rho}p^\mu))\tp^\sigma\\
 &\text{ + 3 permutations},
\end{align}
and one symmetric and traceless four tensor
\begin{align}
	&\hp^{\mu}\hp^{\nu}\hp^{\sigma} \hp^\rho+ \frac{1}{35}|\mathbf{p}|^4 (\Delta^{\mu\nu}\Delta^{\rho \sigma}+\Delta^{\mu\rho}\Delta^{\nu \sigma}+\Delta^{\mu\sigma}\Delta^{\rho \nu} )\\
	 &-\frac{1}{7} [ \Delta^{\mu\nu} | \mathbf{p}|^2 \hp^{\rho}\hp^{\sigma } +
	\Delta^{\mu\rho} | \mathbf{p}|^2 \hp^{\nu}\hp^{\sigma } +\Delta^{\mu\sigma} |\mathbf{p}|^2 \hp^{\rho}\hp^{\nu }\\
 &+\Delta^{\nu \rho} | \mathbf{p}|^2 \hp^{\mu}\hp^{\sigma }+ \Delta^{\nu \sigma} | \mathbf{p}|^2 \hp^{\mu}\hp^{\rho }
	+\Delta^{ \rho \sigma} | \mathbf{p}|^2 \hp^{\mu}\hp^{\nu }
	]\\
 &=\hp^{\mu}\hp^{\nu}\hp^{\sigma} \hp^\rho -[ \frac{1}{7} \Delta^{\mu\nu} |\mathbf{p}|^2(\hp^\rho\hp^\sigma -\frac{1}{3} |\mathbf{p}|^2 \Delta^{\rho\sigma}) + \text{5 perms}] \\
 &+ \frac{1}{15}|\mathbf{p}|^4 \Delta^{\mu\nu} \Delta^{\rho\sigma} +  \text{2 perms}. 
\end{align}

\section{Distribution functions and transformation rules} \label{appendix_spectra_trafo_rules_distrs}
In this section we will give all the expressions to initialize the distribution functions on the freeze-out hypersurface for the thermal part and the transformation functions $A_i (w)$ for the decay map. The distribution functions can be obtained by factoring out the tensor and constant fluid field contributions. Since the contributions are all additive, the vector distribution function after a two-body decay is given by
\begin{equation}
    g^\mu_{b,i} (p,u) = \frac{\nu_a}{\nu_b} \int \frac{\ud k^3}{(2\pi)^3 2 E_k} D^a_{b|c} (p^\nu k_\nu) g^\mu_{a,i}(k,u) \label{eq_transformation_startpoint}
\end{equation}
for a decay $a \to b +c$ and a given contribution type $i$. Using the orthogonality of the different, irreducible representations of $SO(3)$, we can simplify this transformation rule into a scalar integral
\begin{equation}
    f^b_{i,s} (E_p) = B \frac{\nu_a}{\nu_b} \frac{m_a^2}{m_b^2} \frac{1}{2} \int_{-1}^{1} \ud w \; A_s(w) f_{i,s}^a (E(w)) \label{eq_transformation_endpoint}
\end{equation}
as seen before for contribution type $i$ and spin $s$ of the irreducible representation.
For the equilibrium part of the distribution function and its linearizations, we find for the distribution functions
\begin{align}
    f_{s}^\text{eq}&=\feq \\
    f_{s}^\text{eqTemp}&= \feq (1\pm \feq) (\beta E_p-\alpha) \\
    f_{s}^\text{eqChem}&=  \feq(1\pm \feq)  \\
    f_{s}^\text{eqVel}&= \feq(1\pm \feq).
\end{align}
Since the background equilibrium part and its perturbations in temperature and chemical potential are all proportional to $p^\mu$, they share the same decomposition with the same terms appearing in the final spectra expressions. Therefore they also share the same transformation rules
\begin{align}
    A_0^\text{eq} &= A_0^\text{eqTemp} = A_0^\text{eqChem} = \frac{Q(w)}{|\mathbf{p}|},\\
    A_1^\text{eq} &= A_1^\text{eqTemp} = A_1^\text{eqChem} = \frac{E(w)}{E_p},\\
    A_0^\text{eqVel}&= \frac{E(w)^2-m_a^2}{|\mathbf{p}|^2}, \quad A_1^\text{eqVel}=A_0^\text{eq} A_1^\text{eq}, \\
    A_2^\text{eqVel}&= \frac{3}{2} \frac{Q(w)^2}{|\mathbf{p}|^2} - \frac{1}{2} \frac{E(w)^2-m_a^2}{|\mathbf{p}|^2}.
\end{align}
To obtain these transformation rules, in principle one would need to carry out the integration \autoref{eq_transformation_startpoint} for each set of contributions with the same spin to arrive at \autoref{eq_transformation_endpoint}. However, this way of obtaining the transformation rules is very tedious. A more convenient way of obtaining the transformation rules is to start again from \autoref{eq_transformation_startpoint} with the decomposition of $g^\mu$ already applied. To obtain a scalar integration, we contract both sides with the appropriate amount of momenta $p^\mu p^\nu \ldots$. The appearing possible contractions are given by
\begin{align}
    p_\mu \Delta^{\mu\nu} p_\nu &= |\mathbf{p}|^2, \quad k_\mu \Delta^{\mu\nu} k_\nu = E(w)^2-m_a^2, \\
    p_\mu \Delta^{\mu\nu} k_\nu &= Q(w)|\mathbf{p}|.
\end{align}
Since all the appearing quantities only depend on $w$, the integration over the decay operator now is trivial and can be brought into the required form of \autoref{eq_transformation_endpoint}.
The distribution functions for terms proportional to the shear stress tensor are given by
\begin{align}
    f_s^\text{shear}&= \feq (1 \pm \feq) \\
    f_s^\text{shearTemp}&=\feq( 1 \pm \feq) \left( (2 \pm \feq )(\beta E_p-\alpha)\right. \nonumber\\
    &\left.+\beta \partial_\beta \ln \left(\frac{e+p}{\beta^2} \right)\right) \\
    f_s^\text{shearChem}&= \feq (1 \pm \feq)\left( (2 \pm \feq)\right. \nonumber \\
    &\left.-\partial_\alpha \ln(e+p)\right)\\
    f_s^\text{shearVel}&= \feq (1 \pm \feq) (2 \pm \feq)\\
    f_s^\text{shearShear}&= \feq(1 \pm \feq)
\end{align}
Similar to the equilibrium case, the background shear part and its perturbations in temperature and chemical potential share the same tensor structure with the same contractions. Therefore their transformation rules are also the same
\begin{align}
    A_0^\text{shear} &= A_0^\text{shearTemp}= A_0^\text{shearChem} = 0,\\
    A_1^\text{shear} &= A_1^\text{shearTemp}= A_1^\text{shearChem} = A_0^\text{eqVel} A_0^\text{eq},\\
    A_2^\text{shear} &= A_2^\text{shearTemp}= A_2^\text{shearChem} = A_2^\text{eqVel} A_1^\text{eq},\\
    A_3^\text{shear} &= A_3^\text{shearTemp}= A_3^\text{shearChem}\nonumber\\
    &= \frac{5}{2} \frac{Q(w)^3}{|\mathbf{p}|^3} - \frac{3}{2} \frac{Q(w)}{|\mathbf{p}|}\frac{E(w)^2-m_a^2}{|\mathbf{p}|^2},\\
    A_0^\text{shearShear}&=\frac{E(w)}{E_p} \frac{E_p^2 E(w)^2 + \map^2(E(w)^2-m_a^2) }{E_p^4+\map^4}\\
    A_1^\text{shearShear}&= \frac{Q(w)}{\map} \frac{5E_p^2 E(w)^2 + \map^2 (E(w)^2-m_a^2)}{5 E_p^4 + \map^4},\\
    A_2^\text{shearShear}&=A_1^\text{eq} A_1^\text{eqVel},\quad
    A_3^\text{shearShear}=A_3^\text{shear},\\
    A_0^\text{shearVel}&=\left( A_0^\text{eqVel}\right)^2, \quad A_1^\text{shearVel}=A_0^\text{eqVel}A_1^\text{eqVel},\\
    A_2^\text{shearVel}&=A_0^\text{eqVel}A_2^\text{eqVel}, \quad A_3^\text{shearVel}=A_1^\text{eq}A_3^\text{shear},\\
    A_4^\text{shearVel}&=\frac{35Q(w)^4-30 Q(w)^2 (E(w)^2-m_a^2)+ }{8 \map^4}\nonumber\\
    &+\frac{3(E(w)^2-m_a^2)^2}{8 \map^4}
\end{align}
The distribution functions for terms proportional to the bulk pressure are given by
\begin{align}
    f_s^\text{bulk}&=\feq(1 \pm \feq) \left( \beta E_p (1/3 -c_s^2) - \frac{1}{3} \frac{\beta m^2}{E_p } \right) \frac{\tau_B}{\zeta}\\
    f_s^\text{bulkTemp}&=\feq(1 \pm \feq) \left( \beta E_p (1/3 -c_s^2) - \frac{1}{3} \frac{\beta m^2}{E_p } \right) \frac{\tau_B}{\zeta}\nonumber\\
    &\times \left[ (2\pm \feq)(\beta E_p-\alpha )\right.\nonumber \\
    &\left. - \beta \partial_\beta \ln \left( \beta E_p (1/3 -c_s^2) - \frac{1}{3} \frac{\beta m^2}{E_p } \right)\right.\nonumber \\
    &\left. - \beta \partial_\beta \ln \left( \frac{\tau_B}{\zeta} \right) \right]\\
    f_s^\text{bulkChem}&=\feq(1 \pm \feq) \left( \beta E_p (1/3 -c_s^2) - \frac{1}{3} \frac{\beta m^2}{E_p } \right) \frac{\tau_B}{\zeta}\nonumber\\
    &\times \left[ (2\pm \feq)\right.\nonumber \\
    &\left. + \partial_\alpha \ln \left( \beta E_p (1/3 -c_s^2) - \frac{1}{3} \frac{\beta m^2}{E_p } \right)\right.\nonumber \\
    &\left.+ \partial_\alpha \ln \left( \frac{\tau_B}{\zeta} \right) \right]\\
    f_s^\text{bulkVel}&=\feq(1 \pm \feq) \left[ (2 \pm \feq)\right.\nonumber\\
    &\left.\times\left(\beta E_p (1/3 -c_s^2) - \frac{1}{3} \frac{\beta m^2}{E_p } \right)\right.\nonumber\\
    &\left.- (1/3-c_s^2)-\frac{1}{3} \frac{m^2}{E_p^2} \right] \frac{\tau_B}{\zeta}\\
    f_s^\text{bulkBulk}&=\feq(1 \pm \feq) \left( \beta E_p (1/3 -c_s^2) - \frac{1}{3} \frac{\beta m^2}{E_p } \right) \frac{\tau_B}{\zeta}.
\end{align}
In the case of bulk corrections, we can make use of the fact that most of the terms have the same tensor structure as the equilibrium contributions. The transformation rules are given by
\begin{align}
    A_s^\text{bulk}&=A_s^\text{bulkTemp}=A_s^\text{bulkChem}=A_s^\text{bulkBulk}=A_s^\text{eq},\\
    A_s^\text{bulkVel}&=A_s^\text{eqVel}.
\end{align}

For the diffusion contribution to the spectra, the distribution functions are given by
\begin{align}
    f_s^\text{diff}&= \frac{\feq (1\pm\feq)}{\kappa}\left[ \frac{n_B}{e+p} - \frac{Q_B}{E_P}\right]\\
    f_s^\text{diffTemp}&= \frac{\feq (1\pm\feq)}{\kappa}\left[ \frac{n_B}{e+p} - \frac{Q_B}{E_P}\right]\nonumber\\
    &\times \left((2\pm\feq)(\beta E_p -\alpha) - \beta \partial_\beta \ln\left( \frac{n_B}{e+p} - \frac{Q_B}{E_P}\right)\right.\nonumber\\
    &\left.+ \beta \partial_\beta \ln\left(\kappa\right) \right)\\
    f_s^\text{diffChem}&= \frac{\feq (1\pm\feq)}{\kappa}\left[ \frac{n_B}{e+p} - \frac{Q_B}{E_P}\right] \nonumber\\
    &\times \left( (1\pm\feq) - \partial_\alpha \ln\left(\frac{n_B}{e+p}\right)+\partial_\alpha \ln(\kappa)\right)\\
    f_s^\text{diffVel}&=\frac{\feq (1\pm\feq)}{\kappa} \left[(2 \pm \feq ) \left( \frac{n_B}{e+p} - \frac{Q_B}{E_P}\right)\right.\nonumber\\
    & +\left. \frac{Q_B}{E_p^2}\right]\\
    f_s^\text{diffDiff}&= \frac{\feq (1\pm\feq)}{\kappa}\left[ \frac{n_B}{e+p} - \frac{Q_B}{E_P}\right]
\end{align}
Since the diffusion current is also orthogonal to the background fluid velocity, we again can express most transformation rules in terms of previous ones
\begin{align}
    A_s^\text{diff}&=A_s^\text{DiffTemp}=A_s^\text{DiffChem}=A_s^\text{eqVel},\\
    A_0^\text{DiffVel}&= A_1^\text{eq}A_0^\text{eqVel}, \quad A_1^\text{diffVel}= A_0^\text{eq}A_0^\text{eqVel}, \\
    A_2^\text{diffVel}&= A_1^\text{eq}A_2^\text{eqVel},\quad
    A_3^\text{diffVel}= A_3^\text{shear},\\
    A_0^\text{diffDiff}&=\frac{E(w)^2(3E_p^2+\map^2)-m_a^2 \map^2}{3E_p^4+\map^4}\\
    A_1^\text{diffDiff}&=A_0^\text{eq}A_1^\text{eq}, \quad A_2^\text{diffDiff}=A_2^\text{eqVel}
\end{align}

\section{Background kernel with resonances} \label{sec_analytic_res_kernel_bg}

The kernel expressions for the rapidity and azimuthal integrations including feed down from resonance decays which can be precomputed read as
\begin{align}
    K_1^\text{eq} &=\int \ud \phi \ud \eta \; \bar{\gamma } \left(f_0^{\text{eq}}-f_1^{\text{eq}}\right) \bar{E}_p+f_1^{\text{eq}} \cosh (\eta ) m_T \\
    K_2^\text{eq} &=\int \ud \phi \ud \eta \; \bar{\gamma } \bar{v} \left(f_0^{\text{eq}}-f_1^{\text{eq}}\right) \bar{E}_p+f_1^{\text{eq}} p_T \cos (\phi ) \\
    K_1^\text{bulk} &=\int \ud \phi \ud \eta \; \bar{\gamma } \left(f_0^{\text{bulk}}-f_1^{\text{bulk}}\right) \bar{E}_p+f_1^{\text{bulk}} \cosh (\eta ) m_T \\
    K_2^\text{bulk} &=\int \ud \phi \ud \eta \;\bar{\gamma } \bar{v} \left(f_0^{\text{bulk}}-f_1^{\text{bulk}}\right) \bar{E}_p+f_1^{\text{bulk}} p_T \cos (\phi ) 
    \end{align}
    %shear kernels
    \begin{align}
    K_1^\text{shear} &=\int \ud \phi \ud \eta \;  \frac{1}{5} \bar{\gamma } \left(\bar{\gamma } \bar{v} \cosh (\eta ) m_T \left(2 p^2 \bar{v} \left(f_1^{\text{shear}}-f_3^{\text{shear}}\right)\right.\right.\nonumber\\
    &\left.\left.+10 \bar{\gamma } f_2^{\text{shear}} p_T
   \cos (\phi ) \bar{E}_p-15 \bar{\gamma }^2 \bar{v} f_3^{\text{shear}} p_T^2 \cos ^2(\phi )\right.\right.\nonumber\\
    &\left.\left.+5 \bar{v} f_3^{\text{shear}} p_T^2 \sin ^2(\phi )\right)+5 \bar{\gamma }^2 \bar{v}^2
   \cosh ^2(\eta ) m_T^2\right.\nonumber\\
    &\left.\times \left(3 \bar{\gamma } \bar{v} f_3^{\text{shear}} p_T \cos (\phi )-f_2^{\text{shear}} \bar{E}_p\right)\right.\nonumber\\
    &\left.-5 \bar{\gamma }^3 \bar{v}^4 f_3^{\text{shear}} \cosh
   ^3(\eta ) m_T^3\right.\nonumber\\
    &\left.+p_T \left(\bar{\gamma } \bar{v} \cos (\phi ) \left(2 p^2 \left(f_3^{\text{shear}}-f_1^{\text{shear}}\right)\right.\right.\right.\nonumber\\
    &\left.\left.\left.-5 f_3^{\text{shear}} p_T^2 \sin ^2(\phi )\right)-5
   \bar{\gamma }^2 f_2^{\text{shear}} p_T \cos ^2(\phi ) \bar{E}_p\right.\right.\nonumber\\
    &\left.\left.+5 f_2^{\text{shear}} p_T \sin ^2(\phi ) \bar{E}_p+5 \bar{\gamma }^3 \bar{v} f_3^{\text{shear}} p_T^2 \cos ^3(\phi
   )\right)\right) 
   \end{align}
   \begin{align}
   %K2
    K_2^\text{shear} &= \int \ud \phi \ud \eta \; \frac{1}{5} \bar{\gamma } \left(\bar{\gamma } \bar{v} \cosh (\eta ) m_T\right.\nonumber\\
    &\left.\times \left(15 \bar{\gamma }^2 f_3^{\text{shear}} p_T^2 \cos ^2(\phi )\right.\right.\nonumber\\
    &\left.\left.+10 \bar{\gamma } \bar{v} f_2^{\text{shear}}
   p_T \cos (\phi ) \bar{E}_p+2 p^2 \left(f_1^{\text{shear}}-f_3^{\text{shear}}\right)\right.\right.\nonumber\\
    &\left.\left.-5 f_3^{\text{shear}} p_T^2 \sin ^2(\phi )\right)-5 \left(\bar{\gamma }^2-1\right) \cosh ^2(\eta
   ) m_T^2\right.\nonumber\\
    &\left.\times \left(3 \bar{\gamma } f_3^{\text{shear}} p_T \cos (\phi )+\bar{v} f_2^{\text{shear}} \bar{E}_p\right)\right.\nonumber\\
    &\left.+5 \bar{\gamma }^3 \bar{v}^3 f_3^{\text{shear}} \cosh ^3(\eta )
   m_T^3+p_T \left(\bar{\gamma } \cos (\phi )\right.\right.\nonumber\\
    &\left.\left.\times \left(2 p^2 \left(f_3^{\text{shear}}-f_1^{\text{shear}}\right)+5 f_3^{\text{shear}} p_T^2 \sin ^2(\phi )\right)\right.\right.\nonumber\\
    &\left.\left.-5 \bar{\gamma }^3 f_3^{\text{shear}} p_T^2 \cos ^3(\phi )-5 \bar{\gamma }^2 \bar{v} f_2^{\text{shear}} p_T \cos ^2(\phi ) \bar{E}_p\right.\right.\nonumber\\
    &\left.\left.+5 \bar{v} f_2^{\text{shear}} p_T \sin ^2(\phi ) \bar{E}_p\right)\right) 
    \end{align}
   \begin{align}
   %K3
    K_3^\text{shear} &= \int \ud \phi \ud \eta \; \frac{1}{5} \bar{\gamma } \left(\bar{\gamma } \bar{v} \cosh (\eta ) m_T\right.\nonumber\\
    &\left.\times \left(5 \bar{v} f_3^{\text{shear}} \sinh ^2(\eta ) m_T^2+2 p^2 \bar{v}
   \left(f_1^{\text{shear}}-f_3^{\text{shear}}\right)\right.\right.\nonumber\\
    &\left.\left.+10 \bar{\gamma } f_2^{\text{shear}} p_T \cos (\phi ) \bar{E}_p-15 \bar{\gamma }^2 \bar{v} f_3^{\text{shear}} p_T^2 \cos ^2(\phi)\right)\right.\nonumber\\
    &\left.+5 \sinh ^2(\eta ) m_T^2 \left(f_2^{\text{shear}} \bar{E}_p-\bar{\gamma } \bar{v} f_3^{\text{shear}} p_T \cos (\phi )\right)\right.\nonumber\\
    &\left.+5 \left(\bar{\gamma }^2-1\right) \cosh ^2(\eta) m_T^2 \left(3 \bar{\gamma } \bar{v} f_3^{\text{shear}} p_T \cos (\phi )\right.\right.\nonumber\\
    &\left.\left.-f_2^{\text{shear}} \bar{E}_p\right)-5 \bar{\gamma }^3 \bar{v}^4 f_3^{\text{shear}} \cosh ^3(\eta )
   m_T^3\right.\nonumber\\
    &\left.+\bar{\gamma } p_T \cos (\phi ) \left(2 p^2 \bar{v} \left(f_3^{\text{shear}}-f_1^{\text{shear}}\right)\right.\right.\nonumber\\
    &\left.\left.-5 \bar{\gamma } f_2^{\text{shear}} p_T \cos (\phi ) \bar{E}_p+5
   \bar{\gamma }^2 \bar{v} f_3^{\text{shear}} p_T^2 \cos ^2(\phi )\right)\right) 
   \end{align}
   \begin{align}
   %K4
    K_4^\text{shear} &= \int \ud \phi \ud \eta \; -\frac{1}{5} \bar{\gamma } \left(\bar{\gamma } \bar{v} \cosh (\eta ) m_T \right.\nonumber\\
    &\left.\times\left(-15 \bar{\gamma }^2 f_3^{\text{shear}} p_T^2 \cos ^2(\phi )-10 \bar{\gamma } \bar{v} f_2^{\text{shear}}p_T \cos (\phi ) \bar{E}_p\right.\right.\nonumber\\
    &\left.\left.+5 f_3^{\text{shear}} \sinh ^2(\eta ) m_T^2+2 p^2 \left(f_3^{\text{shear}}-f_1^{\text{shear}}\right)\right)\right.\nonumber\\
    &\left.-5 \sinh ^2(\eta ) m_T^2 \left(\bar{\gamma }
   f_3^{\text{shear}} p_T \cos (\phi )+\bar{v} f_2^{\text{shear}} \bar{E}_p\right)\right.\nonumber\\
    &\left.+5 \left(\bar{\gamma }^2-1\right) \cosh ^2(\eta ) m_T^2 \left(3 \bar{\gamma } f_3^{\text{shear}} p_T
   \cos (\phi )\right.\right.\nonumber\\
    &\left.\left.+\bar{v} f_2^{\text{shear}} \bar{E}_p\right)-5 \bar{\gamma }^3 \bar{v}^3 f_3^{\text{shear}} \cosh ^3(\eta ) m_T^3\right.\nonumber\\
    &\left.+\bar{\gamma } p_T \cos (\phi ) \left(5 \bar{\gamma
   }^2 f_3^{\text{shear}} p_T^2 \cos ^2(\phi )\right.\right.\nonumber\\
    &\left.\left.+5 \bar{\gamma } \bar{v} f_2^{\text{shear}} p_T \cos (\phi ) \bar{E}_p+2 p^2\left(f_1^{\text{shear}}-f_3^{\text{shear}}\right)\right)\right)
   \end{align}
   %diff kernels
   \begin{align}
    K_1^\text{diff} &= \int \ud \phi \ud \eta \;  -\bar{\gamma } \bar{v} \cosh (\eta ) m_T \left(f_1^{\text{diff}} \bar{E}_p+2 \bar{\gamma } \bar{v} f_2^{\text{diff}} p_T \cos (\phi )\right)\nonumber\\
    &+\bar{\gamma }^2 \bar{v}^3
   f_2^{\text{diff}} \cosh ^2(\eta ) m_T^2+\frac{1}{3} p^2 \bar{v} \left(f_0^{\text{diff}}-f_2^{\text{diff}}\right)\nonumber\\
   &+\bar{\gamma } f_1^{\text{diff}} p_T \cos (\phi )
   \bar{E}_p +\bar{\gamma }^2 \bar{v} f_2^{\text{diff}} p_T^2 \cos ^2(\phi ) \\
    K_2^\text{diff} &=\int \ud \phi \ud \eta \; -\bar{\gamma } \bar{v} \cosh (\eta ) m_T \left(2 \bar{\gamma } f_2^{\text{diff}} p_T \cos (\phi )+\bar{v} f_1^{\text{diff}} \bar{E}_p\right)\nonumber\\
    &+\bar{\gamma }^2 \bar{v}^2
   f_2^{\text{diff}} \cosh ^2(\eta ) m_T^2+\bar{\gamma }^2 f_2^{\text{diff}} p_T^2 \cos ^2(\phi )\nonumber\\
   &+\bar{\gamma } \bar{v} f_1^{\text{diff}} p_T \cos (\phi ) \bar{E}_p+\frac{1}{3} p^2
   \left(f_0^{\text{diff}}-f_2^{\text{diff}}\right)
\end{align}

\section{Perturbation kernel with resonances} \label{sec_analytic_res_kernel_pert}
In this section we give the the full expression for the perturbation spectra, expanding on the shorthand notation introduced in \autoref{eq:shorthand}. Additionally we give the expressions for the appearing kernels. The perturbation spectrum reduced to the integration in the $\tau-r$-plane is given by
\begin{align}
    &\frac{\delta \ud N}{ p_T \ud p_T \ud \phi_P \ud \eta_P} = \frac{\nu}{(2\pi)^3} \int_0^1 \ud c  \tau (c) r (c) \nonumber \\
	&\left( \frac{\partial r}{\partial c} \left[-\frac{\delta \beta }{\bar{\beta}} K_1^\text{eqTemp}+(\delta \alpha - \Bar{\alpha}\frac{\delta \beta }{\bar{\beta}}) K_1^\text{eqChem}\right.\right.\nonumber\\
 &\left.\left.+\bar{\beta}\bgamma \delta v_1 K_1^\text{eqVel}+\bar{\beta} \delta v_2 K_3^\text{eqVel}+\bar{\beta} \delta v_3 K_5^\text{eqVel}\right.\right.\nonumber
 \end{align}
\begin{align}
 &\left.\left. -\bpi_B \frac{\delta \beta }{\bar{\beta}} K_1^\text{bulkTemp} +\bpi_B (\delta \alpha - \Bar{\alpha}\frac{\delta \beta }{\bar{\beta}}) K_1^\text{bulkChem} \right.\right.\nonumber\\
 &\left.\left.+ \bpi_B \bgamma \delta v_1 K_1^\text{bulkVel} + \bpi_B \delta v_2 K_3^\text{bulkVel} \right.\right.\nonumber\\
 &\left.\left. + \bpi_B \delta v_3 K_5^\text{bulkVel}+  \delta \pi_B K_1^\text{bulkBulk}  \right.\right.\nonumber\\
 &\left.\left. - \frac{\delta \beta}{\bbeta} \bnu \bgamma  K_1^\text{diffTemp} + (\delta \alpha-\alpha \frac{\delta \beta}{\bbeta}) \bnu \bgamma  K_1^\text{diffChem}\right.\right.\nonumber\\
 &\left.\left. + \bnu \bgamma \delta v_1 K_1^\text{diffVel} +  \bnu \delta v_2  K_3^\text{diffVel} +  \bnu \delta v_3  K_5^\text{diffVel}  \right.\right.\nonumber\\
 &\left.\left. +  \bgamma \delta \nu_1 K_1^\text{diffDiff} +  \delta \nu_2  K_3^\text{diffDiff} +  \delta \nu_3  K_5^\text{diffDiff} \right.\right.\nonumber\\
 &\left.\left. -\bpi^{22} \frac{\delta \beta}{\bbeta} K_1^\text{shearTemp}-\bpi^{33} \frac{\delta \beta}{\bbeta} K_3^\text{shearTemp} \right.\right.\nonumber\\
 &\left.\left. +\bpi^{22} \left(\delta \alpha-\balpha \frac{\delta \beta}{\bbeta}\right) K_1^\text{shearChem}\right.\right.\nonumber\\
 &\left.\left.+\bpi^{33} \left(\delta \alpha-\balpha \frac{\delta \beta}{\bbeta}\right) K_3^\text{shearChem} \right.\right.\nonumber\\
 &\left.\left. + \bgamma \bpi^{22} \delta v_1 \left(K_1^\text{shearShear} + K_1^\text{shearVel} \right)\right.\right.\nonumber\\
 &\left.\left. +  \bpi^{22} \delta v_2 \left(K_3^\text{shearShear} + K_3^\text{shearVel} \right)\right.\right.\nonumber\\
 &\left.\left. +  \bpi^{22} \delta v_3 \left(K_5^\text{shearShear} + K_5^\text{shearVel} \right)\right.\right.\nonumber\\
 &\left.\left. + \bgamma \bpi^{33} \delta v_1 \left(K_7^\text{shearShear} + K_7^\text{shearVel}\right)\right.\right.\nonumber\\
 &\left.\left. +  \bpi^{33} \delta v_2 \left(K_9^\text{shearShear} + K_9^\text{shearVel} \right)\right.\right.\nonumber\\
 &\left.\left. +  \bpi^{33} \delta v_3 \left(K_{11}^\text{shearShear} + K_{11}^\text{shearVel} \right) \right.\right.\nonumber\\
 &\left.\left.+ \delta \pi^{12} K_{13}^\text{shearShear} + \delta \pi^{13} K_{15}^\text{shearShear}\right.\right.\nonumber\\
 &\left.\left.+ \delta \pi^{23} K_{17}^\text{shearShear} + \delta \pi^{22} K_{19}^\text{shearShear}\right.\right.\nonumber\\
 &\left.\left.+ \delta \pi^{33} K_{21}^\text{shearShear} \right]\right. \nonumber\\
 &\left. - \frac{\partial \tau}{\partial c} \left[-\frac{\delta \beta }{\bar{\beta}} K_2^\text{eqTemp} +(\delta \alpha - \Bar{\alpha}\frac{\delta \beta }{\bar{\beta}}) K_1^\text{eqChem} \right.\right.\nonumber\\
 &\left.\left.+ \bar{\beta}\bgamma \delta v_1 K_2^\text{eqVel}+\bar{\beta} \delta v_2 K_4^\text{eqVel}+\bar{\beta} \delta v_3 K_6^\text{eqVel} \right.\right.\nonumber\\
  &\left.\left. -\bpi_B \frac{\delta \beta }{\bar{\beta}} K_2^\text{bulkTemp} +\bpi_B (\delta \alpha - \Bar{\alpha}\frac{\delta \beta }{\bar{\beta}}) K_2^\text{bulkChem} \right.\right.\nonumber\\
 &\left.\left.+ \bpi_B \bgamma  \delta v_1 K_2^\text{bulkVel} + \bpi_B \delta  v_2 K_4^\text{bulkVel} \right.\right.\nonumber\\
 &\left.\left. + \bpi_B \delta v_3 K_6^\text{bulkVel} +  \delta \pi_B K_2^\text{bulkBulk} \right.\right.\nonumber\\
 &\left.\left. - \frac{\delta \beta}{\bbeta} \bnu \bgamma  K_2^\text{diffTemp} + (\delta \alpha-\alpha \frac{\delta \beta}{\bbeta}) \bnu \bgamma  K_2^\text{diffChem} \right.\right.\nonumber\\
 &\left.\left. + \bnu \bgamma \delta v_1 K_2^\text{diffVel} +  \bnu \delta v_2  K_4^\text{diffVel} +  \bnu \delta v_3  K_6^\text{diffVel} \right.\right.\nonumber\\
 &\left.\left. +  \bgamma \delta \nu_1 K_2^\text{diffDiff} +  \delta \nu_2  K_4^\text{diffDiff} +  \delta \nu_3  K_6^\text{diffDiff}\right.\right.\nonumber\\
 &\left.\left. -\bpi^{22} \frac{\delta \beta}{\bbeta} K^2_\text{shearTemp}-\bpi^{33} \frac{\delta \beta}{\bbeta} K^4_\text{shearTemp} \right.\right.\nonumber\\
 &\left.\left. +\bpi^{22} \left(\delta \alpha-\balpha \frac{\delta \beta}{\bbeta}\right) K_2^\text{shearChem}\right.\right.\nonumber\\
 &\left.\left.+\bpi^{33} \left(\delta \alpha-\balpha \frac{\delta \beta}{\bbeta}\right) K_4^\text{shearChem}\right.\right.\nonumber\\
 &\left.\left. + \bgamma \bpi^{22} \delta v_1 \left(K_2^\text{shearShear} + K_2^\text{shearVel}\right)\right.\right.\nonumber\\
 &\left.\left. +  \bpi^{22} \delta v_2 \left(K_4^\text{shearShear} + K_4^\text{shearVel} \right)\right.\right.\nonumber\\
 &\left.\left. +  \bpi^{22} \delta v_3 \left(K_6^\text{shearShear} + K_6^\text{shearVel} \right) \right.\right.\nonumber\\
 &\left.\left. + \bgamma \bpi^{33} \delta v_1 \left(K_8^\text{shearShear} + K_8^\text{shearVel}\right)\right.\right.\nonumber
 \end{align}
\begin{align}
 &\left.\left. +  \bpi^{33} \delta v_2 \left(K_{10}^\text{shearShear} + K_{10}^\text{shearVel} \right)\right.\right.\nonumber\\
 &\left.\left. +  \bpi^{33} \delta v_3 \left(K_{12}^\text{shearShear} + K_{12}^\text{shearVel} \right)\right.\right.\nonumber\\
 &\left.\left.+ \delta \pi^{12} K_{14}^\text{shearShear} + \delta \pi^{13} K_{16}^\text{shearShear}\right.\right.\nonumber\\
 &\left.\left.+ \delta \pi^{23} K_{18}^\text{shearShear} + \delta \pi^{22} K_{20}^\text{shearShear}\right.\right.\nonumber\\
 &\left.\left.+ \delta \pi^{33} K_{22}^\text{shearShear} \right]  \right).
\end{align}
The perturbation kernel for perturbations around the equilibrium distribution are given by
\begin{align}
    K_1^\text{eqTemp} &=\int \ud \phi \ud \eta \; \left(\bar{\gamma } \left(f_0^{\text{eqTemp}}-f_1^{\text{eqTemp}}\right) \bar{E}_p\right.\nonumber \\
    &\left.+f_1^{\text{eqTemp}} \cosh (\eta ) m_T \right) \cos(m\phi) \cosh(k\eta) \\
    K_2^\text{eqTemp} &=\int \ud \phi \ud \eta \; \left(\bar{\gamma } \bar{v} \left(f_0^{\text{eqTemp}}-f_1^{\text{eqTemp}}\right) \bar{E}_p\right.\nonumber \\
    &\left.+f_1^{\text{eqTemp}} p_T \cos (\phi )\right) \cos(m\phi) \cosh(k\eta) \\
    K_1^\text{eqChem} &=\int \ud \phi \ud \eta \; \left(\bar{\gamma } \left(f_0^{\text{eqChem}}-f_1^{\text{eqChem}}\right) \bar{E}_p\right.\nonumber \\
    &\left.+f_1^{\text{eqChem}} \cosh (\eta ) m_T\right) \cos(m\phi) \cosh(k\eta) \\
    K_2^\text{eqChem} &=\int \ud \phi \ud \eta \;\left(\bar{\gamma } \bar{v} \left(f_0^{\text{eqChem}}-f_1^{\text{eqChem}}\right) \bar{E}_p\right.\nonumber \\
    &\left.+f_1^{\text{eqChem}} p_T \cos (\phi )\right) \cos(m\phi) \cosh(k\eta) \\
    K_1^\text{eqVel} &= \int \ud \phi \ud \eta \;\left(-\bar{\gamma } \bar{v} \cosh (\eta ) m_T \right.\nonumber \\
    &\left.\times \left(f_1^{\text{eqVel}} \bar{E}_p+2 \bar{\gamma } \bar{v} f_2^{\text{eqVel}} p_T \cos (\phi )\right)\right.\nonumber \\
    &\left.+\bar{\gamma }^2 \bar{v}^3 f_2^{\text{eqVel}} \cosh ^2(\eta ) m_T^2+\frac{1}{3} p^2 \bar{v} \left(f_0^{\text{eqVel}}-f_2^{\text{eqVel}}\right)\right.\nonumber \\
   &\left.+\bar{\gamma } f_1^{\text{eqVel}} p_T \cos (\phi ) \bar{E}_p+\bar{\gamma }^2 \bar{v} f_2^{\text{eqVel}} p_T^2 \cos ^2(\phi )\right)\nonumber\\
   &\times\cos(m\phi) \cosh(k\eta) 
   \end{align}
   \begin{align}
    K_2^\text{eqVel} &= \int \ud \phi \ud \eta \; \left(-\bar{\gamma } \bar{v} \cosh (\eta ) m_T\right.\nonumber\\
    &\left.\times \left(2 \bar{\gamma } f_2^{\text{eqVel}} p_T \cos (\phi )+\bar{v} f_1^{\text{eqVel}} \bar{E}_p\right)\right.\nonumber\\
    &\left.+\bar{\gamma }^2 \bar{v}^2 f_2^{\text{eqVel}} \cosh ^2(\eta ) m_T^2+\bar{\gamma }^2 f_2^{\text{eqVel}} p_T^2 \cos ^2(\phi )\right.\nonumber\\
    &\left.+\bar{\gamma } \bar{v} f_1^{\text{eqVel}} p_T \cos (\phi )\bar{E}_p+\frac{1}{3} p^2 \left(f_0^{\text{eqVel}}-f_2^{\text{eqVel}}\right)\right)\nonumber \\
    &\times  \cos(m\phi) \cosh(k\eta)
    \end{align}
   \begin{align}
    K_3^\text{eqVel} &= \int \ud \phi \ud \eta \;-\bar{\gamma } p_T \sin (\phi )\nonumber\\
    &\times \left(-\bar{\gamma } \bar{v}^2 f_2^{\text{eqVel}} \cosh (\eta ) m_T+f_1^{\text{eqVel}} \bar{E}_p\right.\nonumber\\
    &\left.+\bar{\gamma } \bar{v} f_2^{\text{eqVel}} p_T
   \cos (\phi )\right)\sin(m\phi) \cosh(k\eta) 
   \end{align}
   \begin{align}
    K_4^\text{eqVel} &= \int \ud \phi \ud \eta \; -\bar{\gamma } p_T \sin (\phi )\nonumber\\
    &\times \left(-\bar{\gamma } \bar{v} f_2^{\text{eqVel}} \cosh (\eta ) m_T+\bar{\gamma } f_2^{\text{eqVel}} p_T \cos (\phi )\right.\nonumber\\
    &\left.+\bar{v}  f_1^{\text{eqVel}} \bar{E}_p\right)\sin(m\phi) \cosh(k\eta) 
    \end{align}
   \begin{align}
    K_5^\text{eqVel} &= \int \ud \phi \ud \eta \; \bar{\gamma } \sinh (\eta ) m_T\nonumber\\
    &\times \left(\bar{\gamma } \bar{v}^2 f_2^{\text{eqVel}} \cosh (\eta ) m_T+f_1^{\text{eqVel}} \left(-\bar{E}_p\right)\right.\nonumber\\
    &\left.-\bar{\gamma } \bar{v} f_2^{\text{eqVel}}
   p_T \cos (\phi )\right)\cos(m\phi) \sinh(k\eta) 
   \end{align}
   \begin{align}
    K_6^\text{eqVel} &= \int \ud \phi \ud \eta \; \bar{\gamma } \sinh (\eta ) m_T \nonumber\\
    &\times \left(\bar{\gamma } \bar{v} f_2^{\text{eqVel}} \cosh (\eta ) m_T-\bar{\gamma } f_2^{\text{eqVel}} p_T \cos (\phi )\right.\nonumber\\
    &\left.-\bar{v} f_1^{\text{eqVel}}
   \bar{E}_p\right) \cos(m\phi) \sinh(k\eta)
\end{align}
The perturbation kernel for perturbations around the bulk correction term are given by
\begin{align}
    K_1^\text{bulkTemp} &=\int \ud \phi \ud \eta \;\left(\bar{\gamma } \left(f_0^{\text{bulkTemp}}-f_1^{\text{bulkTemp}}\right) \bar{E}_p\right.\nonumber \\
    &\left.+f_1^{\text{bulkTemp}} \cosh (\eta ) m_T\right)\cos(m\phi) \cosh(k\eta) \\
    K_2^\text{bulkTemp} &=\int \ud \phi \ud \eta \; \left(\bar{\gamma } \bar{v} \left(f_0^{\text{bulkTemp}}-f_1^{\text{bulkTemp}}\right) \bar{E}_p\right.\nonumber\\
    &\left.+f_1^{\text{bulkTemp}} p_T \cos (\phi )\right)\cos(m\phi) \cosh(k\eta) \\
    K_1^\text{bulkChem} &=\int \ud \phi \ud \eta \; \left(\bar{\gamma } \left(f_0^{\text{bulkChem}}-f_1^{\text{bulkChem}}\right) \bar{E}_p\right.\nonumber \\
    &\left.+f_1^{\text{bulkChem}} \cosh (\eta ) m_T\right)\cos(m\phi) \cosh(k\eta) \\
    K_2^\text{bulkChem} &=\int \ud \phi \ud \eta \;\left(\bar{\gamma } \bar{v} \left(f_0^{\text{bulkChem}}-f_1^{\text{bulkChem}}\right) \bar{E}_p\right.\nonumber\\
    &+\left.f_1^{\text{bulkChem}} p_T \cos (\phi )\right)\cos(m\phi) \cosh(k\eta) \\
    K_1^\text{bulkBulk} &=\int \ud \phi \ud \eta \; \left(\bar{\gamma } \left(f_0^{\text{bulkBulk}}-f_1^{\text{bulkBulk}}\right) \bar{E}_p\right.\nonumber\\
    &\left.+f_1^{\text{bulkBulk}} \cosh (\eta ) m_T\right)\cos(m\phi) \cosh(k\eta) \\
    K_2^\text{bulkBulk} &=\int \ud \phi \ud \eta \; \left(\bar{\gamma } \bar{v} \left(f_0^{\text{bulkBulk}}-f_1^{\text{bulkBulk}}\right) \bar{E}_p\right.\nonumber\\
    &\left.+f_1^{\text{bulkBulk}} p_T \cos (\phi )\right)\cos(m\phi) \cosh(k\eta)
    \end{align}
   \begin{align}
    K_1^\text{bulkVel} &= \int \ud \phi \ud \eta \; \left(-\bar{\gamma } \bar{v} \cosh (\eta ) m_T \right.\nonumber\\
    &\left.\times\left(f_1^{\text{bulkVel}} \bar{E}_p+2 \bar{\gamma } \bar{v} f_2^{\text{bulkVel}} p_T \cos (\phi )\right)\right.\nonumber\\
    &\left.+\bar{\gamma }^2 \bar{v}^3 f_2^{\text{bulkVel}} \cosh ^2(\eta ) m_T^2\right.\nonumber\\
   &\left.+\frac{1}{3} p^2 \bar{v} \left(f_0^{\text{bulkVel}}-f_2^{\text{bulkVel}}\right)\right.\nonumber\\
   &\left.+\bar{\gamma } f_1^{\text{bulkVel}} p_T \cos(\phi ) \bar{E}_p\right.\nonumber\\
   &\left.+\bar{\gamma }^2 \bar{v} f_2^{\text{bulkVel}} p_T^2 \cos ^2(\phi ) \right)\nonumber\\
   &\times\cos(m\phi) \cosh(k\eta)
   \end{align}
   \begin{align}
    K_2^\text{bulkVel} &= \int \ud \phi \ud \eta \; \left(-\bar{\gamma } \bar{v} \cosh (\eta ) m_T\right.\nonumber\\
    &\left.\times \left(2 \bar{\gamma } f_2^{\text{bulkVel}} p_T \cos (\phi )+\bar{v} f_1^{\text{bulkVel}} \bar{E}_p\right)\right.\nonumber\\
    &\left.+\bar{\gamma }^2 \bar{v}^2  f_2^{\text{bulkVel}} \cosh ^2(\eta ) m_T^2+\bar{\gamma }^2 f_2^{\text{bulkVel}} p_T^2 \cos ^2(\phi )\right.\nonumber\\
   &\left.+\bar{\gamma } \bar{v} f_1^{\text{bulkVel}} p_T \cos (\phi )   \bar{E}_p\right.\nonumber\\
   &\left.+\frac{1}{3} p^2 \left(f_0^{\text{bulkVel}}-f_2^{\text{bulkVel}}\right)\right) \nonumber\\
   &\times\cos(m\phi) \cosh(k\eta)
   \end{align}
   \begin{align}
    K_3^\text{bulkVel} &= \int \ud \phi \ud \eta \;  \left(-\bar{\gamma } p_T \sin (\phi ) \right.\nonumber\\
    &\left.\times \left(-\bar{\gamma } \bar{v}^2 f_2^{\text{bulkVel}} \cosh (\eta ) m_T+f_1^{\text{bulkVel}} \bar{E}_p\right.\right.\nonumber\\
    &\left.\left.+\bar{\gamma } \bar{v}  f_2^{\text{bulkVel}} p_T \cos (\phi )\right)\right) \sin(m\phi) \cosh(k\eta)
   \end{align}
   \begin{align}
    K_4^\text{bulkVel} &= \int \ud \phi \ud \eta \;  -\bar{\gamma } p_T \sin (\phi )\nonumber\\
    &\times \left(-\bar{\gamma } \bar{v} f_2^{\text{bulkVel}} \cosh (\eta ) m_T+\bar{\gamma } f_2^{\text{bulkVel}} p_T \cos (\phi )\right.\nonumber\\
    &\left.+\bar{v} f_1^{\text{bulkVel}} \bar{E}_p\right) \sin(m\phi) \cosh(k\eta)
   \end{align}
   \begin{align}
    K_5^\text{bulkVel} &=  \int \ud \phi \ud \eta \; \bar{\gamma } \sinh (\eta ) m_T \nonumber\\
    &\times\left(\bar{\gamma } \bar{v}^2 f_2^{\text{bulkVel}} \cosh (\eta ) m_T+f_1^{\text{bulkVel}} \left(-\bar{E}_p\right)\right.\nonumber\\
    &\left.-\bar{\gamma } \bar{v}  f_2^{\text{bulkVel}} p_T \cos (\phi )\right) \cos(m\phi) \sinh(k\eta)
   \end{align}
   \begin{align}
    K_6^\text{bulkVel} &=  \int \ud \phi \ud \eta \;\bar{\gamma } \sinh (\eta ) m_T \nonumber\\
    &\times\left(\bar{\gamma } \bar{v} f_2^{\text{bulkVel}} \cosh (\eta ) m_T-\bar{\gamma } f_2^{\text{bulkVel}} p_T \cos (\phi )\right.\nonumber\\
    &\left.-\bar{v} f_1^{\text{bulkVel}} \bar{E}_p\right)\cos(m\phi) \sinh(k\eta)
\end{align}
The perturbation kernel for perturbations around the shear correction term are given by
\begin{align}
%shearTemp1
    K_1^\text{shearTemp} &= \int \ud \phi \ud \eta \; \frac{1}{5} \bar{\gamma } \left(\bar{\gamma } \bar{v} \cosh (\eta ) m_T \left(2 p^2 \bar{v} \right.\right.\nonumber\\
    &\left.\left. \times\left(f_1^{\text{shearTemp}}-f_3^{\text{shearTemp}}\right)\right.\right.\nonumber\\
    &\left.\left.+10 \bar{\gamma }
   f_2^{\text{shearTemp}} p_T \cos (\phi ) \bar{E}_p\right.\right.\nonumber\\
    &\left.\left.-15 \bar{\gamma }^2 \bar{v} f_3^{\text{shearTemp}} p_T^2 \cos ^2(\phi )\right.\right.\nonumber\\
    &\left.\left.+5 \bar{v} f_3^{\text{shearTemp}} p_T^2 \sin ^2(\phi)\right)\right.\nonumber\\
    &\left.+5 \left(\bar{\gamma }^2-1\right) \cosh ^2(\eta ) m_T^2 \left(3 \bar{\gamma } \bar{v} f_3^{\text{shearTemp}} p_T \cos (\phi )\right.\right.\nonumber\\
    &\left.\left.-f_2^{\text{shearTemp}} \bar{E}_p\right)-5\bar{\gamma }^3 \bar{v}^4 f_3^{\text{shearTemp}} \cosh ^3(\eta ) m_T^3\right.\nonumber\\
    &\left.+p_T \left(\bar{\gamma } \bar{v} \cos (\phi ) \left(2 p^2
   \left(f_3^{\text{shearTemp}}-f_1^{\text{shearTemp}}\right)\right.\right.\right.\nonumber\\
    &\left.\left.\left.-5 f_3^{\text{shearTemp}} p_T^2 \sin ^2(\phi )\right)\right.\right.\nonumber\\
    &\left.\left.-5 \bar{\gamma }^2 f_2^{\text{shearTemp}} p_T \cos ^2(\phi )\bar{E}_p\right.\right.\nonumber\\
    &\left.\left.+5 f_2^{\text{shearTemp}} p_T \sin ^2(\phi ) \bar{E}_p\right.\right.\nonumber\\
    &\left.\left.+5 \bar{\gamma }^3 \bar{v} f_3^{\text{shearTemp}} p_T^2 \cos ^3(\phi )\right)\right) \nonumber\\
    &\times \cos(m\phi) \cosh(k\eta)
    \end{align}
   \begin{align}
   %shearTemp2
    K_2^\text{shearTemp} &= \int \ud \phi \ud \eta \; \frac{1}{5} \bar{\gamma } \left(\bar{\gamma } \bar{v} \cosh (\eta ) m_T\right.\nonumber\\
    &\left.\times \left(15 \bar{\gamma }^2 f_3^{\text{shearTemp}} p_T^2 \cos ^2(\phi )\right.\right.\nonumber\\
    &\left.\left.+10 \bar{\gamma } \bar{v} f_2^{\text{shearTemp}} p_T \cos (\phi ) \bar{E}_p\right.\right.\nonumber\\
    &\left.\left.+2 p^2 \left(f_1^{\text{shearTemp}}-f_3^{\text{shearTemp}}\right)\right.\right.\nonumber\\
    &\left.\left.-5 f_3^{\text{shearTemp}} p_T^2 \sin ^2(\phi )\right)\right.\nonumber\\
    &\left.-5\left(\bar{\gamma }^2-1\right) \cosh ^2(\eta ) m_T^2\right.\nonumber\\
    &\left.\times \left(3 \bar{\gamma } f_3^{\text{shearTemp}} p_T \cos (\phi )+\bar{v} f_2^{\text{shearTemp}} \bar{E}_p\right)\right.\nonumber\\
    &\left.+5 \bar{\gamma}^3 \bar{v}^3 f_3^{\text{shearTemp}} \cosh ^3(\eta ) m_T^3\right.\nonumber\\
    &\left.+p_T \left(\bar{\gamma } \cos (\phi ) \left(2 p^2 \left(f_3^{\text{shearTemp}}-f_1^{\text{shearTemp}}\right)\right.\right.\right.\nonumber\\
    &\left.\left.\left.+5f_3^{\text{shearTemp}} p_T^2 \sin ^2(\phi )\right)\right.\right.\nonumber\\
    &\left.\left.-5 \bar{\gamma }^3 f_3^{\text{shearTemp}} p_T^2 \cos ^3(\phi )\right.\right.\nonumber\\
    &\left.\left.-5 \bar{\gamma }^2 \bar{v} f_2^{\text{shearTemp}} p_T \cos ^2(\phi) \bar{E}_p\right.\right.\nonumber\\
    &\left.\left.+5 \bar{v} f_2^{\text{shearTemp}} p_T \sin ^2(\phi ) \bar{E}_p\right)\right)\nonumber\\
    &\times \cos(m\phi) \cosh(k\eta)
    \end{align}
    \begin{align}
    K_3^\text{shearTemp} &= \int \ud \phi \ud \eta \; \frac{1}{5} \bar{\gamma } \left(\bar{\gamma } \bar{v} \cosh (\eta ) m_T\right.\nonumber\\
    &\left.\times \left(5 \bar{v} f_3^{\text{shearTemp}} \sinh ^2(\eta ) m_T^2\right.\right.\nonumber\\
    &\left.\left.+2 p^2 \bar{v}\left(f_1^{\text{shearTemp}}-f_3^{\text{shearTemp}}\right)\right.\right.\nonumber\\
    &\left.\left.+10 \bar{\gamma } f_2^{\text{shearTemp}} p_T \cos (\phi ) \bar{E}_p\right.\right.\nonumber\\
    &\left.\left.-15 \bar{\gamma }^2 \bar{v} f_3^{\text{shearTemp}} p_T^2 \cos ^2(\phi )\right)\right.\nonumber\\
    &\left.+5 \sinh ^2(\eta ) m_T^2 \left(f_2^{\text{shearTemp}} \bar{E}_p\right.\right.\nonumber\\
    &\left.\left.-\bar{\gamma } \bar{v} f_3^{\text{shearTemp}} p_T \cos (\phi )\right).\right.\nonumber\\
    &\left.+5 \left(\bar{\gamma}^2-1\right) \cosh ^2(\eta ) m_T^2 .\right.\nonumber\\
    &\left.\times\left(3 \bar{\gamma } \bar{v} f_3^{\text{shearTemp}} p_T \cos (\phi )-f_2^{\text{shearTemp}} \bar{E}_p\right).\right.\nonumber\\
    &\left.-5 \bar{\gamma }^3 \bar{v}^4 f_3^{\text{shearTemp}} \cosh ^3(\eta ) m_T^3.\right.\nonumber\\
    &\left.+\bar{\gamma } p_T \cos (\phi ) \left(2 p^2 \bar{v} \left(f_3^{\text{shearTemp}}-f_1^{\text{shearTemp}}\right)\right.\right.\nonumber\\
    &\left.\left.-5 \bar{\gamma }   f_2^{\text{shearTemp}} p_T \cos (\phi ) \bar{E}_p\right.\right.\nonumber\\
    &\left.\left.+5 \bar{\gamma }^2 \bar{v} f_3^{\text{shearTemp}} p_T^2 \cos ^2(\phi )\right)\right)\nonumber\\
    &\times \cos(m\phi) \cosh(k\eta)
    \end{align}
    \begin{align}
    K_4^\text{shearTemp} &= \int \ud \phi \ud \eta \;-\frac{1}{5} \bar{\gamma } \left(\bar{\gamma } \bar{v} \cosh (\eta ) m_T\right.\nonumber\\
    &\left.\times \left(-15 \bar{\gamma }^2 f_3^{\text{shearTemp}} p_T^2 \cos ^2(\phi )\right.\right.\nonumber\\
    &\left.\left.-10 \bar{\gamma } \bar{v} f_2^{\text{shearTemp}} p_T \cos (\phi ) \bar{E}_p\right.\right.\nonumber\\
    &\left.\left.+5 f_3^{\text{shearTemp}} \sinh ^2(\eta ) m_T^2\right.\right.\nonumber\\
    &\left.\left.+2 p^2 \left(f_3^{\text{shearTemp}}-f_1^{\text{shearTemp}}\right)\right)\right.\nonumber\\
    &\left.-5 \sinh^2(\eta ) m_T^2 \left(\bar{\gamma } f_3^{\text{shearTemp}} p_T \cos (\phi )\right.\right.\nonumber\\
    &\left.\left.+\bar{v} f_2^{\text{shearTemp}} \bar{E}_p\right)+5 \left(\bar{\gamma }^2-1\right) \cosh ^2(\eta ) m_T^2\right.\nonumber\\
    &\left.\times \left(3 \bar{\gamma } f_3^{\text{shearTemp}} p_T \cos (\phi )+\bar{v} f_2^{\text{shearTemp}} \bar{E}_p\right)\right.\nonumber\\
    &\left.-5 \bar{\gamma }^3 \bar{v}^3 f_3^{\text{shearTemp}} \cosh ^3(\eta )  m_T^3+\bar{\gamma } p_T \cos (\phi ) \right.\nonumber\\
    &\left.\times\left(5 \bar{\gamma }^2 f_3^{\text{shearTemp}} p_T^2 \cos ^2(\phi )\right.\right.\nonumber\\
    &\left.\left.+5 \bar{\gamma } \bar{v} f_2^{\text{shearTemp}} p_T \cos (\phi ) \bar{E}_p\right.\right.\nonumber\\
    &\left.\left.+2 p^2 \left(f_1^{\text{shearTemp}}-f_3^{\text{shearTemp}}\right)\right)\right)\nonumber\\
    &\times \cos(m\phi) \cosh(k\eta)
    \end{align}
    %shearchem1
    \begin{align}
     K_1^\text{shearChem} &= \int \ud \phi \ud \eta \; \frac{1}{5} \bar{\gamma } \left(\bar{\gamma } \bar{v} \cosh (\eta ) m_T \left(2 p^2 \bar{v} \right.\right.\nonumber\\
    &\left.\left. \times\left(f_1^{\text{shearChem}}-f_3^{\text{shearChem}}\right)\right.\right.\nonumber\\
    &\left.\left.+10 \bar{\gamma }
   f_2^{\text{shearChem}} p_T \cos (\phi ) \bar{E}_p\right.\right.\nonumber\\
    &\left.\left.-15 \bar{\gamma }^2 \bar{v} f_3^{\text{shearChem}} p_T^2 \cos ^2(\phi )\right.\right.\nonumber\\
    &\left.\left.+5 \bar{v} f_3^{\text{shearChem}} p_T^2 \sin ^2(\phi)\right)\right.\nonumber\\
    &\left.+5 \left(\bar{\gamma }^2-1\right) \cosh ^2(\eta ) m_T^2 \left(3 \bar{\gamma } \bar{v} f_3^{\text{shearChem}} p_T \cos (\phi )\right.\right.\nonumber\\
    &\left.\left.-f_2^{\text{shearChem}} \bar{E}_p\right)-5\bar{\gamma }^3 \bar{v}^4 f_3^{\text{shearChem}} \cosh ^3(\eta ) m_T^3\right.\nonumber\\
    &\left.+p_T \left(\bar{\gamma } \bar{v} \cos (\phi ) \left(2 p^2
   \left(f_3^{\text{shearChem}}-f_1^{\text{shearChem}}\right)\right.\right.\right.\nonumber\\
    &\left.\left.\left.-5 f_3^{\text{shearChem}} p_T^2 \sin ^2(\phi )\right)\right.\right.\nonumber\\
    &\left.\left.-5 \bar{\gamma }^2 f_2^{\text{shearChem}} p_T \cos ^2(\phi )\bar{E}_p\right.\right.\nonumber\\
    &\left.\left.+5 f_2^{\text{shearChem}} p_T \sin ^2(\phi ) \bar{E}_p\right.\right.\nonumber\\
    &\left.\left.+5 \bar{\gamma }^3 \bar{v} f_3^{\text{shearChem}} p_T^2 \cos ^3(\phi )\right)\right) \nonumber\\
    &\times \cos(m\phi) \cosh(k\eta)
    \end{align}
    %shearchem2
    \begin{align}
     K_2^\text{shearChem} &= \int \ud \phi \ud \eta \; \frac{1}{5} \bar{\gamma } \left(\bar{\gamma } \bar{v} \cosh (\eta ) m_T\right.\nonumber\\
    &\left.\times \left(15 \bar{\gamma }^2 f_3^{\text{shearChem}} p_T^2 \cos ^2(\phi )\right.\right.\nonumber\\
    &\left.\left.+10 \bar{\gamma } \bar{v} f_2^{\text{shearChem}} p_T \cos (\phi ) \bar{E}_p\right.\right.\nonumber\\
    &\left.\left.+2 p^2 \left(f_1^{\text{shearChem}}-f_3^{\text{shearChem}}\right)\right.\right.\nonumber\\
    &\left.\left.-5 f_3^{\text{shearChem}} p_T^2 \sin ^2(\phi )\right)\right.\nonumber\\
    &\left.-5\left(\bar{\gamma }^2-1\right) \cosh ^2(\eta ) m_T^2\right.\nonumber\\
    &\left.\times \left(3 \bar{\gamma } f_3^{\text{shearChem}} p_T \cos (\phi )+\bar{v} f_2^{\text{shearChem}} \bar{E}_p\right)\right.\nonumber\\
    &\left.+5 \bar{\gamma}^3 \bar{v}^3 f_3^{\text{shearChem}} \cosh ^3(\eta ) m_T^3\right.\nonumber\\
    &\left.+p_T \left(\bar{\gamma } \cos (\phi ) \left(2 p^2 \left(f_3^{\text{shearChem}}-f_1^{\text{shearChem}}\right)\right.\right.\right.\nonumber\\
    &\left.\left.\left.+5f_3^{\text{shearChem}} p_T^2 \sin ^2(\phi )\right)\right.\right.\nonumber\\
    &\left.\left.-5 \bar{\gamma }^3 f_3^{\text{shearChem}} p_T^2 \cos ^3(\phi )\right.\right.\nonumber\\
    &\left.\left.-5 \bar{\gamma }^2 \bar{v} f_2^{\text{shearChem}} p_T \cos ^2(\phi) \bar{E}_p\right.\right.\nonumber\\
    &\left.\left.+5 \bar{v} f_2^{\text{shearChem}} p_T \sin ^2(\phi ) \bar{E}_p\right)\right) \nonumber\\
    &\times \cos(m\phi) \cosh(k\eta)
    \end{align}
    %shearchem3
    \begin{align}
    K_3^\text{shearChem} &= \int \ud \phi \ud \eta \; \frac{1}{5} \bar{\gamma } \left(\bar{\gamma } \bar{v} \cosh (\eta ) m_T\right.\nonumber\\
    &\left.\times \left(5 \bar{v} f_3^{\text{shearChem}} \sinh ^2(\eta ) m_T^2\right.\right.\nonumber\\
    &\left.\left.+2 p^2 \bar{v}\left(f_1^{\text{shearChem}}-f_3^{\text{shearChem}}\right)\right.\right.\nonumber\\
    &\left.\left.+10 \bar{\gamma } f_2^{\text{shearChem}} p_T \cos (\phi ) \bar{E}_p\right.\right.\nonumber\\
    &\left.\left.-15 \bar{\gamma }^2 \bar{v} f_3^{\text{shearChem}} p_T^2 \cos ^2(\phi )\right)\right.\nonumber\\
    &\left.+5 \sinh ^2(\eta ) m_T^2 \left(f_2^{\text{shearChem}} \bar{E}_p\right.\right.\nonumber\\
    &\left.\left.-\bar{\gamma } \bar{v} f_3^{\text{shearChem}} p_T \cos (\phi )\right).\right.\nonumber\\
    &\left.+5 \left(\bar{\gamma}^2-1\right) \cosh ^2(\eta ) m_T^2 .\right.\nonumber\\
    &\left.\times\left(3 \bar{\gamma } \bar{v} f_3^{\text{shearChem}} p_T \cos (\phi )-f_2^{\text{shearChem}} \bar{E}_p\right).\right.\nonumber\\
    &\left.-5 \bar{\gamma }^3 \bar{v}^4 f_3^{\text{shearChem}} \cosh ^3(\eta ) m_T^3.\right.\nonumber\\
    &\left.+\bar{\gamma } p_T \cos (\phi ) \left(2 p^2 \bar{v} \left(f_3^{\text{shearChem}}-f_1^{\text{shearChem}}\right)\right.\right.\nonumber\\
    &\left.\left.-5 \bar{\gamma }   f_2^{\text{shearChem}} p_T \cos (\phi ) \bar{E}_p\right.\right.\nonumber\\
    &\left.\left.+5 \bar{\gamma }^2 \bar{v} f_3^{\text{shearChem}} p_T^2 \cos ^2(\phi )\right)\right)\nonumber\\
    &\times \cos(m\phi) \cosh(k\eta)
    \end{align}
    %shearchem4
    \begin{align}
     K_4^\text{shearChem} &= \int \ud \phi \ud \eta \;-\frac{1}{5} \bar{\gamma } \left(\bar{\gamma } \bar{v} \cosh (\eta ) m_T\right.\nonumber\\
    &\left.\times \left(-15 \bar{\gamma }^2 f_3^{\text{shearChem}} p_T^2 \cos ^2(\phi )\right.\right.\nonumber\\
    &\left.\left.-10 \bar{\gamma } \bar{v} f_2^{\text{shearChem}} p_T \cos (\phi ) \bar{E}_p\right.\right.\nonumber\\
    &\left.\left.+5 f_3^{\text{shearChem}} \sinh ^2(\eta ) m_T^2\right.\right.\nonumber\\
    &\left.\left.+2 p^2 \left(f_3^{\text{shearChem}}-f_1^{\text{shearChem}}\right)\right)\right.\nonumber\\
    &\left.-5 \sinh^2(\eta ) m_T^2 \left(\bar{\gamma } f_3^{\text{shearChem}} p_T \cos (\phi )\right.\right.\nonumber\\
    &\left.\left.+\bar{v} f_2^{\text{shearChem}} \bar{E}_p\right)+5 \left(\bar{\gamma }^2-1\right) \cosh ^2(\eta ) m_T^2\right.\nonumber\\
    &\left.\times \left(3 \bar{\gamma } f_3^{\text{shearChem}} p_T \cos (\phi )+\bar{v} f_2^{\text{shearChem}} \bar{E}_p\right)\right.\nonumber\\
    &\left.-5 \bar{\gamma }^3 \bar{v}^3 f_3^{\text{shearChem}} \cosh ^3(\eta )  m_T^3+\bar{\gamma } p_T \cos (\phi ) \right.\nonumber\\
    &\left.\times\left(5 \bar{\gamma }^2 f_3^{\text{shearChem}} p_T^2 \cos ^2(\phi )\right.\right.\nonumber\\
    &\left.\left.+5 \bar{\gamma } \bar{v} f_2^{\text{shearChem}} p_T \cos (\phi ) \bar{E}_p\right.\right.\nonumber\\
    &\left.\left.+2 p^2 \left(f_1^{\text{shearChem}}-f_3^{\text{shearChem}}\right)\right)\right)\nonumber\\
    &\times \cos(m\phi) \cosh(k\eta)
    \end{align}
  %shearvel1
    \begin{align}
    K_1^\text{shearVel} &= \int \ud \phi \ud \eta \; -\frac{2}{5} p^2 \bar{\gamma } \left(f_1^{\text{shearVel}}-f_3^{\text{shearVel}}\right) \bar{E}_p \nonumber \\
    &\times \left(p_T \cos (\phi )-\bar{v} \cosh (\eta ) m_T\right)\nonumber\\
    &+\frac{1}{21} \bar{v} \left(f_2^{\text{shearVel}}-f_4^{\text{shearVel}}\right) \left(30 \bar{\gamma }^2 \bar{v} \cosh (\eta )\right.\nonumber\\
    &\left. \times m_T p_T \cos (\phi )-15 \left(\bar{\gamma }^2-1\right)   \cosh ^2(\eta ) m_T^2)\right.\nonumber\\
    &\left.-15 \bar{\gamma }^2 p_T^2 \cos ^2(\phi )+4 p^2+3 p_T^2 \sin ^2(\phi )\right)\nonumber\\
    &+f_3^{\text{shearVel}} \bar{E}_p \left(\bar{\gamma } \bar{v} \cosh   (\eta ) m_T-\bar{\gamma } p_T \cos (\phi )\right)\nonumber\\
    &\times \left(-2 \bar{\gamma }^2 \bar{v} \cosh (\eta ) m_T p_T \cos (\phi )\right.\nonumber\\
    &\left.+\left(\bar{\gamma }^2-1\right) \cosh ^2(\eta )   m_T^2+p_T^2 \left(\bar{\gamma }^2 \cos ^2(\phi )\right.\right.\nonumber\\
    &\left.\left.-\sin ^2(\phi )\right)\right)-\bar{\gamma } \bar{v} f_4^{\text{shearVel}} \left(\bar{v} \cosh (\eta ) m_T\right.\nonumber\\
    &\left.-p_T \cos   (\phi )\right) \left(\bar{\gamma } \bar{v} \cosh (\eta ) m_T-\bar{\gamma } p_T \cos (\phi )\right)\nonumber\\
    &\times \left(-2 \bar{\gamma }^2 \bar{v} \cosh (\eta ) m_T p_T \cos (\phi )\right.\nonumber\\
    &\left.+\left(\bar{\gamma }^2-1\right) \cosh ^2(\eta ) m_T^2+p_T^2 \left(\bar{\gamma }^2 \cos ^2(\phi )\right.\right.\nonumber\\
    &\left.\left.-\sin ^2(\phi )\right)\right)-\frac{2}{15} p^4 \bar{v}
   \left(f_0^{\text{shearVel}}-f_4^{\text{shearVel}}\right) 
   \end{align}
  %shearvel2
    \begin{align}
    K_2^\text{shearVel} &= \int \ud \phi \ud \eta \; \frac{2}{5} p^2 \bar{v} \left(f_1^{\text{shearVel}}-f_3^{\text{shearVel}}\right) \bar{E}_p \nonumber\\
    &\times \left(\bar{\gamma } \bar{v} \cosh (\eta ) m_T-\bar{\gamma } p_T \cos (\phi
   )\right)\nonumber\\
    &+\frac{1}{21} \left(f_2^{\text{shearVel}}-f_4^{\text{shearVel}}\right) \left(30 \bar{\gamma }^2 \bar{v} \cosh (\eta ) m_T \right.\nonumber\\
    &\left. \times p_T \cos (\phi )-15 \left(\bar{\gamma   }^2-1\right) \cosh ^2(\eta ) m_T^2\right.\nonumber\\
    &\left.-15 \bar{\gamma }^2 p_T^2 \cos ^2(\phi )+4 p^2+3 p_T^2 \sin ^2(\phi )\right)\nonumber\\
    &+\bar{v} f_3^{\text{shearVel}} \bar{E}_p   \left(\bar{\gamma } \bar{v} \cosh (\eta ) m_T-\bar{\gamma } p_T \cos (\phi )\right)\nonumber\\
    &\times \left(-2 \bar{\gamma }^2 \bar{v} \cosh (\eta ) m_T p_T \cos (\phi )\right.\nonumber\\
    &\left.+\left(\bar{\gamma }^2-1\right) \cosh ^2(\eta ) m_T^2+p_T^2 \left(\bar{\gamma }^2 \cos ^2(\phi )\right.\right.\nonumber\\
    &\left.\left.-\sin ^2(\phi )\right)\right)-\bar{\gamma } f_4^{\text{shearVel}} \left(\bar{v} \cosh (\eta ) m_T\right.\nonumber\\
    &\left.-p_T \cos (\phi )\right) \left(\bar{\gamma } \bar{v} \cosh (\eta ) m_T-\bar{\gamma } p_T \cos (\phi )\right)\nonumber\\
    &\times \left(-2 \bar{\gamma }^2  \bar{v} \cosh (\eta ) m_T p_T \cos (\phi )\right.\nonumber\\
    &\left.+\left(\bar{\gamma }^2-1\right) \cosh ^2(\eta ) m_T^2+p_T^2 \left(\bar{\gamma }^2 \cos ^2(\phi )\right.\right.\nonumber\\
    &\left.\left.-\sin ^2(\phi )\right)\right)-\frac{2}{15} p^4 \left(f_0^{\text{shearVel}}-f_4^{\text{shearVel}}\right)
   \end{align}
  %shearvel3
    \begin{align}
    K_3^\text{shearVel} &= \int \ud \phi \ud \eta \;\frac{1}{5} \bar{\gamma } p_T \sin (\phi ) \left(5 f_3^{\text{shearVel}} \bar{E}_p \right.\nonumber\\
    &\left.\times \left(-2 \bar{\gamma }^2 \bar{v} \cosh (\eta ) m_T p_T \cos (\phi )\right.\right.\nonumber\\
    &\left.\left. +\left(\bar{\gamma  }^2-1\right) \cosh ^2(\eta ) m_T^2+p_T^2 \left(\bar{\gamma }^2 \cos ^2(\phi )\right.\right.\right.\nonumber\\
    &\left.\left.\left.-\sin ^2(\phi )\right)\right)-5 \bar{\gamma } \bar{v} f_4^{\text{shearVel}} \left(\bar{v} \cosh (\eta ) m_T\right.\right.\nonumber\\
    &\left.\left.-p_T \cos (\phi )\right) \left(-2 \bar{\gamma }^2 \bar{v} \cosh (\eta ) m_T p_T \cos (\phi )\right.\right.\nonumber\\
    &\left.\left.+\left(\bar{\gamma }^2-1\right) \cosh ^2(\eta )  m_T^2+p_T^2 \left(\bar{\gamma }^2 \cos ^2(\phi )\right.\right.\right.\nonumber\\
    &\left.\left.\left.-\sin ^2(\phi )\right)\right)-2 p^2 \left(f_1^{\text{shearVel}}-f_3^{\text{shearVel}}\right) \bar{E}_p\right)  \end{align}
  %shearvel4
    \begin{align}
    K_4^\text{shearVel} &= \int \ud \phi \ud \eta \; \frac{1}{5} \bar{\gamma } p_T \sin (\phi ) \left(5 \bar{v} f_3^{\text{shearVel}} \bar{E}_p\right.\nonumber\\
    &\left.\times \left(-2 \bar{\gamma }^2 \bar{v} \cosh (\eta ) m_T p_T \cos (\phi   )\right.\right.\nonumber\\
    &\left.\left.+\left(\bar{\gamma }^2-1\right) \cosh ^2(\eta ) m_T^2+p_T^2 \left(\bar{\gamma }^2 \cos ^2(\phi )\right.\right.\right.\nonumber\\
    &\left.\left.\left.-\sin ^2(\phi )\right)\right)-5 \bar{\gamma } f_4^{\text{shearVel}}   \left(\bar{v} \cosh (\eta ) m_T\right.\right.\nonumber\\
    &\left.\left.-p_T \cos (\phi )\right) \left(-2 \bar{\gamma }^2 \bar{v} \cosh (\eta ) m_T p_T \cos (\phi )\right.\right.\nonumber\\
    &\left.\left.+\left(\bar{\gamma }^2-1\right) \cosh^2(\eta ) m_T^2+p_T^2 \left(\bar{\gamma }^2 \cos ^2(\phi )\right.\right.\right.\nonumber\\
    &\left.\left.\left.-\sin ^2(\phi )\right)\right)-2 p^2 \bar{v} \left(f_1^{\text{shearVel}}-f_3^{\text{shearVel}}\right)  \bar{E}_p\right) 
   \end{align}
  %shearvel5
    \begin{align}
    K_5^\text{shearVel} &= \int \ud \phi \ud \eta \; \frac{1}{7} \bar{\gamma } \sinh (\eta ) m_T \left(7 f_3^{\text{shearVel}} \bar{E}_p \right.\nonumber\\
    &\left. \left(-2 \bar{\gamma }^2 \bar{v} \cosh (\eta ) m_T p_T \cos (\phi )\right.\right.\nonumber\\
    &\left.\left.+\left(\bar{\gamma}^2-1\right) \cosh ^2(\eta ) m_T^2+p_T^2 \left(\bar{\gamma }^2 \cos ^2(\phi )\right.\right.\right.\nonumber\\
    &\left.\left.\left.-\sin ^2(\phi )\right)\right)-2 \bar{\gamma } \bar{v}   \left(f_2^{\text{shearVel}}-f_4^{\text{shearVel}}\right) \right.\nonumber\\
    &\left.\times \left(\bar{v} \cosh (\eta ) m_T-p_T \cos (\phi )\right)\right.\nonumber\\
    &\left.-7 \bar{\gamma } \bar{v} f_4^{\text{shearVel}}
   \left(\bar{v} \cosh (\eta ) m_T-p_T \cos (\phi )\right)\right. \nonumber\\
   &\left.\times \left(-2 \bar{\gamma }^2 \bar{v} \cosh (\eta ) m_T p_T \cos (\phi )\right.\right. \nonumber\\
   &\left.\left.+\left(\bar{\gamma }^2-1\right) \cosh   ^2(\eta ) m_T^2+p_T^2 \left(\bar{\gamma }^2 \cos ^2(\phi )\right.\right.\right. \nonumber\\
   &\left.\left.\left.-\sin ^2(\phi )\right)\right)\right) 
   \end{align}
  %shearvel6
    \begin{align}
    K_6^\text{shearVel} &= \int \ud \phi \ud \eta \;  \frac{1}{7} \bar{\gamma } \sinh (\eta ) m_T \left(7 \bar{v} f_3^{\text{shearVel}} \bar{E}_p \right. \nonumber\\
   &\left.\times\left(-2 \bar{\gamma }^2 \bar{v} \cosh (\eta ) m_T p_T \cos (\phi   )\right.\right. \nonumber\\
   &\left.\left.+\left(\bar{\gamma }^2-1\right) \cosh ^2(\eta ) m_T^2+p_T^2 \left(\bar{\gamma }^2 \cos ^2(\phi )\right.\right.\right. \nonumber\\
   &\left.\left.\left.-\sin ^2(\phi )\right)\right)+2 \bar{\gamma }
   \left(f_2^{\text{shearVel}}-f_4^{\text{shearVel}}\right)\right.\nonumber\\
   &\left.\times\left(p_T \cos (\phi )-\bar{v} \cosh (\eta ) m_T\right)-7 \bar{\gamma } f_4^{\text{shearVel}}\right.\nonumber\\
   &\left.\times \left(\bar{v}  \cosh (\eta ) m_T-p_T \cos (\phi )\right)\right. \nonumber\\
   &\times\left.\left(-2 \bar{\gamma }^2 \bar{v} \cosh (\eta ) m_T p_T \cos (\phi )\right. \right.\nonumber\\
   &\left.\left.+\left(\bar{\gamma }^2-1\right)   \cosh ^2(\eta )   m_T^2\right.\right. \nonumber\\
   &\left.\left.+p_T^2 \left(\bar{\gamma }^2 \cos ^2(\phi )-\sin ^2(\phi )\right)\right)\right)
   \end{align}
  %shearvel7
    \begin{align}
    K_7^\text{shearVel} &= \int \ud \phi \ud \eta \; -\frac{2}{5} p^2 \bar{\gamma } \left(f_1^{\text{shearVel}}-f_3^{\text{shearVel}}\right) \bar{E}_p \nonumber\\
   &\times \left(p_T \cos (\phi )-\bar{v} \cosh (\eta ) m_T\right) \nonumber\\
   &+\frac{1}{21}   \bar{v} \left(f_2^{\text{shearVel}}-f_4^{\text{shearVel}}\right) \left(30 \bar{\gamma }^2 \bar{v} \cosh (\eta ) m_T\right. \nonumber\\
   &\left.\times p_T \cos (\phi )-15 \left(\bar{\gamma }^2-1\right)   \cosh ^2(\eta ) m_T^2\right. \nonumber\\
   &\left.-15 \bar{\gamma }^2 p_T^2 \cos ^2(\phi )+3 \sinh ^2(\eta ) m_T^2+4 p^2\right)\nonumber\\
   &+f_3^{\text{shearVel}} \bar{E}_p \left(\bar{\gamma } \bar{v} \cosh (\eta ) m_T-\bar{\gamma } p_T \cos (\phi )\right) \nonumber\\
   &\times \left(-2 \bar{\gamma }^2 \bar{v} \cosh (\eta ) m_T p_T \cos (\phi )\right. \nonumber\\
   &\left.+\left(\bar{\gamma }^2-1\right) \cosh ^2(\eta ) m_T^2+\bar{\gamma }^2 p_T^2 \cos ^2(\phi )\right. \nonumber\\
   &\left.-\sinh ^2(\eta ) m_T^2\right)-\bar{\gamma } \bar{v} f_4^{\text{shearVel}} \left(\bar{v} \cosh (\eta ) m_T\right. \nonumber\\
   &\left.-p_T \cos (\phi )\right) \left(\bar{\gamma } \bar{v} \cosh (\eta ) m_T-\bar{\gamma } p_T \cos (\phi \right) \nonumber\\
   &\times\left(-2 \bar{\gamma }^2 \bar{v} \cosh (\eta ) m_T p_T \cos (\phi )\right. \nonumber\\
   &\left.+\left(\bar{\gamma }^2-1\right) \cosh ^2(\eta ) m_T^2+\bar{\gamma }^2 p_T^2 \cos ^2(\phi )\right. \nonumber\\
   &\left.-\sinh ^2(\eta ) m_T^2\right)-\frac{2}{15} p^4 \bar{v}   \left(f_0^{\text{shearVel}}-f_4^{\text{shearVel}}\right) 
   \end{align}
  %shearvel8
    \begin{align}
    K_8^\text{shearVel} &= \int \ud \phi \ud \eta \; \frac{2}{5} p^2 \bar{v} \left(f_1^{\text{shearVel}}-f_3^{\text{shearVel}}\right) \bar{E}_p \nonumber \\
    &\times \left(\bar{\gamma } \bar{v} \cosh (\eta ) m_T-\bar{\gamma } p_T \cos (\phi )\right)\nonumber\\
    &+\frac{1}{21} \left(f_2^{\text{shearVel}}-f_4^{\text{shearVel}}\right) \left(30 \bar{\gamma }^2 \bar{v} \cosh (\eta ) m_T \right. \nonumber\\
   &\left. \times  p_T \cos (\phi )-15 \left(\bar{\gamma   }^2-1\right) \cosh ^2(\eta ) m_T^2\right. \nonumber\\
   &\left. -15 \bar{\gamma }^2 p_T^2 \cos ^2(\phi )+3 \sinh 2(\eta ) m_T^2+4 p^2\right)\nonumber\\
   &+\bar{v} f_3^{\text{shearVel}} \bar{E}_p   \left(\bar{\gamma } \bar{v} \cosh (\eta ) m_T-\bar{\gamma } p_T \cos (\phi )\right)\nonumber\\
   &\times \left(-2 \bar{\gamma }^2 \bar{v} \cosh (\eta ) m_T p_T \cos (\phi)\right. \nonumber\\
   &\left.+\left(\bar{\gamma }^2-1\right) \cosh ^2(\eta ) m_T^2+\bar{\gamma }^2 p_T^2 \cos ^2(\phi )\right. \nonumber\\
   &\left.-\sinh ^2(\eta ) m_T^2\right)-\bar{\gamma } f_4^{\text{shearVel}}   \left(\bar{v} \cosh (\eta ) m_T\right.  \nonumber\\
   &\left.-p_T \cos (\phi )\right) \left(\bar{\gamma } \bar{v} \cosh (\eta ) m_T-\bar{\gamma } p_T \cos (\phi )\right) \nonumber\\
   &\times \left(-2 \bar{\gamma }^2  \bar{v} \cosh (\eta ) m_T p_T \cos (\phi )\right.\nonumber\\
   &\left.+\left(\bar{\gamma }^2-1\right) \cosh ^2(\eta ) m_T^2+\bar{\gamma }^2 p_T^2 \cos ^2(\phi )\right. \nonumber\\
   &\left.-\sinh ^2(\eta )  m_T^2\right)-\frac{2}{15} p^4 \left(f_0^{\text{shearVel}}-f_4^{\text{shearVel}}\right)
   \end{align}
  %shearvel9
    \begin{align}
    K_9^\text{shearVel} &= \int \ud \phi \ud \eta \; \frac{1}{7} \bar{\gamma } p_T \sin (\phi ) \left(7 f_3^{\text{shearVel}} \bar{E}_p\right.\nonumber\\
   &\left.\times \left(-2 \bar{\gamma }^2 \bar{v} \cosh (\eta ) m_T p_T \cos (\phi)\right.\right.\nonumber\\
   &\left.\left.+\left(\bar{\gamma }^2-1\right) \cosh^2(\eta ) m_T^2+\bar{\gamma }^2 p_T^2 \cos^2(\phi)\right.\right.\nonumber\\
   &\left.\left.-\sinh^2(\eta )m_T^2\right)+2 \bar{\gamma } \bar{v}\left(f_2^{\text{shearVel}}-f_4^{\text{shearVel}}\right)\right.\nonumber\\
   &\left.\times \left(p_T \cos (\phi )-\bar{v} \cosh (\eta ) m_T\right)-7 \bar{\gamma } \bar{v} f_4^{\text{shearVel}} \right.\nonumber\\
   &\left.\times \left(\bar{v} \cosh (\eta ) m_T-p_T \cos (\phi)\right) \left(-2 \bar{\gamma }^2 \bar{v} \cosh (\eta ) \right. \right. \nonumber\\
   &\left.\left. \times m_T p_T \cos (\phi )+\left(\bar{\gamma }^2-1\right) \cosh^2(\eta ) m_T^2\right. \right. \nonumber\\
   &\left.\left.+\bar{\gamma }^2 p_T^2 \cos ^2(\phi )-\sinh ^2(\eta ) m_T^2\right)\right) 
   \end{align}
  %shearvel10
   \begin{align}
   K_{10}^\text{shearVel} &= \int \ud \phi \ud \eta \; \frac{1}{7} \bar{\gamma } p_T \sin (\phi ) \left(7 \bar{v} f_3^{\text{shearVel}} \bar{E}_p\right.\nonumber\\
   &\left. \times \left(-2 \bar{\gamma }^2 \bar{v} \cosh (\eta ) m_T p_T \cos (\phi )\right.\right.\nonumber\\
   &\left.\left.+\left(\bar{\gamma }^2-1\right) \cosh ^2(\eta ) m_T^2+\bar{\gamma }^2 p_T^2 \cos ^2(\phi )\right.\right.\nonumber\\
   &\left.\left.-\sinh ^2(\eta ) m_T^2\right)+2 \bar{\gamma } \left(f_2^{\text{shearVel}}-f_4^{\text{shearVel}}\right) \right.\nonumber\\
   &\left.\times \left(p_T \cos (\phi )-\bar{v} \cosh (\eta ) m_T\right)-7 \bar{\gamma } f_4^{\text{shearVel}} \right.\nonumber\\
   &\left. \times\left(\bar{v}\cosh (\eta ) m_T-p_T \cos (\phi )\right) \right.\nonumber\\
   &\left.\times \left(-2 \bar{\gamma }^2 \bar{v} \cosh (\eta ) m_T p_T \cos (\phi )\right.\right.\nonumber\\
   &\left.\left.+\left(\bar{\gamma }^2-1\right) \cosh ^2(\eta )   m_T^2+\bar{\gamma }^2 p_T^2 \cos ^2(\phi )\right.\right.\nonumber\\
   &\left.\left.-\sinh ^2(\eta ) m_T^2\right)\right) 
  \end{align}
  %shearvel11
    \begin{align}
    K_{11}^\text{shearVel} &= \int \ud \phi \ud \eta \; \frac{1}{5} \bar{\gamma } \sinh (\eta ) m_T \left(5 f_3^{\text{shearVel}} \bar{E}_p \right. \nonumber\\&\left.\times \left(-2 \bar{\gamma }^2 \bar{v} \cosh (\eta ) m_T p_T \cos (\phi )\right.\right. \nonumber\\&\left.\left.+\left(\bar{\gamma
   }^2-1\right) \cosh ^2(\eta ) m_T^2+\bar{\gamma }^2 p_T^2 \cos ^2(\phi )\right.\right.\nonumber\\
   &\left.\left.-\sinh ^2(\eta ) m_T^2\right)-5 \bar{\gamma } \bar{v} f_4^{\text{shearVel}}\right.\nonumber\\
   &\left.\times \left(\bar{v} \cosh  (\eta ) m_T-p_T \cos (\phi )\right) \right.\nonumber\\
   &\left.\times \left(-2 \bar{\gamma }^2 \bar{v} \cosh (\eta ) m_T p_T \cos (\phi )\right.\right.\nonumber\\
   &\left.\left.+\left(\bar{\gamma }^2-1\right) \cosh ^2(\eta ) m_T^2+\bar{\gamma }^2 p_T^2 \cos ^2(\phi )\right.\right.\nonumber\\
   &\left.\left.-\sinh ^2(\eta ) m_T^2\right)\right.\nonumber\\
   &\left.-2 p^2 \left(f_1^{\text{shearVel}}-f_3^{\text{shearVel}}\right) \bar{E}_p\right) 
\end{align}
  %shearvel12
    \begin{align}
    K_{12}^\text{shearVel} &= \int \ud \phi \ud \eta \; \frac{1}{5} \bar{\gamma } \sinh (\eta ) m_T \left(5 \bar{v} f_3^{\text{shearVel}} \bar{E}_p \right. \nonumber\\
   &\left.\times \left(-2 \bar{\gamma }^2 \bar{v} \cosh (\eta ) m_T p_T \cos (\phi )\right.\right. \nonumber\\
   &\left.\left.+\left(\bar{\gamma }^2-1\right) \cosh ^2(\eta ) m_T^2+\bar{\gamma }^2 p_T^2 \cos ^2(\phi )\right.\right.\nonumber\\
   &\left.\left.-\sinh ^2(\eta ) m_T^2\right)-5 \bar{\gamma } f_4^{\text{shearVel}}\right.\nonumber\\
   &\left.\times \left(\bar{v} \cosh (\eta ) m_T-p_T \cos (\phi )\right) \right.\nonumber\\
   &\left.\times \left(-2 \bar{\gamma }^2 \bar{v} \cosh (\eta ) m_T p_T \cos (\phi )\right.\right.\nonumber\\
   &\left.\left.+\left(\bar{\gamma }^2-1\right) \cosh^2(\eta ) m_T^2+\bar{\gamma }^2 p_T^2 \cos ^2(\phi )\right.\right.\nonumber\\
   &\left.\left.-\sinh ^2(\eta ) m_T^2\right)\right.\nonumber\\
   &\left.-2 p^2 \bar{v} \left(f_1^{\text{shearVel}}-f_3^{\text{shearVel}}\right)
   \bar{E}_p\right) 
   \end{align}
   \begin{align}
    K_1^\text{shearShear} &= \int \ud \phi \ud \eta \; -\frac{2}{15} \cosh (\eta  k) \cos (m \phi ) \nonumber\\
    &\times \left(5 \bar{\gamma } f_2^{\text{shearShear}} \bar{E}_p \right.\nonumber\\
    &\left.\times\left(-6 \bar{\gamma }^3 \bar{v} \cosh (\eta ) m_T p_T \cos (\phi ) \right.\right.\nonumber\\
    &\left.\left.\times \left(-3\bar{\gamma }^2+\bar{\gamma } \bar{v}^5-3 \left(\bar{\gamma }-1\right) \bar{\gamma } \bar{v}^3+2 \bar{\gamma } \bar{v}+2\right)\right.\right.\nonumber\\
    &\left.\left.-3 \left(\bar{\gamma }^2-1\right) \bar{\gamma } \cosh ^2(\eta ) m_T^2 \left(3 \bar{\gamma }^2-\bar{\gamma } \bar{v}^5\right.\right.\right.\nonumber\\
    &\left.\left.\left.+3 \left(\bar{\gamma }-1\right) \bar{\gamma } \bar{v}^3-2 \bar{\gamma } \bar{v}-2\right).\right.\right.\nonumber\\
    &\left.\left.+p^2 \left(3  \bar{\gamma }^3-2 \bar{\gamma }-3 \bar{\gamma }^2 \bar{v}^3+3 \bar{\gamma }^3 \bar{v}-3 \bar{\gamma }^2 \bar{v}.\right.\right.\right.\nonumber\\
    &\left.\left.\left.-3 \bar{\gamma } \bar{v}+\bar{v}^3+\bar{v}\right)+3 \bar{\gamma }^3   p_T^2 \cos ^2(\phi ).\right.\right.\nonumber\\
    &\left.\left.\times \left(-3 \bar{\gamma }^2+\bar{\gamma } \bar{v}^5-3 \left(\bar{\gamma }-1\right) \bar{\gamma } \bar{v}^3+2 \bar{\gamma } \bar{v}+2\right)\right).\right.\nonumber\\
    &\left.+3 f_1^{\text{shearShear}} \left(p^2 \left(-6 \bar{\gamma }^4+9 \bar{\gamma }^2+3 \bar{\gamma }.\right.\right.\right.\nonumber\\
    &\left.\left.\left.+\bar{\gamma }^5 \left(\bar{v}^3+2 \bar{v}+3\right)+\bar{\gamma }^3 \left(\bar{v}-6\right)-3\right).\right.\right.\nonumber\\
    &\left.\left.+5 \bar{E}_p^2 \left(-6 \bar{\gamma }^4-3 \bar{\gamma }^3+7 \bar{\gamma }^2+\bar{\gamma }^5 \left(\bar{v}^3+2 \bar{v}+3\right)-2\right)\right)\right.\nonumber\\
    &\left.\times\left(\bar{v} \cosh (\eta ) m_T-p_T \cos (\phi )\right)\right.\nonumber\\
    &\left.-3 \bar{\gamma }^6 f_3^{\text{shearShear}} \left(\bar{v} \cosh (\eta ) m_T-p_T \cos (\phi \right)\right.\nonumber\\
    &\left.\times  \left(p^2 \left(-6 \bar{\gamma }^4+9 \bar{\gamma }^2+3 \bar{\gamma}\right.\right.\right.\nonumber\\
    &\left.\left.\left.+\bar{\gamma }^5 \left(\bar{v}^3+2 \bar{v}+3\right)+\bar{\gamma }^3 \left(\bar{v}-6\right)-3\right)\bar{\gamma}^{-6}\right.\right.\nonumber\\
    &\left.\left. -5 \bar{v} \left(\bar{\gamma }+\left(\bar{\gamma }-1\right) \bar{v}^3-\bar{v}\right) \left(p_T \cos (\phi )\right.\right.\right.\nonumber\\
    &\left.\left.\left.-\bar{v} \cosh (\eta ) m_T\right){}^2\right)-5 \bar{\gamma } f_0^{\text{shearShear}} \bar{E}_p \left(p^2 \right.\right.\nonumber\\
    &\left.\left.\times\left(3 \bar{\gamma }^3-2 \bar{\gamma }-3 \bar{\gamma }^2 \bar{v}^3+3 \bar{\gamma }^3 \bar{v}-3 \bar{\gamma }^2 \bar{v}-3 \bar{\gamma } \bar{v}\right.\right.\right.\nonumber\\
    &\left.\left.\left.+\bar{v}^3+\bar{v}\right)+3 \bar{v} \bar{E}_p^2 \left(-2 \bar{\gamma }^2+\bar{\gamma }^3 \left(\bar{v}+1\right)+1\right)\right)\right)
   \end{align}
   \begin{align}
    K_2^\text{shearShear} &= \int \ud \phi \ud \eta \;  \frac{2}{15 \bar{\gamma }} \cosh (\eta  k) \cos (m \phi ) \nonumber\\
    &\times \left(5 f_2^{\text{shearShear}} \bar{E}_p \left(-6 \bar{\gamma }^2 \bar{v} \cosh (\eta ) m_T p_T \cos (\phi )\right. \right.\nonumber\\
   &\left.\left.\times\left(-6 \bar{\gamma }^4-4  \bar{\gamma }^3+7 \bar{\gamma }^2+\bar{\gamma }+3 \bar{\gamma }^5 \left(\bar{v}+1\right)-2\right)\right.\right.\nonumber\\
   &\left.\left.+3 \left(\bar{\gamma }^2-1\right) \cosh ^2(\eta ) m_T^2 \left(-6 \bar{\gamma  }^4-4 \bar{\gamma }^3+7 \bar{\gamma }^2\right.\right.\right.\nonumber\\
   &\left.\left.\left.+\bar{\gamma }+3 \bar{\gamma }^5 \left(\bar{v}+1\right)-2\right)-\left(p^2 \left(-6 \bar{\gamma }^4+7 \bar{\gamma }^2\right.\right.\right.\right.\nonumber\\
   &\left.\left.\left.\left.+\bar{\gamma }+3   \bar{\gamma }^5 \left(\bar{v}^3+1\right)+\bar{\gamma }^3 \left(3 \bar{v}-4\right)-2\right)\right)\right.\right.\nonumber\\
   &\left.\left.+3 \bar{\gamma }^2 p_T^2 \cos ^2(\phi ) \left(-6 \bar{\gamma }^4-4 \bar{\gamma   }^3+7 \bar{\gamma }^2\right.\right.\right.\nonumber\\
   &\left.\left.\left.+\bar{\gamma }+3 \bar{\gamma }^5 \left(\bar{v}+1\right)-2\right)\right)-3 \bar{\gamma } f_1^{\text{shearShear}}\right.\nonumber\\
   &\left.\times \left(p^2 \bar{\gamma }^2 \left(3 \bar{\gamma   }^3-\bar{\gamma }^2 \bar{v}^5+\left(2 \bar{\gamma }^3-3 \bar{\gamma }^2-1\right) \bar{v}^3\right.\right.\right.\nonumber\\
   &\left.\left.\left.+\left(\bar{\gamma }^3-2 \bar{\gamma }^2-\bar{\gamma }-1\right) \bar{v}\right)+5 \bar{v}   \bar{E}_p^2 \left(-6 \bar{\gamma }^4-2 \bar{\gamma }^3\right.\right.\right.\nonumber\\
   &\left.\left.\left.+5 \bar{\gamma }^2+3 \bar{\gamma }^5 \left(\bar{v}+1\right)-1\right)\right) \left(\bar{v} \cosh (\eta ) m_T\right.\right.\nonumber\\
   &\left.\left.-p_T \cos (\phi  )\right)-3 \bar{\gamma }^7 f_3^{\text{shearShear}} \left(\bar{v} \cosh (\eta ) m_T\right.\right.\nonumber\\
   &\left.\left.-p_T \cos (\phi )\right) \left(5 \left(\bar{\gamma }+\left(\bar{\gamma }-1\right)  \bar{v}^3-\bar{v}\right) \left(p_T \cos (\phi )\right.\right.\right.\nonumber\\
   &\left.\left.\left.-\bar{v} \cosh (\eta ) m_T\right){}^2+p^2 \bar{\gamma }^{-4} \left(-3 \bar{\gamma }^3+\bar{\gamma }^2 \bar{v}^5\right.\right.\right.\nonumber\\
   &\left.\left.\left.+\left(-2 \bar{\gamma }^3+3   \bar{\gamma }^2+1\right) \bar{v}^3+\left(-\bar{\gamma }^3+2 \bar{\gamma }^2+\bar{\gamma }+1\right) \bar{v}\right)\right)\right.\nonumber\\
   &\left.+5 f_0^{\text{shearShear}} \bar{E}_p   \left(p^2 \left(-6 \bar{\gamma }^4+7 \bar{\gamma }^2\right.\right.\right.\nonumber\\
   &\left.\left.\left.+\bar{\gamma }+3 \bar{\gamma }^5 \left(\bar{v}^3+1\right)\right.\right.\right.\nonumber\\
   &\left.\left.\left.+\bar{\gamma }^3 \left(3 \bar{v}-4\right)-2\right)+3  \left(\bar{\gamma }^2-1\right) \bar{E}_p^2 \left(-2 \bar{\gamma }^2\right.\right.\right.\nonumber\\
   &\left.\left.\left.+\bar{\gamma }^3 \left(\bar{v}+1\right)+1\right)\right)\right)
   \end{align}
   \begin{align}
    K_3^\text{shearShear} &= \int \ud \phi \ud \eta \;  2 \bar{\gamma } p_T \sin (\phi ) \cosh (\eta  k) \sin (m \phi )\nonumber\\
    &\times\left(5 \bar{v} \cosh (\eta ) m_T \left(2 \bar{v} f_3^{\text{shearShear}} p_T \cos (\phi )\right.\right.\nonumber\\
   &\left.\left.+\bar{\gamma }
   \left(\bar{v}-1\right)^2 \left(\bar{v}+1\right) f_2^{\text{shearShear}} \bar{E}_p\right)\right.\nonumber\\
   &\left.-5 \bar{v}^3 f_3^{\text{shearShear}} \cosh ^2(\eta ) m_T^2+\left(\bar{v}-1\right)  \left(\bar{v}+1\right)^2  \right.\nonumber\\
   &\left.\times\left(f_1^{\text{shearShear}}\left(p^2 \left(\bar{\gamma }^2 \left(\bar{v}-1\right)+1\right)\right.\right.\right.\nonumber\\
   &\left.\left.\left.+5 \bar{\gamma }^2 \left(\bar{v}-1\right)   \bar{E}_p^2\right)\right.\right.\nonumber\\
   &\left.\left.+p^2 f_3^{\text{shearShear}} \left(-\left(\bar{\gamma }^2 \left(\bar{v}-1\right)\right)-1\right)\right)\right.\nonumber\\
   &\left.-5 \bar{\gamma } \left(\bar{v}-1\right)^2
   \left(\bar{v}+1\right) f_2^{\text{shearShear}} p_T \cos (\phi ) \bar{E}_p\right.\nonumber\\
   &\left.-5 \bar{v} f_3^{\text{shearShear}} p_T^2 \cos ^2(\phi )\right)\nonumber\\
   &/\left(5 \left(\bar{v}-1\right)
   \left(\bar{v}+1\right)^2\right)
       \end{align}
   \begin{align}
    K_4^\text{shearShear} &= \int \ud \phi \ud \eta \;  -2 p_T \sin (\phi ) \cosh (\eta  k) \sin (m \phi ) \nonumber\\
    &\times \left(5 \bar{\gamma }^2 \bar{v} \cosh (\eta ) m_T \left(\bar{\gamma } \left(\bar{v}-1\right)^2 \left(\bar{v}+1\right)\right.\right.\nonumber\\
   &\left.\left. \times f_2^{\text{shearShear}} \bar{E}_p-2 f_3^{\text{shearShear}} p_T \cos (\phi )\right)\right.\nonumber\\
   &\left.+5 \left(\bar{\gamma }^2-1\right) f_3^{\text{shearShear}} \cosh ^2(\eta ) m_T^2\right.\nonumber\\
   &\left.+\bar{\gamma }^2 \left(\bar{\gamma }^2 \left(\bar{v}^2-1\right)^2 \left(f_1^{\text{shearShear}} \left(p^2-5 \bar{v} \bar{E}_p^2\right)\right.\right.\right.\nonumber\\
   &\left.\left.\left.-p^2 f_3^{\text{shearShear}}\right)-5 \bar{\gamma } \left(\bar{v}-1\right)^2 \left(\bar{v}+1\right) f_2^{\text{shearShear}} \right.\right.\nonumber\\
   &\left.\left. \times p_T \cos (\phi ) \bar{E}_p+5 f_3^{\text{shearShear}} p_T^2 \cos ^2(\phi )\right)\right)\nonumber \\
   &\times/\left(5 \bar{\gamma }
   \left(\bar{v}-1\right) \left(\bar{v}+1\right)^2\right)
       \end{align}
   \begin{align}
    K_5^\text{shearShear} &= \int \ud \phi \ud \eta \;  0
       \end{align}
   \begin{align}
    K_6^\text{shearShear} &= \int \ud \phi \ud \eta \;  0
       \end{align}
       \begin{align}
   K_7^\text{shearShear} &= \int \ud \phi \ud \eta \;-2 \cosh (\eta  k) \cos (m \phi ) \left(5 f_2^{\text{shearShear}} \right. \nonumber\\
   &\left.\times \bar{E}_p \left(-6 \bar{\gamma }^2 \bar{v}   \cosh (\eta ) m_T p_T \cos (\phi ) \left(-6 \bar{\gamma }^4-4 \bar{\gamma }^3\right. \right.\right.\nonumber\\
   &\left.\left.\left.+7 \bar{\gamma }^2+\bar{\gamma }+3   \bar{\gamma }^5 \left(\bar{v}+1\right)-2\right)\right.\right.\nonumber\\
   &\left.\left.+3 \left(\bar{\gamma }^2-1\right) \cosh ^2(\eta )\right. \right.\nonumber\\
   &\left.\left.\times  m_T^2 \left(-6   \bar{\gamma }^4-4 \bar{\gamma }^3+7 \bar{\gamma }^2+\bar{\gamma }+3 \bar{\gamma }^5   \left(\bar{v}+1\right)-2\right)\right. \right.\nonumber\\
   &\left.\left.-\left(p^2 \left(-6 \bar{\gamma }^4+7 \bar{\gamma }^2+\bar{\gamma }+3
   \bar{\gamma }^5 \left(\bar{v}^3+1\right)\right. \right.\right.\right.\nonumber\\
   &\left.\left.\left.\left.+\bar{\gamma }^3 \left(3 \bar{v}-4\right)-2\right)\right)+3 \bar{\gamma   }^2 p_T^2 \cos ^2(\phi ) \left(-6 \bar{\gamma }^4\right. \right.\right.\nonumber\\
   &\left.\left.\left.-4 \bar{\gamma }^3+7 \bar{\gamma }^2+\bar{\gamma }+3   \bar{\gamma }^5 \left(\bar{v}+1\right)-2\right)\right)\right. \nonumber\\
   &\left.-3 \bar{\gamma } f_1^{\text{shearShear}} \left(p^2   \bar{\gamma }^2 \left(3 \bar{\gamma }^3-\bar{\gamma }^2 \bar{v}^5\right. \right.\right.\nonumber\\
   &\left.\left.\left.+\left(2 \bar{\gamma }^3-3 \bar{\gamma   }^2-1\right) \bar{v}^3+\left(\bar{\gamma }^3-2 \bar{\gamma }^2-\bar{\gamma }-1\right) \bar{v}\right)\right. \right.\nonumber\\
   &\left.\left.+5 \bar{v}   \bar{E}_p^2 \left(-6 \bar{\gamma }^4-2 \bar{\gamma }^3+5 \bar{\gamma }^2+3 \bar{\gamma }^5   \left(\bar{v}+1\right)-1\right)\right)\right. \nonumber\\
   &\left.\times \left(\bar{v} \cosh (\eta ) m_T-p_T \cos (\phi )\right)\right. \nonumber\\
   &\left.-3 p^2   \bar{\gamma }^3 f_3^{\text{shearShear}} \left(-3 \bar{\gamma }^3+\bar{\gamma }^2 \bar{v}^5\right.\right. \nonumber\\
   &\left.\left.+\left(-2 \bar{\gamma   }^3+3 \bar{\gamma }^2+1\right) \bar{v}^3+\left(-\bar{\gamma }^3+2 \bar{\gamma }^2+\bar{\gamma }+1\right)\right. \right.\nonumber\\
   &\left.\left.\times   \bar{v}\right) \left(\bar{v} \cosh (\eta ) m_T-p_T \cos (\phi )\right)-15 \bar{\gamma }^7\right. \nonumber\\
   &\left.  \times f_3^{\text{shearShear}} \left(-\bar{\gamma }-\left(\left(\bar{\gamma }-1\right) \bar{v}^3\right)+\bar{v}\right)   \left(p_T \cos (\phi )\right.\right. \nonumber\\
   &\left.\left.-\bar{v} \cosh (\eta ) m_T\right){}^3+5 f_0^{\text{shearShear}} \bar{E}_p \left(p^2   \left(-6 \bar{\gamma }^4\right. \right. \right. \nonumber\\
   &\left.\left.+7 \bar{\gamma }^2+\bar{\gamma }+3 \bar{\gamma }^5 \left(\bar{v}^3+1\right)+\bar{\gamma   }^3 \left(3 \bar{v}-4\right)-2\right)\right. \nonumber\\
   &\left.\left.+3 \left(\bar{\gamma }^2-1\right) \bar{E}_p^2 \left(-2 \bar{\gamma
   }^2+\bar{\gamma }^3 \left(\bar{v}+1\right)+1\right)\right)\right)/\left(15 \bar{\gamma }\right)  
   \end{align}
        \begin{align}
    K_8^\text{shearShear} &= \int \ud \phi \ud \eta \;2 \cosh (\eta  k) \cos (m \phi ) \left(5 f_2^{\text{shearShear}}\right.\nonumber\\
   &\left.\times  \bar{E}_p \left(-6 \bar{\gamma }^2 \bar{v}
   \cosh (\eta ) m_T p_T \cos (\phi ) \left(-6 \bar{\gamma }^4\right.\right.\right.\nonumber\\
   &\left.\left.\left.-4 \bar{\gamma }^3+7 \bar{\gamma }^2+\bar{\gamma }+3   \bar{\gamma }^5 \left(\bar{v}+1\right)-2\right)\right.\right.\nonumber\\
   &\left.\left.+3 \left(\bar{\gamma }^2-1\right) \cosh ^2(\eta ) m_T^2 \left(-6   \bar{\gamma }^4-4 \bar{\gamma }^3+7 \bar{\gamma }^2\right.\right.\right.\nonumber\\
   &\left.\left.\left.+\bar{\gamma }+3 \bar{\gamma }^5  \left(\bar{v}+1\right)-2\right)-\left(p^2 \left(-6 \bar{\gamma }^4+7 \bar{\gamma }^2\right.\right.\right.\right.\nonumber\\
   &\left.\left.\left.\left.+\bar{\gamma }+3  \bar{\gamma }^5 \left(\bar{v}^3+1\right)+\bar{\gamma }^3 \left(3 \bar{v}-4\right)-2\right)\right)\right.\right.\nonumber\\
   &\left.\left.+3 \bar{\gamma  }^2 p_T^2 \cos ^2(\phi ) \left(-6 \bar{\gamma }^4-4 \bar{\gamma }^3+7 \bar{\gamma }^2+\bar{\gamma }\right.\right.\right.\nonumber\\
   &\left.\left.\left.+3   \bar{\gamma }^5 \left(\bar{v}+1\right)-2\right)\right)-3 \bar{\gamma } f_1^{\text{shearShear}} \left(p^2   \bar{\gamma }^2 \right.\right.\nonumber\\
   &\left.\left.\times\left(3 \bar{\gamma }^3-\bar{\gamma }^2 \bar{v}^5+\left(2 \bar{\gamma }^3-3 \bar{\gamma   }^2-1\right) \bar{v}^3\right.\right.\right.\nonumber\\
   &\left.\left.\left.+\left(\bar{\gamma }^3-2 \bar{\gamma }^2-\bar{\gamma }-1\right) \bar{v}\right)+5 \bar{v}  \bar{E}_p^2 \left(-6 \bar{\gamma }^4-2 \bar{\gamma }^3\right.\right.\right.\nonumber\\
   &\left.\left.\left.+5 \bar{\gamma }^2+3 \bar{\gamma }^5  \left(\bar{v}+1\right)-1\right)\right) \left(\bar{v} \cosh (\eta ) m_T\right.\right.\nonumber\\
   &\left.\left.-p_T \cos (\phi )\right)-3 p^2  \bar{\gamma }^3 f_3^{\text{shearShear}} \left(-3 \bar{\gamma }^3+\bar{\gamma }^2 \bar{v}^5\right.\right.\nonumber\\
   &\left.\left.+\left(-2 \bar{\gamma}^3+3 \bar{\gamma }^2+1\right) \bar{v}^3+\left(-\bar{\gamma }^3+2 \bar{\gamma }^2+\bar{\gamma }+1\right) \bar{v}\right) \right.\nonumber\\
   &\left.\times\left(\bar{v} \cosh (\eta ) m_T-p_T \cos (\phi )\right)-15 \bar{\gamma }^7   f_3^{\text{shearShear}}\right.\nonumber\\
   &\left.\times \left(-\bar{\gamma }-\left(\left(\bar{\gamma }-1\right) \bar{v}^3\right)+\bar{v}\right)   \left(p_T \cos (\phi )\right.\right.\nonumber\\
   &\left.\left.-\bar{v} \cosh (\eta ) m_T\right){}^3+5 f_0^{\text{shearShear}} \bar{E}_p \left(p^2   \left(-6 \bar{\gamma }^4\right.\right.\right.\nonumber\\
   &\left.\left.\left.+7 \bar{\gamma }^2+\bar{\gamma }+3 \bar{\gamma }^5 \left(\bar{v}^3+1\right)+\bar{\gamma  }^3 \left(3 \bar{v}-4\right)-2\right)\right.\right.\nonumber\\
   &\left.\left.+3 \left(\bar{\gamma }^2-1\right) \bar{E}_p^2 \left(-2 \bar{\gamma  }^2+\bar{\gamma }^3 \left(\bar{v}+1\right)+1\right)\right)\right)/\left(15 \bar{\gamma }\right)
    \end{align}
   \begin{align}
    K_9^\text{shearShear} &= \int \ud \phi \ud \eta \;  0
       \end{align}
   \begin{align}
    K_{10}^\text{shearShear} &= \int \ud \phi \ud \eta \; 0
       \end{align}
   \begin{align}
    K_{11}^\text{shearShear} &= \int \ud \phi \ud \eta \;2 \sinh (\eta ) m_T \cosh (\eta  k) \cos (m \phi )\nonumber\\
    &\times \left(5 \bar{\gamma }^2 \bar{v} \cosh (\eta ) m_T \left(\bar{\gamma } \left(\bar{v}-1\right)^2   \left(\bar{v}+1\right)\right.\right.\nonumber\\
   &\left.\left.\times  f_2^{\text{shearShear}} \bar{E}_p-2  f_3^{\text{shearShear}} p_T \cos (\phi )\right)\right.\nonumber\\
   &\left.+5 \left(\bar{\gamma  }^2-1\right) f_3^{\text{shearShear}} \cosh ^2(\eta ) m_T^2\right.\nonumber\\
   &\left.+\bar{\gamma }^2   \left(\bar{\gamma }^2 \left(\bar{v}^2-1\right)^2   \left(f_1^{\text{shearShear}} \left(p^2-5 \bar{v} \bar{E}_p^2\right)\right.\right.\right.\nonumber\\
   &\left.\left.\left.-p^2   f_3^{\text{shearShear}}\right)-5 \bar{\gamma } \left(\bar{v}-1\right)^2   \left(\bar{v}+1\right) \right.\right.\nonumber\\
   &\left.\left.\times f_2^{\text{shearShear}} p_T \cos (\phi )   \bar{E}_p\right.\right.\nonumber\\
   &\left.\left.+5 f_3^{\text{shearShear}} p_T^2 \cos ^2(\phi )\right)\right)\nonumber\\
   &/\left(5   \bar{\gamma } \left(\bar{v}-1\right) \left(\bar{v}+1\right)^2\right)
       \end{align}
   \begin{align}
    K_{12}^\text{shearShear} &= \int \ud \phi \ud \eta \;-2 \sinh (\eta ) m_T \cosh (\eta  k) \cos (m \phi ) \nonumber\\
    &\times\left(5 \bar{\gamma   }^2 \bar{v} \cosh (\eta ) m_T \left(\bar{\gamma } \left(\bar{v}-1\right)^2   \left(\bar{v}+1\right) \right.\right.\nonumber\\
   &\left.\left.\times f_2^{\text{shearShear}} \bar{E}_p-2 f_3^{\text{shearShear}} p_T \cos (\phi )\right)\right.\nonumber\\
   &\left.+5 \left(\bar{\gamma}^2-1\right) f_3^{\text{shearShear}} \cosh ^2(\eta ) m_T^2\right.\nonumber\\
   &\left.+\bar{\gamma }^2 \left(\bar{\gamma }^2 \left(\bar{v}^2-1\right)^2\left(f_1^{\text{shearShear}} \left(p^2-5 \bar{v} \bar{E}_p^2\right)\right.\right.\right.\nonumber\\
   &\left.\left.\left.-p^2   f_3^{\text{shearShear}}\right)-5 \bar{\gamma } \left(\bar{v}-1\right)^2   \left(\bar{v}+1\right) f_2^{\text{shearShear}}\right.\right.\nonumber\\
   &\left.\left.\times p_T \cos (\phi ) \bar{E}_p+5 f_3^{\text{shearShear}} p_T^2 \cos ^2(\phi )\right)\right)\nonumber\\
   &/\left(5   \bar{\gamma } \left(\bar{v}-1\right) \left(\bar{v}+1\right)^2\right)
    \end{align}
   \begin{align}
    K_{13}^\text{shearShear} &= \int \ud \phi \ud \eta \;-\frac{2}{5} \bar{\gamma } p_T \sin (\phi ) \cosh (\eta  k) \cos (m \phi ))\nonumber\\
    &\times \left(-5 \bar{\gamma } \bar{v} \cosh (\eta ) m_T \left(f_2^{\text{shearShear}} \bar{E}_p\right.\right.\nonumber\\
    &\left.\left.+2 \bar{\gamma }  \bar{v} f_3^{\text{shearShear}} p_T \cos (\phi )\right)+5 \bar{\gamma }^2 \bar{v}^3 f_3^{\text{shearShear}} \right.\nonumber\\
    &\left.\times\cosh ^2(\eta ) m_T^2+p^2 \bar{v}   \left(f_1^{\text{shearShear}}-f_3^{\text{shearShear}}\right)\right.\nonumber\\
    &\left.+5 \bar{\gamma } f_2^{\text{shearShear}} p_T \cos (\phi ) \bar{E}_p\right.\nonumber\\
    &\left.+5 \bar{\gamma }^2 \bar{v} f_3^{\text{shearShear}}  p_T^2 \cos ^2(\phi )\right)
    \end{align}
   \begin{align}
    K_{14}^\text{shearShear} &= \int \ud \phi \ud \eta \;-\frac{2}{5} \bar{\gamma } p_T \sin (\phi ) \cosh (\eta  k) \cos (m \phi )\nonumber\\
    &\times\left(-5 \bar{\gamma } \bar{v} \cosh (\eta ) m_T \left(2 \bar{\gamma } f_3^{\text{shearShear}} p_T \cos(\phi)\right.\right.\nonumber\\
    &\left.\left.+\bar{v} f_2^{\text{shearShear}} \bar{E}_p\right)+5 \bar{\gamma }^2 \bar{v}^2 f_3^{\text{shearShear}} \right.\nonumber\\
    &\left.\times \cosh ^2(\eta ) m_T^2+5 \bar{\gamma }^2 f_3^{\text{shearShear}} p_T^2  \cos ^2(\phi )\right.\nonumber\\
    &\left.+5 \bar{\gamma } \bar{v} f_2^{\text{shearShear}} p_T \cos (\phi ) \bar{E}_p\right.\nonumber\\
    &\left.+p^2 \left(f_1^{\text{shearShear}}-f_3^{\text{shearShear}}\right)\right)
    \end{align}
   \begin{align}
    K_{15}^\text{shearShear} &= \int \ud \phi \ud \eta \;-\frac{2}{5} \bar{\gamma } \sinh (\eta ) m_T \cosh (\eta  k) \cos (m \phi)\nonumber\\
    &\times \left(-5 \bar{\gamma } \bar{v} \cosh (\eta ) m_T \left(f_2^{\text{shearShear}} \bar{E}_p \right.\right.\nonumber\\
    &\left.\left.+2 \bar{\gamma } \bar{v} f_3^{\text{shearShear}} p_T \cos (\phi )\right)+5 \bar{\gamma }^2 \bar{v}^3 f_3^{\text{shearShear}}\right.\nonumber\\
    &\left.\times \cosh ^2(\eta ) m_T^2+p^2 \bar{v}   \left(f_1^{\text{shearShear}}-f_3^{\text{shearShear}}\right)\right.\nonumber\\
    &\left.+5 \bar{\gamma } f_2^{\text{shearShear}} p_T \cos (\phi ) \bar{E}_p\right.\nonumber\\
    &\left.+5 \bar{\gamma }^2 \bar{v}   f_3^{\text{shearShear}} p_T^2 \cos ^2(\phi)\right)
    \end{align}
   \begin{align}
    K_{16}^\text{shearShear} &= \int \ud \phi \ud \eta \;-\frac{2}{5} \bar{\gamma } \sinh (\eta ) m_T \cosh (\eta  k) \cos (m \phi ) \nonumber\\
    &\times\left(-5 \bar{\gamma } \bar{v} \cosh (\eta ) m_T \left(2 \bar{\gamma } f_3^{\text{shearShear}}
   p_T \cos (\phi )\right.\right.\nonumber\\
    &\left.\left.+\bar{v} f_2^{\text{shearShear}} \bar{E}_p\right)+5 \bar{\gamma }^2 \bar{v}^2 f_3^{\text{shearShear}} \right.\nonumber\\
    &\left.\times\cosh ^2(\eta ) m_T^2+5 \bar{\gamma }^2 f_3^{\text{shearShear}} p_T^2 \cos ^2(\phi )\right.\nonumber\\
    &\left.+5 \bar{\gamma } \bar{v} f_2^{\text{shearShear}} p_T \cos (\phi ) \bar{E}_p\right.\nonumber\\
    &\left.+p^2 \left(f_1^{\text{shearShear}}-f_3^{\text{shearShear}}\right)\right)
    \end{align}
   \begin{align}
    K_{17}^\text{shearShear} &= \int \ud \phi \ud \eta \; 2 \bar{\gamma } \sinh (\eta ) m_T p_T \sin (\phi ) \cosh (\eta  k)\nonumber\\
    &\times \cos (m \phi ) \left(\bar{\gamma } \bar{v} f_3^{\text{shearShear}} \left(p_T \cos (\phi )\right.\right.\nonumber\\
    &\left.\left.-\bar{v} \cosh (\eta ) m_T\right)+f_2^{\text{shearShear}} \bar{E}_p\right)
    \end{align}
   \begin{align}
    K_{18}^\text{shearShear} &= \int \ud \phi \ud \eta \; 2 \bar{\gamma } \sinh (\eta ) m_T p_T \sin (\phi ) \cosh (\eta  k) \nonumber\\
    &\times \cos (m \phi ) \left(-\bar{\gamma } \bar{v} f_3^{\text{shearShear}} \cosh (\eta ) m_T\right.\nonumber\\
    &\left.+\bar{\gamma }
   f_3^{\text{shearShear}} p_T \cos (\phi )+\bar{v} f_2^{\text{shearShear}} \bar{E}_p\right)
    \end{align}
   \begin{align}
    K_{19}^\text{shearShear} &= \int \ud \phi \ud \eta \; \frac{1}{15} \bar{\gamma } \cosh (\eta  k) \cos (m \phi )\nonumber\\
    &\times \left(5 f_2^{\text{shearShear}} \bar{E}_p \left(-2 \bar{\gamma }^3 \bar{v}^2 \left(p^2\right.\right.\right.\nonumber\\
   &\left.\left.\left.-3 \bar{\gamma }^2 \left(p_T \cos (\phi   )-\bar{v} \cosh (\eta ) m_T\right){}^2\right)\right.\right.\nonumber\\
   &\left.\left.-3 \bar{\gamma }^2 \left(\bar{\gamma }^4 \left(p_T \cos (\phi )-\bar{v} \cosh (\eta ) m_T\right){}^2\right.\right.\right.\nonumber\\
   &\left.\left.\left.-\frac{1}{3} p^2 \bar{\gamma
   }^2\right)-\left(\bar{\gamma }^2-1\right)^2 \left(3 \bar{\gamma }^2 \left(p_T \cos (\phi )\right.\right.\right.\right.\nonumber\\
   &\left.\left.\left.\left.-\bar{v} \cosh (\eta ) m_T\right){}^2-p^2\right)-p^2\right.\right.\nonumber\\
   &\left.\left.+3 p_T^2 \sin ^2(\phi )\right)+6 p^2
   \left(\bar{\gamma }-1\right)^2 \bar{\gamma }^2 \left(\bar{\gamma }+1\right)\right.\nonumber\\
   &\left.\times \bar{v} f_1^{\text{shearShear}} \left(\bar{v} \cosh (\eta ) m_T-p_T \cos (\phi )\right)\right.\nonumber\\
   &\left.+6 p^2   \bar{\gamma } \left(2 \bar{\gamma }^4-2 \bar{\gamma }^3-2 \bar{\gamma }^2+2 \bar{\gamma }+1\right) \right.\nonumber\\
   &\left.\times\bar{v} f_1^{\text{shearShear}} \left(\bar{v} \cosh (\eta ) m_T-p_T \cos (\phi   )\right)\right.\nonumber\\
   &\left.+6 p^2 \bar{\gamma } \left(3 \bar{\gamma }^4-3 \bar{\gamma }^3-3 \bar{\gamma }^2+3 \bar{\gamma }+1\right) \right.\nonumber\\
   &\left.\times\bar{v} f_3^{\text{shearShear}} \left(p_T \cos (\phi )-\bar{v}   \cosh (\eta ) m_T\right)\right.\nonumber\\
   &\left.+15 \left(\bar{\gamma }-1\right) \bar{\gamma }^2 \bar{v} f_1^{\text{shearShear}} \bar{E}_p^2 \left(\bar{\gamma }^2 \left(\bar{v}^2+1\right)-1\right)\right.\nonumber\\
   &\left.\times\left(\bar{v} \cosh (\eta ) m_T-p_T \cos (\phi )\right)+30 \bar{\gamma }^2 \left(2 \bar{\gamma }^3\right.\right.\nonumber\\
   &\left.\left.-2 \bar{\gamma }^2-\bar{\gamma }+1\right) \bar{v} f_1^{\text{shearShear}}
   \bar{E}_p^2 \left(\bar{v} \cosh (\eta ) m_T\right.\right.\nonumber\\
   &\left.\left.-p_T \cos (\phi )\right)-30 \bar{\gamma }^3 \left(2 \bar{\gamma }^3-2 \bar{\gamma }^2-\bar{\gamma }+1\right)\right.\nonumber\\
   &\left.\times \bar{v}^2   f_2^{\text{shearShear}} \bar{E}_p \left(p_T \cos (\phi )-\bar{v} \cosh (\eta ) m_T\right){}^2\right.\nonumber\\
   &\left.+15 \bar{\gamma } \bar{v} f_3^{\text{shearShear}} \left(p_T \cos (\phi )-\bar{v} \cosh   (\eta ) m_T\right) \right.\nonumber\\
   &\left.\times\left(\bar{\gamma }^5 \left(-\bar{v}^2\right) \left(\bar{\gamma } \bar{v}^2-1\right) \left(p_T \cos (\phi )\right.\right.\right.\nonumber\\
   &\left.\left.\left.-\bar{v} \cosh (\eta )   m_T\right){}^2-\left(\bar{\gamma }^3-\bar{\gamma }^2+1\right) \bar{\gamma }^3 \left(p_T \cos (\phi )\right.\right.\right.\nonumber\\
   &\left.\left.\left.-\bar{v} \cosh (\eta ) m_T\right){}^2+p_T^2 \sin ^2(\phi )\right)-10 p^2   f_0^{\text{shearShear}} \right.\nonumber\\
   &\left.\times\bar{E}_p \left(-3 \bar{\gamma }^2+\bar{\gamma }^4 \left(\bar{v}^2+2\right)-\bar{\gamma }^3 \left(\bar{v}^2+2\right)\right.\right.\nonumber\\
   &\left.\left.+\bar{\gamma }   \left(\bar{v}^2+2\right)+1\right)+10 p^2 f_2^{\text{shearShear}} \bar{E}_p \left(-2 \bar{\gamma }^2\right.\right.\nonumber\\
   &\left.\left.+\bar{\gamma }^4 \left(\bar{v}^2+1\right)-\bar{\gamma }^3   \left(\bar{v}^2+1\right)+\bar{\gamma } \left(\bar{v}^2+1\right)+1\right)\right.\nonumber\\
   &\left.-15 \left(\bar{\gamma }-1\right) \bar{\gamma } f_0^{\text{shearShear}} \bar{E}_p^3 \left(\bar{\gamma }^2
   \left(\bar{v}^2+1\right)-1\right)\right)
    \end{align}
   \begin{align}
    K_{20}^\text{shearShear} &= \int \ud \phi \ud \eta \; \frac{1}{15} \bar{\gamma } \cosh (\eta  k) \cos (m \phi )\nonumber\\
    &\times\left(5 \bar{v} f_2^{\text{shearShear}} \bar{E}_p \left(-2 \bar{\gamma }^3 \bar{v}^2 \left(p^2\right.\right.\right.\nonumber\\
   &\left.\left.\left.-3 \bar{\gamma }^2 \left(p_T   \cos (\phi )-\bar{v} \cosh (\eta ) m_T\right){}^2\right)\right.\right.\nonumber\\
   &\left.\left.-3 \bar{\gamma }^2 \left(\bar{\gamma }^4 \left(p_T \cos (\phi )-\bar{v} \cosh (\eta ) m_T\right){}^2-\frac{1}{3} p^2   \bar{\gamma }^2\right)\right.\right.\nonumber\\
   &\left.\left.-\left(\bar{\gamma }^2-1\right)^2 \left(3 \bar{\gamma }^2 \left(p_T \cos (\phi )-\bar{v} \cosh (\eta ) m_T\right){}^2\right.\right.\right.\nonumber\\
   &\left.\left.\left.-p^2\right)-p^2+3 p_T^2 \sin ^2(\phi   )\right)+6 p^2 \bar{\gamma }^2 f_1^{\text{shearShear}} \right.\nonumber\\
   &\left.\times\left(\bar{\gamma }^3 \left(\bar{v}^4+1\right)-\bar{\gamma }^2 \left(\bar{v}^2+1\right)+1\right) \left(\bar{v} \cosh (\eta )  m_T\right.\right.\nonumber\\
   &\left.\left.-p_T \cos (\phi )\right)-6 p^2 \left(\bar{\gamma }-1\right)^2 \bar{\gamma }^2 \left(\bar{\gamma }+1\right) f_1^{\text{shearShear}}\right.\nonumber\\
   &\left.\times \left(p_T \cos (\phi )-\bar{v} \cosh (\eta )   m_T\right)+6 p^2 \bar{\gamma }^2 f_3^{\text{shearShear}}\right.\nonumber\\
   &\left.\times \left(-\bar{\gamma }+\bar{\gamma }^3 \left(\bar{v}^4+2\right)-\bar{\gamma }^2 \left(\bar{v}^2+2\right)+2\right) \left(p_T  \cos (\phi )\right.\right.\nonumber\\
   &\left.\left.-\bar{v} \cosh (\eta ) m_T\right)+15 \left(\bar{\gamma }-1\right) \bar{\gamma }^2 f_1^{\text{shearShear}} \bar{E}_p^2 \left(\bar{\gamma }^2 \right.\right.\nonumber\\
   &\left.\left.\times  \left(\bar{v}^2+1\right)-1\right) \left(\bar{v} \cosh (\eta ) m_T-p_T \cos (\phi )\right)\right.\nonumber\\
   &\left.+30 \bar{\gamma }^2 \left(2 \bar{\gamma }^3-2 \bar{\gamma }^2-\bar{\gamma }+1\right)   \bar{v}^2 f_1^{\text{shearShear}} \bar{E}_p^2 \right.\nonumber\\
   &\left.\times\left(\bar{v} \cosh (\eta ) m_T-p_T \cos (\phi )\right)-30 \bar{\gamma }^3 \left(2 \bar{\gamma }^3\right.\right.\nonumber\\
   &\left.\left.-2 \bar{\gamma }^2-\bar{\gamma
   }+1\right) \bar{v} f_2^{\text{shearShear}} \bar{E}_p \left(p_T \cos (\phi )\right.\right.\nonumber\\
   &\left.\left.-\bar{v} \cosh (\eta ) m_T\right){}^2-15 \bar{\gamma } f_3^{\text{shearShear}} \right.\nonumber\\
   &\left.\times\left(\bar{v} \cosh (\eta   ) m_T-p_T \cos (\phi )\right) \left(\bar{\gamma }^5 \left(-\bar{v}^2\right) \right.\right.\nonumber\\
   &\left.\left.\times\left(\bar{\gamma } \bar{v}^2-1\right) \left(p_T \cos (\phi )-\bar{v} \cosh (\eta )   m_T\right){}^2\right.\right.\nonumber\\
   &\left.\left.-\left(\bar{\gamma }^3-\bar{\gamma }^2+1\right) \bar{\gamma }^3 \left(p_T \cos (\phi )-\bar{v} \cosh (\eta ) m_T\right){}^2\right.\right.\nonumber\\
   &\left.\left.+p_T^2 \sin ^2(\phi )\right)-10 p^2   \bar{\gamma } \bar{v} f_0^{\text{shearShear}} \bar{E}_p \left(-3 \bar{\gamma }^2-\right.\right.\nonumber\\
   &\left.\left.\bar{\gamma }+\bar{\gamma }^3 \left(\bar{v}^2+2\right)+2\right)+10 p^2 \bar{\gamma } \bar{v}   f_2^{\text{shearShear}} \bar{E}_p \right.\nonumber\\
   &\left.\times\left(-2 \bar{\gamma }^2+\bar{\gamma }^3 \left(\bar{v}^2+1\right)+1\right)-15 \left(\bar{\gamma }-1\right)\right.\nonumber\\
   &\left.\times \bar{\gamma } \bar{v}   f_0^{\text{shearShear}} \bar{E}_p^3 \left(\bar{\gamma }^2 \left(\bar{v}^2+1\right)-1\right)\right)
    \end{align}
   \begin{align}
    K_{21}^\text{shearShear} &= \int \ud \phi \ud \eta \; \frac{1}{15} \bar{\gamma } \cosh (\eta  k) \cos (m \phi )\nonumber\\
    &\times \left(-5 f_2^{\text{shearShear}} \bar{E}_p \left(6 \bar{\gamma }^2 \bar{v} \cosh (\eta ) m_T\right.\right.\nonumber\\
   &\left.\left.\times p_T \cos (\phi ) \left(-2 \bar{\gamma }^4+2 \bar{\gamma }^2+2 \bar{\gamma }^3 \bar{v}^2-1\right)\right.\right.\nonumber\\
   &\left.\left.-3 \bar{\gamma }^2 \bar{v}^2 \cosh ^2(\eta ) m_T^2 \left(-2 \bar{\gamma }^4+2 \bar{\gamma }^2+2 \bar{\gamma }^3 \bar{v}^2\right.\right.\right.\nonumber\\
   &\left.\left.\left.-1\right)+\bar{\gamma }^2 \left(2 p^2 \left(-\bar{\gamma }^2+\bar{\gamma } \bar{v}^2+1\right)\right.\right.\right.\nonumber\\
   &\left.\left.\left.+3 p_T^2 \cos ^2(\phi ) \left(2\bar{\gamma }^4-2 \bar{\gamma }^2-2 \bar{\gamma }^3 \bar{v}^2+1\right)\right)\right.\right.\nonumber\\
   &\left.\left.-3 \sinh ^2(\eta ) m_T^2\right)+6 p^2 \left(\bar{\gamma }-1\right)^2 \bar{\gamma }^2 \left(\bar{\gamma }+1\right)\right.\nonumber\\
   &\left.\times \bar{v} f_1^{\text{shearShear}} \left(\bar{v} \cosh (\eta ) m_T-p_T \cos (\phi )\right)\right.\nonumber\\
   &\left.+6 p^2 \bar{\gamma } \left(2 \bar{\gamma }^4-2\bar{\gamma }^3-2 \bar{\gamma }^2+2 \bar{\gamma   }+1\right)\right.\nonumber\\
   &\left. \bar{v} f_1^{\text{shearShear}} \left(\bar{v} \cosh (\eta ) m_T-p_T \cos (\phi )\right)\right.\nonumber\\
   &\left.+6 p^2 \bar{\gamma } \left(3 \bar{\gamma }^4-3 \bar{\gamma }^3-3 \bar{\gamma }^2+3 \bar{\gamma }+1\right)\right.\nonumber\\
   &\left.\times \bar{v} f_3^{\text{shearShear}} \left(p_T \cos (\phi )-\bar{v} \cosh (\eta ) m_T\right)\right.\nonumber\\
   &\left.+15 \left(\bar{\gamma }-1\right) \bar{\gamma }^2 \bar{v}  f_1^{\text{shearShear}} \bar{E}_p^2 \left(\bar{\gamma }^2 \left(\bar{v}^2+1\right)\right.\right.\nonumber\\
   &\left.\left.-1\right) \left(\bar{v} \cosh (\eta ) m_T-p_T \cos (\phi )\right)\right.\nonumber\\
   &\left.+30 \bar{\gamma }^2 \left(2 \bar{\gamma }^3-2 \bar{\gamma }^2-\bar{\gamma }+1\right) \bar{v} f_1^{\text{shearShear}}\right.\nonumber\\
   &\left.\times \bar{E}_p^2 \left(\bar{v} \cosh (\eta ) m_T-p_T \cos (\phi )\right)\right.\nonumber\\
   &\left.-30 \bar{\gamma }^3   \left(2 \bar{\gamma }^3-2 \bar{\gamma }^2-\bar{\gamma }+1\right) \bar{v}^2 f_2^{\text{shearShear}}\right.\nonumber\\
   &\left.\times  \bar{E}_p \left(p_T \cos (\phi )-\bar{v} \cosh (\eta ) m_T\right){}^2\right.\nonumber\\
   &\left.+15   \bar{\gamma } \bar{v} f_3^{\text{shearShear}} \left(p_T \cos (\phi )-\bar{v} \cosh (\eta ) m_T\right)\right.\nonumber\\
   &\left.\times \left(\bar{\gamma }^5 \left(-\bar{v}^2\right) \left(\bar{\gamma }   \bar{v}^2-1\right) \left(p_T \cos (\phi )\right.\right.\right.\nonumber\\
   &\left.\left.\left.-\bar{v} \cosh (\eta ) m_T\right){}^2-\left(\bar{\gamma }^3-\bar{\gamma }^2+1\right) \bar{\gamma }^3\right.\right.\nonumber\\
   &\left.\left.\times \left(p_T \cos (\phi )-\bar{v} \cosh   (\eta ) m_T\right){}^2+\sinh ^2(\eta ) m_T^2\right)\right.\nonumber\\
   &\left.-10 p^2 f_0^{\text{shearShear}} \bar{E}_p \left(-3 \bar{\gamma }^2+\bar{\gamma }^4 \left(\bar{v}^2+2\right)\right.\right.\nonumber\\
   &\left.\left.-\bar{\gamma }^3   \left(\bar{v}^2+2\right)+\bar{\gamma } \left(\bar{v}^2+2\right)+1\right)+10 p^2 f_2^{\text{shearShear}} \right.\nonumber\\
   &\left.\times\bar{E}_p \left(-2 \bar{\gamma }^2+\bar{\gamma }^4   \left(\bar{v}^2+1\right)-\bar{\gamma }^3 \left(\bar{v}^2+1\right)\right.\right.\nonumber\\
   &\left.\left.+\bar{\gamma } \left(\bar{v}^2+1\right)+1\right)-15 \left(\bar{\gamma }-1\right) \bar{\gamma }
   f_0^{\text{shearShear}} \right.\nonumber\\
   &\left.\times \bar{E}_p^3 \left(\bar{\gamma }^2 \left(\bar{v}^2+1\right)-1\right)\right)
    \end{align}
   \begin{align}
    K_{22}^\text{shearShear} &= \int \ud \phi \ud \eta \; \frac{1}{15} \bar{\gamma } \cosh (\eta  k) \cos (m \phi ) \nonumber\\
    &\times \left(5 \bar{v} f_2^{\text{shearShear}} \bar{E}_p \left(-6 \bar{\gamma }^2 \bar{v} \cosh (\eta ) m_T\right.\right.\nonumber\\
   &\left.\left.\times p_T \cos (\phi ) \left(-2 \bar{\gamma }^4+2 \bar{\gamma }^2+2 \bar{\gamma }^3 \bar{v}^2-1\right)\right.\right.\nonumber\\
   &\left.\left.+3 \bar{\gamma }^2 \bar{v}^2 \cosh ^2(\eta ) m_T^2 \left(-2 \bar{\gamma }^4+2 \bar{\gamma }^2\right.\right.\right.\nonumber\\
   &\left.\left.\left.+2\bar{\gamma }^3 \bar{v}^2-1\right)+\bar{\gamma }^2 \left(-2 p^2 \left(-\bar{\gamma }^2+\bar{\gamma } \bar{v}^2+1\right)\right.\right.\right.\nonumber\\
   &\left.\left.\left.-3 p_T^2 \cos ^2(\phi ) \left(2 \bar{\gamma }^4-2 \bar{\gamma
   }^2-2 \bar{\gamma }^3 \bar{v}^2+1\right)\right)\right.\right.\nonumber\\
   &\left.\left.+3 \sinh ^2(\eta ) m_T^2\right)+6 p^2 \bar{\gamma }^2 f_1^{\text{shearShear}}\right.\nonumber\\
   &\left.\times \left(\bar{\gamma }^3 \left(\bar{v}^4+1\right)-\bar{\gamma }^2 \left(\bar{v}^2+1\right)+1\right)\right.\nonumber\\
   &\left.\times  \left(\bar{v} \cosh (\eta ) m_T-p_T \cos (\phi )\right)\right.\nonumber\\
   &\left.-6 p^2 \left(\bar{\gamma }-1\right)^2 \bar{\gamma   }^2 \left(\bar{\gamma }+1\right) f_1^{\text{shearShear}}\right.\nonumber\\
   &\left.\times  \left(p_T \cos (\phi )-\bar{v} \cosh (\eta ) m_T\right)\right.\nonumber\\
   &\left.+6 p^2 \bar{\gamma }^2 f_3^{\text{shearShear}} \left(-\bar{\gamma }+\bar{\gamma }^3 \left(\bar{v}^4+2\right)\right.\right.\nonumber\\
   &\left.\left.-\bar{\gamma }^2 \left(\bar{v}^2+2\right)+2\right) \left(p_T \cos (\phi )-\bar{v} \cosh (\eta ) m_T\right)\right.\nonumber\\
   &\left.+15 \left(\bar{\gamma }-1\right)\bar{\gamma }^2 f_1^{\text{shearShear}} \bar{E}_p^2 \left(\bar{\gamma }^2 \left(\bar{v}^2+1\right)-1\right)\right.\nonumber\\
   &\left.\times  \left(\bar{v} \cosh (\eta ) m_T-p_T \cos (\phi )\right)\right.\nonumber\\
   &\left.+30 \bar{\gamma }^2 \left(2 \bar{\gamma }^3-2 \bar{\gamma }^2-\bar{\gamma }+1\right) \bar{v}^2 f_1^{\text{shearShear}} \bar{E}_p^2\right.\nonumber\\
   &\left.\times \left(\bar{v} \cosh (\eta ) m_T-p_T \cos (\phi )\right)\right.\nonumber\\
   &\left.-30\bar{\gamma }^3 \left(2 \bar{\gamma }^3-2 \bar{\gamma }^2-\bar{\gamma }+1\right) \bar{v} f_2^{\text{shearShear}} \bar{E}_p\right.\nonumber\\
   &\left.\times \left(p_T \cos (\phi )-\bar{v} \cosh (\eta )
   m_T\right){}^2-15 \bar{\gamma } f_3^{\text{shearShear}}\right.\nonumber\\
   &\left.\times \left(\bar{v} \cosh (\eta ) m_T-p_T \cos (\phi )\right) \left(-\bar{\gamma }^5\bar{v}^2\right.\right.\nonumber\\
   &\left.\left.\times  \left(\bar{\gamma } \bar{v}^2-1\right) \left(p_T \cos (\phi )-\bar{v} \cosh (\eta ) m_T\right){}^2\right.\right.\nonumber\\
   &\left.\left.-\left(\bar{\gamma }^3-\bar{\gamma }^2+1\right) \bar{\gamma }^3 \left(p_T \cos (\phi )-\bar{v} \cosh(\eta ) m_T\right){}^2\right.\right.\nonumber\\
   &\left.\left.+\sinh ^2(\eta ) m_T^2\right)-10 p^2 \bar{\gamma } \bar{v} f_0^{\text{shearShear}} \bar{E}_p \right.\nonumber\\
   &\left.\times\left(-3 \bar{\gamma }^2-\bar{\gamma }+\bar{\gamma }^3 \left(\bar{v}^2+2\right)+2\right)\right.\nonumber\\
   &\left.+10 p^2 \bar{\gamma } \bar{v} f_2^{\text{shearShear}} \bar{E}_p \left(-2 \bar{\gamma }^2+\bar{\gamma }^3 \left(\bar{v}^2\right.\right.\right.\nonumber\\
   &\left.\left.\left.+1\right)+1\right)-15   \left(\bar{\gamma }-1\right) \bar{\gamma } \bar{v} f_0^{\text{shearShear}} \bar{E}_p^3 \right.\nonumber\\
   &\left.\times\left(\bar{\gamma }^2 \left(\bar{v}^2+1\right)-1\right)\right)
\end{align}
The perturbation kernel for perturbations around the diffusion correction term are given by
\begin{align}
    K_1^\text{diffTemp} &= \int \ud \phi \ud \eta \; \left(-\bar{\gamma } \bar{v} \cosh (\eta ) m_T \right.\nonumber\\
    &\left.\times\left(f_1^{\text{diffTemp}} \bar{E}_p+2 \bar{\gamma } \bar{v} f_2^{\text{diffTemp}} p_T \cos (\phi )\right)\right.\nonumber\\
    &\left.+\bar{\gamma }^2 \bar{v}^3 f_2^{\text{diffTemp}} \cosh ^2(\eta ) m_T^2\right.\nonumber\\
    &\left.+\frac{1}{3} p^2 \bar{v} \left(f_0^{\text{diffTemp}}-f_2^{\text{diffTemp}}\right)\right.\nonumber\\
    &\left.+\bar{\gamma } f_1^{\text{diffTemp}} p_T \cos (\phi ) \bar{E}_p\right.\nonumber\\
    &\left.+\bar{\gamma }^2 \bar{v} f_2^{\text{diffTemp}} p_T^2 \cos ^2(\phi ) \right)\nonumber\\
    &\times\cos(m\phi) \cosh(k\eta)
   \end{align}
   \begin{align}
    K_2^\text{diffTemp} &= \int \ud \phi \ud \eta \; \left(-\bar{\gamma } \bar{v} \cosh (\eta ) m_T\right.\nonumber\\
    &\left.\times \left(2 \bar{\gamma } f_2^{\text{diffTemp}} p_T \cos (\phi )+\bar{v} f_1^{\text{diffTemp}} \bar{E}_p\right)\right.\nonumber\\
    &\left.+\bar{\gamma }^2 \bar{v}^2  f_2^{\text{diffTemp}} \cosh ^2(\eta ) m_T^2\right.\nonumber\\
    &\left.+\bar{\gamma }^2 f_2^{\text{diffTemp}} p_T^2 \cos ^2(\phi )\right.\nonumber\\
    &\left.+\bar{\gamma } \bar{v} f_1^{\text{diffTemp}} p_T \cos (\phi ) \bar{E}_p\right.\nonumber\\
    &\left.+\frac{1}{3} p^2 \left(f_0^{\text{diffTemp}}-f_2^{\text{diffTemp}}\right)\right)\nonumber\\
    &\times\cos(m\phi) \cosh(k\eta)
   \end{align}
   \begin{align}
    K_1^\text{diffChem} &= \int \ud \phi \ud \eta \;\left( -\bar{\gamma } \bar{v} \cosh (\eta ) m_T \right.\nonumber\\
    &\left.\times\left(f_1^{\text{diffChem}} \bar{E}_p+2 \bar{\gamma } \bar{v} f_2^{\text{diffChem}} p_T \cos (\phi )\right)\right.\nonumber\\
    &\left.+\bar{\gamma }^2 \bar{v}^3 f_2^{\text{diffChem}} \cosh ^2(\eta ) m_T^2\right.\nonumber\\
    &\left.+\frac{1}{3} p^2 \bar{v} \left(f_0^{\text{diffChem}}-f_2^{\text{diffChem}}\right)\right.\nonumber\\
    &\left.+\bar{\gamma } f_1^{\text{diffChem}} p_T \cos (\phi ) \bar{E}_p\right.\nonumber\\
    &\left.+\bar{\gamma }^2 \bar{v} f_2^{\text{diffChem}} p_T^2 \cos ^2(\phi )\right)\nonumber\\
    &\times\cos(m\phi) \cosh(k\eta)
   \end{align}
   \begin{align}
    K_2^\text{diffChem} &= \int \ud \phi \ud \eta \;\left( -\bar{\gamma } \bar{v} \cosh (\eta ) m_T \right.\nonumber\\
    &\left.\times\left(2 \bar{\gamma } f_2^{\text{diffChem}} p_T \cos (\phi )+\bar{v} f_1^{\text{diffChem}} \bar{E}_p\right)\right.\nonumber\\
    &\left.+\bar{\gamma }^2 \bar{v}^2f_2^{\text{diffChem}} \cosh ^2(\eta ) m_T^2\right.\nonumber\\
    &\left.+\bar{\gamma }^2 f_2^{\text{diffChem}} p_T^2 \cos ^2(\phi )\right.\nonumber\\
    &\left.+\bar{\gamma } \bar{v} f_1^{\text{diffChem}} p_T \cos (\phi )
   \bar{E}_p\right.\nonumber\\
    &\left.+\frac{1}{3} p^2 \left(f_0^{\text{diffChem}}-f_2^{\text{diffChem}}\right) \right)\nonumber\\
    &\times\cos(m\phi) \cosh(k\eta)
   \end{align}
   \begin{align}
    K_1^\text{diffVel} &= \int \ud \phi \ud \eta \; -\frac{1}{5} \bar{\gamma } \bar{v} \cosh (\eta ) m_T \nonumber\\
    &\times \left(3 p^2 \bar{v} \left(f_1^{\text{diffVel}}-f_3^{\text{diffVel}}\right) \right.\nonumber\\
    &\left.+10 \bar{\gamma } f_2^{\text{diffVel}} p_T \cos (\phi
   ) \bar{E}_p+15 \bar{\gamma }^2 \bar{v} f_3^{\text{diffVel}} p_T^2 \cos ^2(\phi )\right)\nonumber\\
   &+\bar{\gamma }^2 \bar{v}^2 \cosh ^2(\eta ) m_T^2 \left(f_2^{\text{diffVel}} \bar{E}_p+3
   \bar{\gamma } \bar{v} f_3^{\text{diffVel}} p_T \cos (\phi )\right)\nonumber\\
   &-\bar{\gamma }^3 \bar{v}^4 f_3^{\text{diffVel}} \cosh ^3(\eta ) m_T^3\nonumber\\
   &+\frac{1}{3} p^2 
   \left(f_0^{\text{diffVel}}-f_2^{\text{diffVel}}\right) \bar{E}_p\nonumber\\
   &+\frac{3}{5} p^2 \bar{\gamma } \bar{v} \left(f_1^{\text{diffVel}}-f_3^{\text{diffVel}}\right) p_T \cos (\phi
   )\nonumber\\
   &+\bar{\gamma }^2 f_2^{\text{diffVel}} p_T^2 \cos ^2(\phi ) \bar{E}_p+\bar{\gamma }^3 \bar{v} f_3^{\text{diffVel}} p_T^3 \cos ^3(\phi ) 
   \end{align}
   \begin{align}
    K_2^\text{diffVel} &= \int \ud \phi \ud \eta \; -\frac{1}{5} \bar{\gamma } \bar{v} \cosh (\eta ) m_T \nonumber\\
    &\times \left(15 \bar{\gamma }^2 f_3^{\text{diffVel}} p_T^2 \cos ^2(\phi )+10 \bar{\gamma } \bar{v} f_2^{\text{diffVel}} p_T \cos (\phi
   ) \bar{E}_p\right.\nonumber\\
    &\left.+3 p^2 \left(f_1^{\text{diffVel}}-f_3^{\text{diffVel}}\right)\right)\nonumber\\
    &+\bar{\gamma }^2 \bar{v}^2 \cosh ^2(\eta ) m_T^2 \left(3 \bar{\gamma } f_3^{\text{diffVel}} p_T \cos
   (\phi )\right.\nonumber\\
    &\left.+\bar{v} f_2^{\text{diffVel}} \bar{E}_p\right)-\bar{\gamma }^3 \bar{v}^3 f_3^{\text{diffVel}} \cosh ^3(\eta ) m_T^3\nonumber\\
    &+\frac{3}{5} p^2 \bar{\gamma }
   \left(f_1^{\text{diffVel}}-f_3^{\text{diffVel}}\right) p_T \cos (\phi )\nonumber\\
   &+\frac{1}{3} p^2 \bar{v} \left(f_0^{\text{diffVel}}-f_2^{\text{diffVel}}\right)
   \bar{E}_p\nonumber\\
   &+\bar{\gamma }^3 f_3^{\text{diffVel}} p_T^3 \cos ^3(\phi )+\bar{\gamma }^2 \bar{v} f_2^{\text{diffVel}} p_T^2 \cos ^2(\phi ) \bar{E}_p  
   \end{align}
   \begin{align}
    K_3^\text{diffVel} &= \int \ud \phi \ud \eta \; -\frac{1}{5} \bar{\gamma } p_T \sin (\phi )\nonumber\\
    &\times  \left(-5 \bar{\gamma } \bar{v} \cosh (\eta ) m_T \left(f_2^{\text{diffVel}} \bar{E}_p\right.\right.\nonumber\\
    &\left.\left.+2 \bar{\gamma } \bar{v} f_3^{\text{diffVel}} p_T \cos
   (\phi )\right)+5 \bar{\gamma }^2 \bar{v}^3 f_3^{\text{diffVel}} \cosh ^2(\eta ) m_T^2\right.\nonumber\\
   &\left.+p^2 \bar{v} \left(f_1^{\text{diffVel}}-f_3^{\text{diffVel}}\right)+5 \bar{\gamma }
   f_2^{\text{diffVel}} p_T \cos (\phi ) \bar{E}_p\right.\nonumber\\
   &\left.+5 \bar{\gamma }^2 \bar{v} f_3^{\text{diffVel}} p_T^2 \cos ^2(\phi )\right) \\
    K_4^\text{diffVel} &= \int \ud \phi \ud \eta \; -\frac{1}{5} \bar{\gamma } p_T \sin (\phi )\nonumber\\
    &\times  \left(-5 \bar{\gamma } \bar{v} \cosh (\eta ) m_T \left(2 \bar{\gamma } f_3^{\text{diffVel}} p_T \cos (\phi )\right.\right.\nonumber\\
    &\left.\left.+\bar{v} f_2^{\text{diffVel}}
   \bar{E}_p\right)+5 \bar{\gamma }^2 \bar{v}^2 f_3^{\text{diffVel}} \cosh ^2(\eta ) m_T^2\right.\nonumber\\
   &\left.+5 \bar{\gamma }^2 f_3^{\text{diffVel}} p_T^2 \cos ^2(\phi )+5 \bar{\gamma } \bar{v}
   f_2^{\text{diffVel}} p_T \cos (\phi ) \bar{E}_p\right.\nonumber\\
   &\left.+p^2 \left(f_1^{\text{diffVel}}-f_3^{\text{diffVel}}\right)\right) 
   \end{align}
   \begin{align}
    K_5^\text{diffVel} &= \int \ud \phi \ud \eta \; -\frac{1}{5} \bar{\gamma } \sinh (\eta ) m_T \nonumber\\
    &\times \left(-5 \bar{\gamma } \bar{v} \cosh (\eta ) m_T \left(f_2^{\text{diffVel}} \bar{E}_p\right.\right.\nonumber\\
    &\left.\left.+2 \bar{\gamma } \bar{v} f_3^{\text{diffVel}} p_T
   \cos (\phi )\right)\right.\nonumber\\
   &\left.+5 \bar{\gamma }^2 \bar{v}^3 f_3^{\text{diffVel}} \cosh ^2(\eta ) m_T^2\right.\nonumber\\
   &\left.+p^2 \bar{v} \left(f_1^{\text{diffVel}}-f_3^{\text{diffVel}}\right)\right.\nonumber\\
   &\left.+5 \bar{\gamma }
   f_2^{\text{diffVel}} p_T \cos (\phi ) \bar{E}_p\right.\nonumber\\
   &\left.+5 \bar{\gamma }^2 \bar{v} f_3^{\text{diffVel}} p_T^2 \cos ^2(\phi )\right) \\
    K_6^\text{diffVel} &= \int \ud \phi \ud \eta \; -\frac{1}{5} \bar{\gamma } \sinh (\eta ) m_T \nonumber\\
    &\times \left(-5 \bar{\gamma } \bar{v} \cosh (\eta ) m_T \left(2 \bar{\gamma } f_3^{\text{diffVel}} p_T \cos (\phi )\right.\right.\nonumber\\
    &\left.\left.+\bar{v} f_2^{\text{diffVel}}
   \bar{E}_p\right)\right.\nonumber\\
   &\left.+5 \bar{\gamma }^2 \bar{v}^2 f_3^{\text{diffVel}} \cosh ^2(\eta ) m_T^2\right.\nonumber\\
   &\left.+5 \bar{\gamma }^2 f_3^{\text{diffVel}} p_T^2 \cos ^2(\phi )\right.\nonumber\\
   &\left.+5 \bar{\gamma } \bar{v}
   f_2^{\text{diffVel}} p_T \cos (\phi ) \bar{E}_p\right.\nonumber\\
   &\left.+p^2 \left(f_1^{\text{diffVel}}-f_3^{\text{diffVel}}\right)\right) 
   \end{align}
   \begin{align}
    K_1^\text{diffDiff} &= \int \ud \phi \ud \eta \; -\bar{\gamma } \bar{v} \cosh (\eta ) m_T \nonumber\\
    &\times\left(f_1^{\text{diffDiff}} \bar{E}_p+2 \bar{\gamma } \bar{v} f_2^{\text{diffDiff}} p_T \cos (\phi )\right)\nonumber\\
    &+\bar{\gamma }^2 \bar{v}^3
   f_2^{\text{diffDiff}} \cosh ^2(\eta ) m_T^2\nonumber\\
    &+\frac{1}{3} p^2 \bar{v} \left(f_0^{\text{diffDiff}}-f_2^{\text{diffDiff}}\right)\nonumber\\
    &+\bar{\gamma } f_1^{\text{diffDiff}} p_T \cos (\phi )
   \bar{E}_p\nonumber\\
    &+\bar{\gamma }^2 \bar{v} f_2^{\text{diffDiff}} p_T^2 \cos ^2(\phi ) \\
    K_2^\text{diffDiff} &= \int \ud \phi \ud \eta \; -\bar{\gamma } \bar{v} \cosh (\eta ) m_T \nonumber\\
    &\times\left(2 \bar{\gamma } f_2^{\text{diffDiff}} p_T \cos (\phi )+\bar{v} f_1^{\text{diffDiff}} \bar{E}_p\right)\nonumber\\
    &+\bar{\gamma }^2 \bar{v}^2
   f_2^{\text{diffDiff}} \cosh ^2(\eta ) m_T^2+\bar{\gamma }^2 f_2^{\text{diffDiff}} p_T^2 \cos ^2(\phi )\nonumber\\
    &+\bar{\gamma } \bar{v} f_1^{\text{diffDiff}} p_T \cos (\phi )
   \bar{E}_p\nonumber\\
    &+\frac{1}{3} p^2 \left(f_0^{\text{diffDiff}}-f_2^{\text{diffDiff}}\right) \\
    K_3^\text{diffDiff} &= \int \ud \phi \ud \eta \; -\bar{\gamma } p_T \sin (\phi )\nonumber\\
    &\times \left(-\bar{\gamma } \bar{v}^2 f_2^{\text{diffDiff}} \cosh (\eta ) m_T+f_1^{\text{diffDiff}} \bar{E}_p\right.\nonumber\\
    &\left.+\bar{\gamma } \bar{v} f_2^{\text{diffDiff}} p_T
   \cos (\phi )\right) \\
    K_4^\text{diffDiff} &= \int \ud \phi \ud \eta \; -\bar{\gamma } p_T \sin (\phi ) \nonumber\\
    &\times\left(-\bar{\gamma } \bar{v} f_2^{\text{diffDiff}} \cosh (\eta ) m_T+\bar{\gamma } f_2^{\text{diffDiff}} p_T \cos (\phi )\right.\nonumber\\
    &\left.+\bar{v}
   f_1^{\text{diffDiff}} \bar{E}_p\right) \\
    K_5^\text{diffDiff} &= \int \ud \phi \ud \eta \; \bar{\gamma } \sinh (\eta ) m_T\nonumber\\
    &\times \left(\bar{\gamma } \bar{v}^2 f_2^{\text{diffDiff}} \cosh (\eta ) m_T+f_1^{\text{diffDiff}} \left(-\bar{E}_p\right)\right.\nonumber\\
    &\left.-\bar{\gamma } \bar{v}
   f_2^{\text{diffDiff}} p_T \cos (\phi )\right) \\
    K_6^\text{diffDiff} &= \int \ud \phi \ud \eta \;  \bar{\gamma } \sinh (\eta ) m_T\nonumber\\
    &\times \left(\bar{\gamma } \bar{v} f_2^{\text{diffDiff}} \cosh (\eta ) m_T-\bar{\gamma } f_2^{\text{diffDiff}} p_T \cos (\phi )\right.\nonumber\\
    &\left. -\bar{v} f_1^{\text{diffDiff}}
   \bar{E}_p\right)
\end{align}

\section{Thermal background kernel} \label{sec_thermal_bg_ker}
In this section we give the expressions for the thermal background spectra. Since both the azimuthal intagrations $A_*\left((1+j)\Tilde{p}_T \bv \bgamma \right)$, as well as the integrations in rapidity $R_*\left((1+j)\Tilde{m}_T  \bgamma \right)$ always depend on the same argument, we will supress it in the following.
\begin{align}
    \Bar{Y}^a_{1} &= A_0 R_c \\
    \Bar{Y}^b_{1} &= A_c R_0 \\
    \Bar{Y}^a_{2} &=2 (j+1) \bar{\gamma }^2 \bar{v} A_c R_{\text{cc}} \tilde{m}_T \tilde{p}_T-(j+1) \bar{\gamma }^2 A_{\text{cc}} R_c \tilde{p}_T^2\nonumber\\
    &-(j+1) \bar{\gamma }^2 \bar{v}^2
   R_{\text{ccc}} \tilde{m}_T^2+(j+1) A_{\text{ss}} R_c \tilde{p}_T^2\\
    \Bar{Y}^b_{2} &= 2 (j+1) \bar{\gamma }^2 \bar{v} A_{\text{cc}} R_c \tilde{m}_T \tilde{p}_T-(j+1) \bar{\gamma }^2 \bar{v}^2 A_c R_{\text{cc}} \tilde{m}_T^2\nonumber\\
    &-(j+1) \bar{\gamma }^2
   A_{\text{ccc}} \tilde{p}_T^2+(j+1) A_c A_{\text{ss}} \tilde{p}_T^2
   \end{align}
    \begin{align}
    \Bar{Y}^a_{3} &= 2 (j+1) \bar{\gamma }^2 \bar{v} A_c R_{\text{cc}} \tilde{m}_T \tilde{p}_T-(j+1) \bar{\gamma }^2 A_{\text{cc}} R_c \tilde{p}_T^2\nonumber\\
    &-(j+1) \bar{\gamma }^2 \bar{v}^2   R_{\text{ccc}} \tilde{m}_T^2+(j+1) R_{\text{ss}} R_c \tilde{m}_T^2\\
    \Bar{Y}^b_{3} &= 2 (j+1) \bar{\gamma }^2 \bar{v} A_{\text{cc}} R_c \tilde{m}_T \tilde{p}_T-(j+1) \bar{\gamma }^2 \bar{v}^2 A_c R_{\text{cc}} \tilde{m}_T^2\nonumber\\
    &-(j+1) \bar{\gamma }^2
   A_{\text{ccc}} \tilde{p}_T^2+(j+1) R_{\text{ss}} A_c \tilde{m}_T^2
   \end{align}
    \begin{align}
    \Bar{Y}^a_{4} &=\frac{(1+j)\tau_B\left(\frac{1}{3}-c_s^2\right)}{\zeta}\left(- \bar{\gamma } \bar{v} A_c R_c  \tilde{p}_T+ \bar{\gamma } A_0  R_{\text{cc}} \tilde{m}_T\right) \nonumber\\
   &-\frac{(j+1) \bbeta m \tau_B}{3 \zeta}\sum_{n=0}^\infty \sum_{k=0}^n \sum_{l=0}^{n-k}  (-1)^{n+k+l} \frac{n!}{k! l! (n-k-l)!} \nonumber\\
   &\times (\bar{\gamma } m_T/m)^{n-k-l} (\bar{\gamma } \bar{v} p_T/m)^l R_{(n-k-l+1)*\text{c}} A_{l*\text{c}} \\
    \Bar{Y}^b_{4} &=\frac{(1+j)\tau_B\left(\frac{1}{3}-c_s^2\right)}{\zeta}\left(- \bar{\gamma } \bar{v} A_{\text{cc}} R_0 \tilde{p}_T+ \bar{\gamma } A_c R_c\tilde{m}_T\right)\nonumber\\
   &-\frac{(j+1) \bbeta m \tau_B}{3 \zeta}\sum_{n=0}^\infty \sum_{k=0}^n \sum_{l=0}^{n-k}  (-1)^{n+k+l} \frac{n!}{k! l! (n-k-l)!} \nonumber\\
   &\times (\bar{\gamma } m_T/m)^{n-k-l} (\bar{\gamma } \bar{v} p_T/m)^l R_{(n-k-l)*\text{c}} A_{(l+1)*\text{c}}
   \end{align}
    \begin{align}
    \Bar{Y}^a_{5} &=-\frac{(j+1) n_B \bar{v} A_0 R_{\text{cc}} m_T}{\kappa 
   (e+p)}+\frac{(j+1) n_B A_c R_c p_T}{\kappa  (e+p)}\\
   &-\frac{(j+1)Q_B}{m\kappa}\sum_{n=0}^\infty \sum_{k=0}^n \sum_{l=0}^{n-k}  (-1)^{n+k+l} \frac{n!}{k! l! (n-k-l)!} \nonumber\\
   &\times\left[\tilde{m}_T (\bar{\gamma }\frac{m_T}{m})^{n-k-l} (\bar{\gamma } \bar{v} \frac{p_T}{m})^l R_{(n-k-l+1)*\text{c}} A_{(l+1)*\text{c}}\right.\nonumber\\
   &- \bv \tilde{m}_T\left. (\bar{\gamma }\frac{m_T}{m})^{n-k-l} (\bar{\gamma } \bar{v} \frac{p_T}{m})^l R_{(n-k-l+2)*\text{c}} A_{(l)*\text{c}}\right]\nonumber\\   
    \Bar{Y}^b_{5} &= -\frac{(j+1) n_B \bar{v} A_c R_{\text{c}} m_T}{\kappa 
   (e+p)}+\frac{(j+1) n_B A_{\text{cc}} R_0 p_T}{\kappa  (e+p)}\\
   &-\frac{(j+1)Q_B}{m\kappa}\sum_{n=0}^\infty \sum_{k=0}^n \sum_{l=0}^{n-k}  (-1)^{n+k+l} \frac{n!}{k! l! (n-k-l)!} \nonumber\\
   &\times\left[\tilde{m}_T (\bar{\gamma }\frac{m_T}{m})^{n-k-l} (\bar{\gamma } \bar{v} \frac{p_T}{m})^l R_{(n-k-l)*\text{c}} A_{(l+2)*\text{c}}\right.\nonumber\\
   &- \bv \tilde{m}_T\left. (\bar{\gamma }\frac{m_T}{m})^{n-k-l} (\bar{\gamma } \bar{v} \frac{p_T}{m})^l R_{(n-k-l+1)*\text{c}} A_{(l+1)*\text{c}}\right]
\end{align}
The corresponding field vector is given by $\Bar{\Phi}=\left(1,\frac{\beta \bpi^{22}}{2w},\frac{\beta \bpi^{33}}{2w},\bpi_B, \bgamma \bnu \right)$.

\section{Thermal perturbation kernel} \label{sec_thermal_pert_ker}
In this section we give the expressions for the thermal perturbation spectra. Since both the azimuthal intagrations $A_*\left(m,(1+j)\Tilde{p}_T \bv \bgamma \right)$, as well as the integrations in rapidity $R_*\left(k,(1+j)\Tilde{m}_T  \bgamma \right)$ always depend on the same argument, we will supress it in the following. Here we also introduce the abbreviations $-\beta\partial_\beta x = x^*$ and $\partial_\alpha x = x^+$, for the thermodynamic quantities $x$.
\begin{align}
    Y_1^a &= -\frac{1}{2} (j+1) \bar{v}^2 \bar{\gamma }^3 A_0 R_{\text{cccc}} \tilde{m}_T^3 \bar{\pi }^t\nonumber\\
    &+\frac{1}{2}(1+j) \mu  \bar{v}^2 \bar{\gamma }^2 A_0 R_{\text{ccc}}
   \tilde{m}_T^2 \bar{\pi }^t\nonumber\\
   &+\bar{v}^2 \bar{\gamma }^2 A_0 R_{\text{ccc}} \tilde{m}_T^2 \bar{\pi }^t-(j+1) \bar{v}^2 \bar{\gamma }^3 A_{\text{cc}} R_{\text{cc}} \tilde{m}_T \tilde{p}_T^2 \bar{\pi   }^t\nonumber\\
   &+\frac{1}{2} (j+1) \bar{v}^3 \bar{\gamma }^3 A_c R_{\text{ccc}} \tilde{m}_T^2 \tilde{p}_T \bar{\pi }^t\nonumber\\
   &+(j+1) \bar{v} \bar{\gamma }^3 A_c
   R_{\text{ccc}} \tilde{m}_T^2 \tilde{p}_T \bar{\pi }^t\nonumber\\
   &-((1+j) \balpha +2) \bar{v} \bar{\gamma }^2 A_c R_{\text{cc}} \tilde{m}_T \tilde{p}_T \bar{\pi }^t\nonumber\\
   &+\frac{ \bar{v}^2 \bar{\gamma }^2 A_0 R_{\text{ccc}} \tilde{m}_T^2 (e^* +p^*) \bar{\pi }^t}{2   (e+p)}+\bar{\gamma } A_0 R_{\text{cc}} \tilde{m}_T \nonumber\\
   &-\frac{ \bar{v} \bar{\gamma }^2 A_c R_{\text{cc}} \tilde{m}_T \tilde{p}_T (e^* +p^*) \bar{\pi }^t}{e+p} -\bar{v} \bar{\gamma } A_c R_c \tilde{p}_T\nonumber\\
   &+\frac{1}{2} (j+1) \bar{\gamma } A_0 R_{\text{ccss}} \tilde{m}_T^3 \bar{\pi
   }^{33}\nonumber\\
   &-\frac{1}{2}(1+j) \balpha  A_0 R_{\text{css}} \tilde{m}_T^2 \bar{\pi }^{33}\nonumber\\
   &-A_0   R_{\text{css}} \tilde{m}_T^2 \bar{\pi }^{33} -\frac{1}{2} (j+1) \bar{v} \bar{\gamma } A_c  R_{\text{css}} \tilde{m}_T^2 \tilde{p}_T \bar{\pi }^{33}\nonumber\\
   &-\frac{ A_0 R_{\text{css}} \tilde{m}_T^2 (e^*+p^*) \bar{\pi }^{33}}{2 (e+p)}\nonumber\\
   &+\frac{1}{2} (j+1) \bar{v} \bar{\gamma }  \left(\bar{\pi }^t \bar{\gamma }^2 A_{\text{ccc}}-\bar{\pi }^{22} A_{\text{css}}\right) R_c \tilde{p}_T^3\nonumber\\
   &+\frac{1}{2}(1+j) \balpha  \left(\bar{\pi }^t \bar{\gamma }^2 A_{\text{cc}}-\bar{\pi }^{22} A_{\text{ss}}\right)   R_c \tilde{p}_T^2\nonumber\\
   &+\left(\bar{\pi }^t \bar{\gamma }^2 A_{\text{cc}}-\bar{\pi }^{22} A_{\text{ss}}\right) R_c \tilde{p}_T^2 -\balpha A_0 R_c\nonumber\\
   &-\frac{1}{2} (j+1) \bar{\gamma }   \left(\bar{\pi }^t \bar{\gamma }^2 A_{\text{cc}}-\bar{\pi }^{22} A_{\text{ss}}\right) R_{\text{cc}} \tilde{m}_T \tilde{p}_T^2 \nonumber\\
   &+\frac{ \left(\bar{\pi }^t \bar{\gamma }^2
   A_{\text{cc}}-\bar{\pi }^{22} A_{\text{ss}}\right) R_c \tilde{p}_T^2 (e^*+p^*)}{2 (e+p)}\nonumber\\
   &+\bgamma \bpi_B \tau_B (c_s^2)^* (A_c R_c \Tilde{p}_T -\Tilde{m}_TA_0 R_{\text{cc}} )\nonumber\\
   &+\sum_{n=0}^\infty \sum_{k=0}^n \sum_{l=0}^{n-k}  (-1)^{n+k+l} \frac{n!}{k! l! (n-k-l)!} \nonumber\\
   &\times(\frac{m_T}{m})^{n-k-l} (\bar{v} \frac{p_T}{m})^l\left[\frac{\bpi_B \tau_B m \beta}{3 \zeta} \right.\nonumber\\
    &\times \big((1+j) \Tilde{m}_T A_{l*\text{c}}   R_{(n-k-l+2)*\text{c}} \nonumber\\
    &-(1+j) \Tilde{p}_T \bv A_{(l+1)*\text{c}}   R_{(n-k-l+1)*\text{c}} \nonumber\\
    &- \bgamma^{-1} A_{l*\text{c}}   R_{(n-k-l+1)*\text{c}} \nonumber\\
    &-(1+j) \balpha \bgamma^{-1} A_{l*\text{c}}   R_{(n-k-l+1)*\text{c}} \nonumber\\
    &+ \tau_B^*/\tau_B \bgamma^{-1} A_{l*\text{c}}   R_{(n-k-l+1)*\text{c}} \nonumber\\
     &-\zeta^*/\zeta \bgamma^{-1} A_{l*\text{c}}   R_{(n-k-l+1)*\text{c}}\big) \nonumber\\
     &+\frac{\bpi_B \tau_B (3 c_s^2-1)}{3\bbeta m} \big(- \bgamma \Tilde{m}_T^2 A_{l*\text{c}}   R_{(n-k-l+3)*\text{c}} \nonumber\\
     &+2 \bgamma \bv \Tilde{m}_T \Tilde{p}_T A_{(l+1)*\text{c}}   R_{(n-k-l+2)*\text{c}} \nonumber\\
     &- \bgamma \bv^2 \Tilde{m}_T^2 A_{(l+2)*\text{c}}   R_{(n-k-l+1)*\text{c}} \nonumber
     \end{align}
\begin{align}
     &+(1+j) \bgamma^2  \Tilde{m}_T^3  A_{l*\text{c}}   R_{(n-k-l+4)*\text{c}} \nonumber\\
     &-3 \bgamma^2 \bv (1+j) \Tilde{m}_T^2 \Tilde{p}_T A_{(l+1)*\text{c}}   R_{(n-k-l+3)*\text{c}} \nonumber\\
     &+3 \bgamma^2 \bv^2 \Tilde{m}_T \Tilde{p}_T^2 (1+j) A_{(l+2)*\text{c}}   R_{(n-k-l+2)*\text{c}} \nonumber\\
     &- \bgamma^2 \bv^3 (1+j) \Tilde{p}_T^3 A_{(l+3)*\text{c}}   R_{(n-k-l+1)*\text{c}} \nonumber\\
     &- (1+j)\bgamma \balpha \Tilde{m}_T^2 A_{l*\text{c}}   R_{(n-k-l+3)*\text{c}} \nonumber\\
     &+2 (1+j)\bgamma \bv \balpha \Tilde{m}_T \Tilde{p}_T A_{(l+1)*\text{c}}   R_{(n-k-l+2)*\text{c}} \nonumber\\
     &+2 \bgamma \bv^2 \balpha (1+j) \Tilde{p}_T^2 A_{(l+2)*\text{c}}   R_{(n-k-l+1)*\text{c}} \nonumber\\
     &+ \bgamma \Tilde{m}_T^2 \tau_B^*/\tau_B A_{l*\text{c}}   R_{(n-k-l+3)*\text{c}} \nonumber\\
     &-2 \bgamma \bv \Tilde{m}_T \tau_B^*/\tau_B \Tilde{p}_T A_{(l+1)*\text{c}}   R_{(n-k-l+2)*\text{c}} \nonumber\\
     &-2 \bgamma \bv \Tilde{m}_T \tau_B^*/\tau_B \Tilde{p}_T A_{(l+1)*\text{c}}   R_{(n-k-l+2)*\text{c}} \nonumber\\
     &+ \bgamma \bv^2 \Tilde{p}_T^2 \tau_B^*/\tau_B \Tilde{p}_T A_{(l+2)*\text{c}}   R_{(n-k-l+1)*\text{c}} \nonumber\\
     &+3 \bgamma (c_s^2)^*/(3c_s^2-1) \Tilde{m}_T^2 \Tilde{p}_T A_{l*\text{c}}   R_{(n-k-l+3)*\text{c}} \nonumber\\
     &-6 \bv \bgamma (c_s^2)^*/(3c_s^2-1) \Tilde{m}_T \Tilde{p}_T \Tilde{p}_T A_{(l+1)*\text{c}}   R_{(n-k-l+2)*\text{c}} \nonumber\\
     &+3 \bgamma \bv^2 (c_s^2)^*/(3c_s^2-1) \Tilde{p}_T^2 \Tilde{p}_T A_{(l+2)*\text{c}}   R_{(n-k-l+1)*\text{c}} \nonumber\\
     &-1/3 \bgamma   \Tilde{m}_T^2 \zeta^*/\zeta A_{l*\text{c}}   R_{(n-k-l+3)*\text{c}} \nonumber\\
     &+2/3 \bv \bgamma   \Tilde{m}_T \Tilde{p}_T \zeta^*/\zeta A_{(l+1)*\text{c}}   R_{(n-k-l+2)*\text{c}} \nonumber\\
     &-1/3 \bgamma  \bv^2 \Tilde{p}_T^2 \zeta^*/\zeta A_{(l+2)*\text{c}}   R_{(n-k-l+1)*\text{c}} \nonumber\\
     &+\frac{\kappa^* }{\bbeta m \kappa^2 (e+p)} \big( \beta (e+p) \Tilde{p}_T Q_B \bnu   A_{(l+1)*\text{c}}   R_{(n-k-l+1)*\text{c}} \nonumber \\
     &-\bbeta (e+p) \Tilde{m}_T Q_B \bnu \bv  A_{l*\text{c}}   R_{(n-k-l+2)*\text{c}} \nonumber \\
     &- \Tilde{m}_T \Tilde{p}_T n_B \bgamma \bnu   A_{(l+1)*\text{c}}   R_{(n-k-l+2)*\text{c}} \nonumber \\
    &+ \Tilde{m}_T^2 \bv n_B \bgamma \bnu   A_{l*\text{c}}   R_{(n-k-l+3)*\text{c}} \nonumber \\
     &+ \Tilde{p}_T^2 n_B \bgamma \bnu \bv   A_{(l+2)*\text{c}}   R_{(n-k-l+1)*\text{c}} \nonumber \\
      &- \Tilde{m}_T \Tilde{p}_T n_B \bgamma \bnu \bv^2  A_{(l+1)*\text{c}}   R_{(n-k-l+2)*\text{c}}\big) \nonumber \\
     & +\frac{\bnu}{\bbeta m \kappa}\big( (1+j)Q_B \Tilde{m}_T \Tilde{p}_T \bbeta \bgamma   A_{(l+1)*\text{c}}   R_{(n-k-l+3)*\text{c}} \nonumber \\
     &-(1+j)Q_B \Tilde{m}_T^2  \bbeta  \bv \bgamma A_{l*\text{c}}   R_{(n-k-l+3)*\text{c}} \nonumber \\
     &-(1+j)Q_B  \Tilde{p}_T^2 \bbeta  \bv \bgamma  A_{(l+2)*\text{c}}   R_{(n-k-l+1)*\text{c}} \nonumber \\
     &+(1+j)Q_B \Tilde{m}_T \Tilde{p}_T \bbeta \bv^2 \bgamma   A_{(l+1)*\text{c}}   R_{(n-k-l+2)*\text{c}} \nonumber \\
     &-(1+j)n_B \Tilde{m}_T^2 \Tilde{p}_T \bgamma^2 /(e+p)   A_{(l+1)*\text{c}}   R_{(n-k-l+3)*\text{c}} \nonumber \\
    &+(1+j)n_B \Tilde{m}_T^3  \bv \bgamma^2 /(e+p)   A_{l*\text{c}}   R_{(n-k-l+4)*\text{c}} \nonumber \\
    &+2(1+j)n_B \Tilde{m}_T \Tilde{p}_T^2 \bv \bgamma^2 /(e+p)   A_{(l+2)*\text{c}}   R_{(n-k-l+2)*\text{c}} \nonumber \\
    &-2(1+j)n_B \Tilde{m}_T^2 \Tilde{p}_T \bgamma^2 \bv^2 /(e+p)   A_{(l+1)*\text{c}}   R_{(n-k-l+3)*\text{c}} \nonumber \\
    &-(1+j)n_B  \Tilde{p}_T^3 \bgamma^2 \bv^2 /(e+p)   A_{(l+3)*\text{c}}   R_{(n-k-l+1)*\text{c}} \nonumber \\
    &+(1+j)n_B \Tilde{m}_T \Tilde{p}_T^2 \bgamma^2 \bv^3 /(e+p)   A_{(l+2)*\text{c}}   R_{(n-k-l+2)*\text{c}} \nonumber \\
    &-(1+j) \Tilde{p}_T Q_B \bbeta \balpha   A_{(l+1)*\text{c}}   R_{(n-k-l+1)*\text{c}} \nonumber \\
    &-(1+j)n_B \Tilde{m}_T^2 \Tilde{p}_T \bgamma^2 /(e+p)   A_{(l+1)*\text{c}}   R_{(n-k-l+3)*\text{c}} \nonumber \\
    &+(1+j)Q_B \Tilde{m}_T \bv \bbeta \balpha   A_{l*\text{c}}   R_{(n-k-l+2)*\text{c}} \nonumber \\
    &+(1+j)n_B \Tilde{m}_T \Tilde{p}_T \bgamma \balpha /(e+p)   A_{(l+1)*\text{c}}   R_{(n-k-l+2)*\text{c}} \nonumber \\
    &-(1+j)n_B \Tilde{m}_T^2  \bgamma \bv \balpha /(e+p)   A_{l*\text{c}}   R_{(n-k-l+3)*\text{c}} \nonumber \\
    &-(1+j)n_B \Tilde{m}_T^2 \Tilde{p}_T \bgamma^2 /(e+p)   A_{(l+1)*\text{c}}   R_{(n-k-l+3)*\text{c}} \nonumber \\
    &-(1+j)n_B \Tilde{p}_T^2 \bgamma \bv \balpha /(e+p)   A_{(l+2)*\text{c}}   R_{(n-k-l+1)*\text{c}} \nonumber \\
    &+(1+j)n_B \Tilde{m}_T \Tilde{p}_T \bgamma \bv^2 \balpha /(e+p)   A_{(l+1)*\text{c}}   R_{(n-k-l+2)*\text{c}} \nonumber 
\end{align}
\begin{align}
    &+\left( (e+p)n_B^*-n_B(e^*+p^*) \right) /(e+p)^2 \big(\nonumber\\
    &-\Tilde{m}_T \Tilde{p}_T \bgamma  A_{(l+1)*\text{c}}   R_{(n-k-l+2)*\text{c}} \nonumber \\
    &+\Tilde{m}_T^2 \bv \bgamma  A_{l*\text{c}}   R_{(n-k-l+3)*\text{c}} \nonumber \\
    &+ \Tilde{p}_T^2 \bv \bgamma  A_{(l+2)*\text{c}}   R_{(n-k-l+1)*\text{c}} \nonumber \\
    &-\Tilde{m}_T \Tilde{p}_T \bgamma  \bv^2 A_{(l+1)*\text{c}}   R_{(n-k-l+2)*\text{c}}\big) \big)\big] \nonumber \\
   \end{align}
%R_c goes A_c
\begin{align}
    Y_1^b &= -\frac{1}{2} (j+1) \bar{v}^2 \bar{\gamma }^3 A_c R_{\text{ccc}} \tilde{m}_T^3 \bar{\pi }^t\nonumber\\
    &+\frac{1}{2}(1+j) \mu  \bar{v}^2 \bar{\gamma }^2 A_c R_{\text{cc}}
   \tilde{m}_T^2 \bar{\pi }^t\nonumber\\
   &+\bar{v}^2 \bar{\gamma }^2 A_c R_{\text{cc}} \tilde{m}_T^2 \bar{\pi }^t-(j+1) \bar{v}^2 \bar{\gamma }^3 A_{\text{ccc}} R_{\text{c}} \tilde{m}_T \tilde{p}_T^2 \bar{\pi   }^t\nonumber\\
   &+\frac{1}{2} (j+1) \bar{v}^3 \bar{\gamma }^3 A_{\text{cc}} R_{\text{cc}} \tilde{m}_T^2 \tilde{p}_T \bar{\pi }^t\nonumber\\
   &+(j+1) \bar{v} \bar{\gamma }^3 A_{\text{cc}}
   R_{\text{cc}} \tilde{m}_T^2 \tilde{p}_T \bar{\pi }^t\nonumber\\
   &-((1+j) \balpha +2) \bar{v} \bar{\gamma }^2 A_{\text{cc}} R_{\text{c}} \tilde{m}_T \tilde{p}_T \bar{\pi }^t\nonumber\\
   &+\frac{ \bar{v}^2 \bar{\gamma }^2 A_c R_{\text{cc}} \tilde{m}_T^2 (e^* +p^*) \bar{\pi }^t}{2   (e+p)}+\bar{\gamma } A_c R_{\text{c}} \tilde{m}_T \nonumber\\
   &-\frac{ \bar{v} \bar{\gamma }^2 A_{\text{cc}} R_{\text{c}} \tilde{m}_T \tilde{p}_T (e^* +p^*) \bar{\pi }^t}{e+p} -\bar{v} \bar{\gamma } A_{\text{cc}} R_0 \tilde{p}_T\nonumber\\
   &+\frac{1}{2} (j+1) \bar{\gamma } A_c R_{\text{css}} \tilde{m}_T^3 \bar{\pi
   }^{33}\nonumber\\
   &-\frac{1}{2}(1+j) \balpha  A_c R_{\text{ss}} \tilde{m}_T^2 \bar{\pi }^{33}\nonumber\\
   &-A_c   R_{\text{ss}} \tilde{m}_T^2 \bar{\pi }^{33} -\frac{1}{2} (j+1) \bar{v} \bar{\gamma } A_{\text{cc}}  R_{\text{ss}} \tilde{m}_T^2 \tilde{p}_T \bar{\pi }^{33}\nonumber\\
   &-\frac{ A_c R_{\text{ss}} \tilde{m}_T^2 (e^*+p^*) \bar{\pi }^{33}}{2 (e+p)}\nonumber\\
   &+\frac{1}{2} (j+1) \bar{v} \bar{\gamma }  \left(\bar{\pi }^t \bar{\gamma }^2 A_{\text{cccc}}-\bar{\pi }^{22} A_{\text{ccss}}\right) R_0 \tilde{p}_T^3\nonumber\\
   &+\frac{1}{2}(1+j) \balpha  \left(\bar{\pi }^t \bar{\gamma }^2 A_{\text{ccc}}-\bar{\pi }^{22} A_{\text{css}}\right)   R_0 \tilde{p}_T^2\nonumber\\
   &+\left(\bar{\pi }^t \bar{\gamma }^2 A_{\text{ccc}}-\bar{\pi }^{22} A_{\text{css}}\right) R_0 \tilde{p}_T^2 -\balpha A_c R_0\nonumber\\
   &-\frac{1}{2} (j+1) \bar{\gamma }   \left(\bar{\pi }^t \bar{\gamma }^2 A_{\text{ccc}}-\bar{\pi }^{22} A_{\text{css}}\right) R_{\text{c}} \tilde{m}_T \tilde{p}_T^2 \nonumber\\
   &+\frac{ \left(\bar{\pi }^t \bar{\gamma }^2
   A_{\text{ccc}}-\bar{\pi }^{22} A_{\text{css}}\right) R_0 \tilde{p}_T^2 (e^*+p^*)}{2 (e+p)}\nonumber\\
   &+\bgamma \bpi_B \tau_B (c_s^2)^* (A_{\text{cc}} R_0 \Tilde{p}_T -\Tilde{m}_T A_c R_{\text{c}} )\nonumber\\
   &+\sum_{n=0}^\infty \sum_{k=0}^n \sum_{l=0}^{n-k}  (-1)^{n+k+l} \frac{n!}{k! l! (n-k-l)!} \nonumber\\
   &\times(\frac{m_T}{m})^{n-k-l} (\bar{v} \frac{p_T}{m})^l\left[\frac{\bpi_B \tau_B m \beta}{3 \zeta} \right.\nonumber\\
    &\times \big((1+j) \Tilde{m}_T A_{(l+1)*\text{c}}   R_{(n-k-l+1)*\text{c}} \nonumber\\
    &-(1+j) \Tilde{p}_T \bv A_{(l+2)*\text{c}}   R_{(n-k-l)*\text{c}} \nonumber\\
    &- \bgamma^{-1} A_{(l+1)*\text{c}}   R_{(n-k-l)*\text{c}} \nonumber\\
    &-(1+j) \balpha \bgamma^{-1} A_{(l+1)*\text{c}}   R_{(n-k-l)*\text{c}} \nonumber\\
    &+ \tau_B^*/\tau_B \bgamma^{-1} A_{(l+1)*\text{c}}   R_{(n-k-l)*\text{c}} \nonumber\\
     &-\zeta^*/\zeta \bgamma^{-1} A_{(l+1)*\text{c}}   R_{(n-k-l)*\text{c}}\big) \nonumber
     \end{align}
\begin{align}
     &+\frac{\bpi_B \tau_B (3 c_s^2-1)}{3\bbeta m} \big(- \bgamma \Tilde{m}_T^2 A_{(l+1)*\text{c}}   R_{(n-k-l+2)*\text{c}} \nonumber\\
     &+2 \bgamma \bv \Tilde{m}_T \Tilde{p}_T A_{(l+2)*\text{c}}   R_{(n-k-l+1)*\text{c}} \nonumber\\
     &- \bgamma \bv^2 \Tilde{m}_T^2 A_{(l+3)*\text{c}}   R_{(n-k-l)*\text{c}} \nonumber\\
     &+(1+j) \bgamma^2  \Tilde{m}_T^3  A_{(l+1)*\text{c}}   R_{(n-k-l+3)*\text{c}} \nonumber\\
     &-3 \bgamma^2 \bv (1+j) \Tilde{m}_T^2 \Tilde{p}_T A_{(l+2)*\text{c}}   R_{(n-k-l+2)*\text{c}} \nonumber\\
     &+3 \bgamma^2 \bv^2 \Tilde{m}_T \Tilde{p}_T^2 (1+j) A_{(l+3)*\text{c}}   R_{(n-k-l+1)*\text{c}} \nonumber\\
     &- \bgamma^2 \bv^3 (1+j) \Tilde{p}_T^3 A_{(l+4)*\text{c}}   R_{(n-k-l)*\text{c}} \nonumber\\
     &- (1+j)\bgamma \balpha \Tilde{m}_T^2 A_{(l+1)*\text{c}}   R_{(n-k-l+2)*\text{c}} \nonumber\\
     &+2 (1+j)\bgamma \bv \balpha \Tilde{m}_T \Tilde{p}_T A_{(l+2)*\text{c}}   R_{(n-k-l+1)*\text{c}} \nonumber\\
     &+2 \bgamma \bv^2 \balpha (1+j) \Tilde{p}_T^2 A_{(l+3)*\text{c}}   R_{(n-k-l)*\text{c}} \nonumber\\
     &+ \bgamma \Tilde{m}_T^2 \tau_B^*/\tau_B A_{(l+1)*\text{c}}   R_{(n-k-l+2)*\text{c}} \nonumber\\
     &-2 \bgamma \bv \Tilde{m}_T \tau_B^*/\tau_B \Tilde{p}_T A_{(l+2)*\text{c}}   R_{(n-k-l+1)*\text{c}} \nonumber\\
     &-2 \bgamma \bv \Tilde{m}_T \tau_B^*/\tau_B \Tilde{p}_T A_{(l+2)*\text{c}}   R_{(n-k-l+1)*\text{c}} \nonumber\\
     &+ \bgamma \bv^2 \Tilde{p}_T^2 \tau_B^*/\tau_B \Tilde{p}_T A_{(l+3)*\text{c}}   R_{(n-k-l)*\text{c}} \nonumber\\
     &+3 \bgamma (c_s^2)^*/(3c_s^2-1) \Tilde{m}_T^2 \Tilde{p}_T A_{(l+1)*\text{c}}   R_{(n-k-l+2)*\text{c}} \nonumber\\
     &-6 \bv \bgamma (c_s^2)^*/(3c_s^2-1) \Tilde{m}_T \Tilde{p}_T \Tilde{p}_T A_{(l+2)*\text{c}}   R_{(n-k-l+1)*\text{c}} \nonumber\\
     &+3 \bgamma \bv^2 (c_s^2)^*/(3c_s^2-1) \Tilde{p}_T^2 \Tilde{p}_T A_{(l+3)*\text{c}}   R_{(n-k-l)*\text{c}} \nonumber\\
     &-1/3 \bgamma   \Tilde{m}_T^2 \zeta^*/\zeta A_{(l+1)*\text{c}}   R_{(n-k-l+2)*\text{c}} \nonumber\\
     &+2/3 \bv \bgamma   \Tilde{m}_T \Tilde{p}_T \zeta^*/\zeta A_{(l+2)*\text{c}}   R_{(n-k-l+1)*\text{c}} \nonumber\\
     &-1/3 \bgamma  \bv^2 \Tilde{p}_T^2 \zeta^*/\zeta A_{(l+3)*\text{c}}   R_{(n-k-l)*\text{c}} \nonumber\\
     &+\frac{\kappa^* }{\bbeta m \kappa^2 (e+p)} \big( \beta (e+p) \Tilde{p}_T Q_B \bnu   A_{(l+2)*\text{c}}   R_{(n-k-l)*\text{c}} \nonumber \\
     &-\bbeta (e+p) \Tilde{m}_T Q_B \bnu \bv  A_{(l+1)*\text{c}}   R_{(n-k-l+1)*\text{c}} \nonumber \\
     &- \Tilde{m}_T \Tilde{p}_T n_B \bgamma \bnu   A_{(l+2)*\text{c}}   R_{(n-k-l+1)*\text{c}} \nonumber \\
    &+ \Tilde{m}_T^2 \bv n_B \bgamma \bnu   A_{(l+1)*\text{c}}   R_{(n-k-l+2)*\text{c}} \nonumber \\
     &+ \Tilde{p}_T^2 n_B \bgamma \bnu \bv   A_{(l+3)*\text{c}}   R_{(n-k-l)*\text{c}} \nonumber \\
      &- \Tilde{m}_T \Tilde{p}_T n_B \bgamma \bnu \bv^2  A_{(l+2)*\text{c}}   R_{(n-k-l+1)*\text{c}}\big) \nonumber \\
     & +\frac{\bnu}{\bbeta m \kappa}\big( (1+j)Q_B \Tilde{m}_T \Tilde{p}_T \bbeta \bgamma   A_{(l+2)*\text{c}}   R_{(n-k-l+2)*\text{c}} \nonumber \\
     &-(1+j)Q_B \Tilde{m}_T^2  \bbeta  \bv \bgamma A_{(l+1)*\text{c}}   R_{(n-k-l+2)*\text{c}} \nonumber \\
     &-(1+j)Q_B  \Tilde{p}_T^2 \bbeta  \bv \bgamma  A_{(l+3)*\text{c}}   R_{(n-k-l)*\text{c}} \nonumber \\
     &+(1+j)Q_B \Tilde{m}_T \Tilde{p}_T \bbeta \bv^2 \bgamma   A_{(l+2)*\text{c}}   R_{(n-k-l+1)*\text{c}} \nonumber \\
     &-(1+j)n_B \Tilde{m}_T^2 \Tilde{p}_T \bgamma^2 /(e+p)   A_{(l+2)*\text{c}}   R_{(n-k-l+2)*\text{c}} \nonumber \\
    &+(1+j)n_B \Tilde{m}_T^3  \bv \bgamma^2 /(e+p)   A_{(l+1)*\text{c}}   R_{(n-k-l+3)*\text{c}} \nonumber \\
    &+2(1+j)n_B \Tilde{m}_T \Tilde{p}_T^2 \bv \bgamma^2 /(e+p)   A_{(l+3)*\text{c}}   R_{(n-k-l+1)*\text{c}} \nonumber \\
    &-2(1+j)n_B \Tilde{m}_T^2 \Tilde{p}_T \bgamma^2 \bv^2 /(e+p)   A_{(l+2)*\text{c}}   R_{(n-k-l+2)*\text{c}} \nonumber \\
    &-(1+j)n_B  \Tilde{p}_T^3 \bgamma^2 \bv^2 /(e+p)   A_{(l+2)*\text{c}}   R_{(n-k-l)*\text{c}} \nonumber \\
    &+(1+j)n_B \Tilde{m}_T \Tilde{p}_T^2 \bgamma^2 \bv^3 /(e+p)   A_{(l+3)*\text{c}}   R_{(n-k-l+1)*\text{c}} \nonumber \\
    &-(1+j) \Tilde{p}_T Q_B \bbeta \balpha   A_{(l+2)*\text{c}}   R_{(n-k-l)*\text{c}} \nonumber \\
    &-(1+j)n_B \Tilde{m}_T^2 \Tilde{p}_T \bgamma^2 /(e+p)   A_{(l+2)*\text{c}}   R_{(n-k-l+2)*\text{c}} \nonumber \\
    &+(1+j)Q_B \Tilde{m}_T \bv \bbeta \balpha   A_{(l+1)*\text{c}}   R_{(n-k-l+1)*\text{c}} \nonumber \\
    &+(1+j)n_B \Tilde{m}_T \Tilde{p}_T \bgamma \balpha /(e+p)   A_{(l+2)*\text{c}}   R_{(n-k-l+1)*\text{c}} \nonumber \\
    &-(1+j)n_B \Tilde{m}_T^2  \bgamma \bv \balpha /(e+p)   A_{(l+1)*\text{c}}   R_{(n-k-l+2)*\text{c}} \nonumber 
    \end{align}
\begin{align}
    &-(1+j)n_B \Tilde{m}_T^2 \Tilde{p}_T \bgamma^2 /(e+p)   A_{(l+2)*\text{c}}   R_{(n-k-l+2)*\text{c}} \nonumber \\
    &-(1+j)n_B \Tilde{p}_T^2 \bgamma \bv \balpha /(e+p)   A_{(l+3)*\text{c}}   R_{(n-k-l)*\text{c}} \nonumber \\
    &+(1+j)n_B \Tilde{m}_T \Tilde{p}_T \bgamma \bv^2 \balpha /(e+p)   A_{(l+2)*\text{c}}   R_{(n-k-l+1)*\text{c}} \nonumber \\
    &+\left( (e+p)n_B^*-n_B(e^*+p^*) \right) /(e+p)^2 \big(\nonumber\\
    &-\Tilde{m}_T \Tilde{p}_T \bgamma  A_{(l+2)*\text{c}}   R_{(n-k-l+1)*\text{c}} \nonumber \\
    &+\Tilde{m}_T^2 \bv \bgamma  A_{(l+1)*\text{c}}   R_{(n-k-l+2)*\text{c}} \nonumber \\
    &+ \Tilde{p}_T^2 \bv \bgamma  A_{(l+3)*\text{c}}   R_{(n-k-l)*\text{c}} \nonumber \\
    &-\Tilde{m}_T \Tilde{p}_T \bgamma  \bv^2 A_{(l+2)*\text{c}}   R_{(n-k-l+1)*\text{c}}\big) \big)\big] \nonumber \\
   \end{align}

\begin{align}
    Y_2^a &=-\frac{(j+1) \bar{\pi }_B \bar{\gamma }^2 \left(\bar{v}^2+1\right) A_c \tau _B \left(3 c_s^2-1\right) R_{\text{cc}} \tilde{m}_T \tilde{p}_T}{3 \zeta   } \nonumber\\
    &+\frac{(j+1) \bar{\pi }_B \bar{\gamma }^2 \bar{v} A_{\text{cc}} \tau _B R_c \left(3 c_s^2-1\right) \tilde{p}_T^2}{3 \zeta }\nonumber\\
    &+\frac{\bar{\pi }_B   \bar{\gamma } A_c \tau _B R_c \left(3 c_s^2-1\right) \tilde{p}_T}{3 \zeta }\nonumber\\
    &-\bar{\gamma } \bar{\pi }^t \left(\bar{v}^2+1\right) A_c R_{\text{cc}}   \tilde{m}_T \tilde{p}_T +\bar{\gamma }^2 \bar{\pi }^t A_{\text{cc}} R_c \tilde{p}_T^2\nonumber\\
    &-\frac{3}{2} (j+1) \bar{\gamma }^3 \bar{\pi }^t \bar{v}^2 A_c R_{\text{ccc}} \tilde{m}_T^2 \tilde{p}_T\nonumber\\
    &-\frac{1}{2} (j+1) \bar{\gamma }^3 \bar{\pi }^t A_{\text{ccc}} R_c \tilde{p}_T^3\nonumber\\
    &+\frac{1}{2} \bar{\pi }^{33} (j+1) \bar{\gamma } A_c R_{\text{css}} \tilde{m}_T^2 \tilde{p}_T\nonumber\\
    &+\frac{1}{2} \bar{\pi }^{22} (j+1) \bar{\gamma } A_{\text{css}} R_c \tilde{p}_T^3\nonumber\\
   &+\bar{\gamma } A_c R_c \tilde{p}_T-\frac{1}{2} \bar{\pi }^{22} (j+1) \bar{\gamma } \bar{v} A_{\text{ss}} R_{\text{cc}} \tilde{m}_T  \tilde{p}_T^2\nonumber\\
   &+\frac{3}{2} (j+1) \bar{\gamma }^3 \bar{\pi }^t \bar{v} A_{\text{cc}} R_{\text{cc}} \tilde{m}_T \tilde{p}_T^2\nonumber\\
   &-\frac{\bar{\pi }_B \bar{\gamma } \bar{v} \tau _B \left(3 c_s^2-1\right) A_0 R_{\text{cc}} \tilde{m}_T}{3 \zeta }\nonumber\\
   &+\frac{(j+1) \bar{\pi }_B \bar{\gamma }^2 \bar{v} \tau_B \left(3 c_s^2-1\right) A_0 R_{\text{ccc}} \tilde{m}_T^2}{3 \zeta }\nonumber\\
   &-\frac{1}{2} \bar{\pi }^{33} (j+1) \bar{\gamma } \bar{v} A_0 R_{\text{ccss}} \tilde{m}_T^3 -\bar{\gamma } \bar{v} A_0 R_{\text{cc}} \tilde{m}_T\nonumber\\
   &+\bar{\gamma }^2 \bar{\pi }^t \bar{v} A_0 R_{\text{ccc}}\tilde{m}_T^2+\frac{1}{2} (j+1) \bar{\gamma }^3 \bar{\pi }^t \bar{v}^3 A_0 R_{\text{cccc}} \tilde{m}_T^3\nonumber\\
   &-\frac{2 (j+1) \bnu   \bar{\gamma }^2 \bar{v} A_c n_B R_{\text{cc}} m_T \tilde{p}_T}{\kappa(e+p)}-\frac{\bnu \bar{\gamma } \bar{v} A_c n_B R_c p_T}{\kappa  (e+p)}\nonumber\\
   &+\frac{(j+1) \bnu  \bar{\gamma }^2  A_{\text{cc}} n_B R_c \tilde{p}_T p_T}{\kappa  (e+p)}\nonumber\\
   &+\frac{\bnu \bar{\gamma } n_B A_0 R_{\text{cc}} m_T}{\kappa  (e+p)}+\frac{(j+1) \bnu \bar{\gamma }^2 \bar{v}^2 n_B A_0 R_{\text{ccc}} \tilde{m}_T m_T}{\kappa (e+p)}\nonumber\\
    &+\sum_{n=0}^\infty \sum_{k=0}^n \sum_{l=0}^{n-k}  (-1)^{n+k+l} \frac{n!(1+n)}{ k! l! (n-k-l)!} \nonumber\\
     &\times(\frac{m_T}{m})^{n-k-l} ( \bar{v} \frac{p_T}{m})^l\nonumber
     \end{align}
\begin{align}
    &\times \left[ \frac{Q_B}{m^2 \kappa }\left((1+j) \bnu \bgamma \bv p_T^2 \Tilde{p}_T A_{(l+3)*\text{c}} R_{(n-k-l+1)*\text{c}}\right.\right.\nonumber\\
    &\left.\left. -(1+j) \bnu \bgamma (1+\bv^2) p_T^2  A_{(l+2)*\text{c}} R_{(n-k-l+1)*\text{c}}\right.\right.\nonumber\\  
    &\left.\left. -(1+j)(1+2\bv^2) \bnu\bgamma p_T^2 \Tilde{m}_T  A_{(l+2)*\text{c}} R_{(n-k-l+2)*\text{c}}\right.\right.\nonumber\\
    &\left.\left. +4 \bnu \bv m_T p_T  A_{(l+1)*\text{c}} R_{(n-k-l+2)*\text{c}}\right.\right.\nonumber\\
    &\left.\left. (1+j)\bv^2 \bnu (2+\bv) \bgamma p_T m_T \Tilde{m}_T  A_{(l+1)*\text{c}} R_{(n-k-l+3)*\text{c}}\right.\right.\nonumber\\
    &\left.\left. -(1+\bv^2) \bnu m_T^2 A_{l*\text{c}} R_{(n-k-l+3)*\text{c}}\right.\right.\nonumber\\
    &\left.\left. -(1+j) \bnu \bgamma \bv^2 m_T^2 \Tilde{m}_T A_{l*\text{c}} R_{(n-k-l+4)*\text{c}} \right)\right.  \nonumber\\
    &\left.+\frac{\bpi_B \tau_B }{3\bgamma\zeta} \left( -(1+j)\bv^2 \Tilde{p}_T \Tilde{m}_T A_{(l+1)*\text{c}} R_{(n-k-l+2)*\text{c}} \right.\right. \nonumber\\
    &\left.\left. +(1+j) \bv \bgamma^{-1} \Tilde{p}_T^2 A_{(l+2)*\text{c}} R_{(n-k-l+1)*\text{c}}  \right.\right. \nonumber\\
    &\left.\left. + \bv \Tilde{m}_T A_{l*\text{c}} R_{(n-k-l+2)*\text{c}}  \right.\right. \nonumber\\
     &\left.\left. + (1+j) \bv \Tilde{m}_T^2 A_{l*\text{c}} R_{(n-k-l+3)*\text{c}}  \right.\right. \nonumber\\
      &\left.\left. - \bv \bgamma^{-1} \Tilde{p}_T A_{(l+1)*\text{c}} R_{(n-k-l+1)*\text{c}}  \right) \right]
\end{align}
%R_c goes A_c
\begin{align}
    Y_2^b &=-\frac{(j+1) \bar{\pi }_B \bar{\gamma }^2 \left(\bar{v}^2+1\right) A_{\text{cc}} \tau _B \left(3 c_s^2-1\right) R_{\text{c}} \tilde{m}_T \tilde{p}_T}{3 \zeta} \nonumber\\
    &+\frac{(j+1) \bar{\pi }_B \bar{\gamma }^2 \bar{v} A_{\text{ccc}} \tau _B R_0 \left(3 c_s^2-1\right) \tilde{p}_T^2}{3 \zeta }\nonumber\\
    &+\frac{\bar{\pi }_B   \bar{\gamma } A_{\text{cc}} \tau _B R_0 \left(3 c_s^2-1\right) \tilde{p}_T}{3 \zeta }\nonumber\\
    &-\bar{\gamma } \bar{\pi }^t \left(\bar{v}^2+1\right) A_{\text{cc}} R_{\text{c}}   \tilde{m}_T \tilde{p}_T +\bar{\gamma }^2 \bar{\pi }^t A_{\text{ccc}} R_0 \tilde{p}_T^2\nonumber\\
    &-\frac{3}{2} (j+1) \bar{\gamma }^3 \bar{\pi }^t \bar{v}^2 A_{\text{cc}} R_{\text{cc}} \tilde{m}_T^2 \tilde{p}_T\nonumber\\
    &-\frac{1}{2} (j+1) \bar{\gamma }^3 \bar{\pi }^t A_{\text{cccc}} R_0 \tilde{p}_T^3\nonumber\\
    &+\frac{1}{2} \bar{\pi }^{33} (j+1) \bar{\gamma } A_{\text{cc}} R_{\text{ss}} \tilde{m}_T^2 \tilde{p}_T\nonumber\\
    &+\frac{1}{2} \bar{\pi }^{22} (j+1) \bar{\gamma } A_{\text{ccss}} R_0 \tilde{p}_T^3\nonumber\\
   &+\bar{\gamma } A_{\text{cc}} R_0 \tilde{p}_T-\frac{1}{2} \bar{\pi }^{22} (j+1) \bar{\gamma } \bar{v} A_{\text{css}} R_{\text{c}} \tilde{m}_T  \tilde{p}_T^2\nonumber\\
   &+\frac{3}{2} (j+1) \bar{\gamma }^3 \bar{\pi }^t \bar{v} A_{\text{ccc}} R_{\text{c}} \tilde{m}_T \tilde{p}_T^2\nonumber\\
   &-\frac{\bar{\pi }_B \bar{\gamma } \bar{v} \tau _B \left(3 c_s^2-1\right) A_c R_{\text{c}} \tilde{m}_T}{3 \zeta }\nonumber\\
   &+\frac{(j+1) \bar{\pi }_B \bar{\gamma }^2 \bar{v} \tau_B \left(3 c_s^2-1\right) A_c R_{\text{cc}} \tilde{m}_T^2}{3 \zeta }\nonumber\\
   &-\frac{1}{2} \bar{\pi }^{33} (j+1) \bar{\gamma } \bar{v} A_c R_{\text{css}} \tilde{m}_T^3 -\bar{\gamma } \bar{v} A_c R_{\text{c}} \tilde{m}_T\nonumber\\
   &+\bar{\gamma }^2 \bar{\pi }^t \bar{v} A_c R_{\text{cc}}\tilde{m}_T^2+\frac{1}{2} (j+1) \bar{\gamma }^3 \bar{\pi }^t \bar{v}^3 A_c R_{\text{ccc}} \tilde{m}_T^3\nonumber\\
   &-\frac{2 (j+1) \bnu   \bar{\gamma }^2 \bar{v} A_{\text{cc}} n_B R_{\text{c}} m_T \tilde{p}_T}{\kappa(e+p)}-\frac{\bnu \bar{\gamma } \bar{v} A_{\text{cc}} n_B R_0 p_T}{\kappa  (e+p)}\nonumber\\
   &+\frac{(j+1) \bnu  \bar{\gamma }^2  A_{\text{ccc}} n_B R_0 \tilde{p}_T p_T}{\kappa  (e+p)}\nonumber\\
   &+\frac{\bnu \bar{\gamma } n_B A_c R_{\text{c}} m_T}{\kappa  (e+p)}+\frac{(j+1) \bnu \bar{\gamma }^2 \bar{v}^2 n_B A_c R_{\text{cc}} \tilde{m}_T m_T}{\kappa (e+p)}\nonumber
   \end{align}
\begin{align}
    &+\sum_{n=0}^\infty \sum_{k=0}^n \sum_{l=0}^{n-k}  (-1)^{n+k+l} \frac{n!(1+n)}{ k! l! (n-k-l)!} \nonumber\\
     &\times(\frac{m_T}{m})^{n-k-l} ( \bar{v} \frac{p_T}{m})^l\nonumber\\
    &\times \left[ \frac{Q_B}{m^2 \kappa }\left((1+j) \bnu \bgamma \bv p_T^2 \Tilde{p}_T A_{(l+4)*\text{c}} R_{(n-k-l)*\text{c}}\right.\right.\nonumber\\
    &\left.\left. -(1+j) \bnu \bgamma (1+\bv^2) p_T^2  A_{(l+3)*\text{c}} R_{(n-k-l)*\text{c}}\right.\right.\nonumber\\  
    &\left.\left. -(1+j)(1+2\bv^2) \bnu\bgamma p_T^2 \Tilde{m}_T  A_{(l+3)*\text{c}} R_{(n-k-l+1)*\text{c}}\right.\right.\nonumber\\ %here
    &\left.\left. +4 \bnu \bv m_T p_T  A_{(l+2)*\text{c}} R_{(n-k-l+1)*\text{c}}\right.\right.\nonumber\\
    &\left.\left. (1+j)\bv^2 \bnu (2+\bv) \bgamma p_T m_T \Tilde{m}_T  A_{(l+2)*\text{c}} R_{(n-k-l+2)*\text{c}}\right.\right.\nonumber\\
    &\left.\left. -(1+\bv^2) \bnu m_T^2 A_{(l+1)*\text{c}} R_{(n-k-l+2)*\text{c}}\right.\right.\nonumber\\
    &\left.\left. -(1+j) \bnu \bgamma \bv^2 m_T^2 \Tilde{m}_T A_{(l+1)*\text{c}} R_{(n-k-l+3)*\text{c}} \right)\right.  \nonumber\\
    &\left.+\frac{\bpi_B \tau_B }{3\bgamma\zeta} \left( -(1+j)\bv^2 \Tilde{p}_T \Tilde{m}_T A_{(l+2)*\text{c}} R_{(n-k-l+1)*\text{c}} \right.\right. \nonumber\\
    &\left.\left. +(1+j) \bv \bgamma^{-1} \Tilde{p}_T^2 A_{(l+3)*\text{c}} R_{(n-k-l)*\text{c}}  \right.\right. \nonumber\\
    &\left.\left. + \bv \Tilde{m}_T A_{(l+1)*\text{c}} R_{(n-k-l+1)*\text{c}}  \right.\right. \nonumber\\
     &\left.\left. + (1+j) \bv \Tilde{m}_T^2 A_{(l+1)*\text{c}} R_{(n-k-l+2)*\text{c}}  \right.\right. \nonumber\\
      &\left.\left. - \bv \bgamma^{-1} \Tilde{p}_T A_{(l+2)*\text{c}} R_{(n-k-l)*\text{c}}  \right) \right]
\end{align}
\begin{align}
    Y_3^a &= \frac{(j+1) \bpi_B \bar{\gamma } A_s \tau _B \left(3 c_s^2-1\right) R_{\text{cc}} \tilde{m}_T \tilde{p}_T}{3 \zeta }\nonumber\\
    &-\frac{(j+1) \bpi_B \bar{\gamma } \bar{v} A_{\text{cs}} \tau _B R_c \left(3 c_s^2-1\right) \tilde{p}_T^2}{3 \zeta }\nonumber\\
    &-(j+1) \bpi^t \bar{\gamma }^2 \bar{v} A_{\text{cs}} R_{\text{cc}} \tilde{m}_T \tilde{p}_T^2\nonumber\\
    &-\frac{1}{2} (j+1)  \tilde{p}_T^3 \left(\bar{\pi }^{22} A_{\text{sss}}R_c-\bpi^t   \bar{\gamma }^2 A_{\text{ccs}}R_c\right)\nonumber\\
    &-\frac{1}{2} \bar{\pi }^{33} (1+j) A_s R_{\text{css}} \tilde{m}_T^2 \tilde{p}_T\nonumber\\
    &+\bar{\pi }^{22} \bar{\gamma } \tilde{p}_T \left(\tilde{p}_T A_{\text{cs}} R_c -\tilde{m}_T A_s R_{\text{cc}} \right)\nonumber\\
    &+\frac{1}{2} (j+1) \bpi^t \bar{\gamma }^2 \bar{v}^2 A_s R_{\text{ccc}} \tilde{m}_T^2   \tilde{p}_T\nonumber\\
    &-\frac{\bpi_B A_s \tau _B R_c \left(3 c_s^2-1\right) \tilde{p}_T}{3 \zeta }-A_s R_c \tilde{p}_T\nonumber\\
    &+\frac{(j+1) \bnu   \bar{\gamma } n_B p_T \left(\bar{v} A_s R_{\text{cc}} \tilde{m}_T-A_{\text{cs}} R_c \tilde{p}_T\right)}{\kappa  (e+p)}\nonumber\\
    &+\sum_{n=0}^\infty \sum_{k=0}^n \sum_{l=0}^{n-k}  (-1)^{n+k+l} \frac{n!(1+n)}{ k! l! (n-k-l)!} \nonumber\\
   &\times(\frac{m_T}{m})^{n-k-l} ( \bar{v} \frac{p_T}{m})^l\nonumber\\
   &\times \left[ \frac{Q_B}{m^2 \kappa }\left((1+j)\bnu (1+\bv^2) m_T^2 \Tilde{p}_T A_{((l+1)*\text{c})s} R_{(n-k-l+2)*\text{c}}\right.\right.\nonumber\\
   &\left.\left. -(1+j)\bnu \bv p_T^2 \Tilde{p}_T  A_{((l+2)*\text{c})s} R_{(n-k-l+1)*\text{c}} \right.\right.\nonumber\\
   &\left.\left. -(1+j)\bnu \bv m_T^2 \Tilde{p}_T  A_{(l*\text{c})s} R_{(n-k-l+3)*\text{c}} \right.\right.\nonumber\\
    &\left.\left. -\bnu \bv m_T p_T \bgamma^{-1}  A_{(l*\text{c})s} R_{(n-k-l+2)*\text{c}} \right.\right.\nonumber\\
    &\left.\left. +\bnu p_T^2 \bgamma^{-1}  A_{((l+1)*\text{c})s} R_{(n-k-l+1)*\text{c}} \right)\right.\nonumber
    \end{align}
\begin{align}
    &+\frac{\bpi_B \tau_B }{3\bgamma\zeta} \left.\left((1+j)\tilde{p}_T \tilde{m}_T A_{((l*\text{c})s} R_{(n-k-l+2)*\text{c}}\right.\right.\nonumber\\
    &\left.\left. -(1+j)\tilde{p}_T^2 \bv A_{(((l+1)*\text{c})s} R_{(n-k-l+1)*\text{c}}\right.\right.\nonumber\\
    &\left.\left. +\tilde{p}_T \bgamma^{-1} A_{((l*\text{c})s} R_{(n-k-l+1)*\text{c}}  \right)\right]
    \end{align}
    %delta v2 b R_c goes A_c
\begin{align}
    Y_3^b &=\frac{(j+1) \bpi_B \bar{\gamma } A_{\text{cs}} \tau _B \left(3 c_s^2-1\right) R_{\text{c}} \tilde{m}_T \tilde{p}_T}{3 \zeta }\nonumber\\
    &-\frac{(j+1) \bpi_B \bar{\gamma } \bar{v} A_{\text{ccs}} \tau _B R_0 \left(3 c_s^2-1\right) \tilde{p}_T^2}{3 \zeta }\nonumber\\
    &-(j+1) \bpi^t \bar{\gamma }^2 \bar{v} A_{\text{ccs}} R_{\text{c}} \tilde{m}_T \tilde{p}_T^2\nonumber\\
    &-\frac{1}{2} (j+1)  \tilde{p}_T^3 \left(\bar{\pi }^{22} A_{\text{csss}}R_0-\bpi^t   \bar{\gamma }^2 A_{\text{cccs}}R_0\right)\nonumber\\
    &-\frac{1}{2} \bar{\pi }^{33} (1+j) A_{\text{cs}} R_{\text{ss}} \tilde{m}_T^2 \tilde{p}_T\nonumber\\
    &+\bar{\pi }^{22} \bar{\gamma } \tilde{p}_T \left(\tilde{p}_T A_{\text{ccs}} R_0 -\tilde{m}_T A_{\text{cs}} R_{\text{c}} \right)\nonumber\\
    &+\frac{1}{2} (j+1) \bpi^t \bar{\gamma }^2 \bar{v}^2 A_{\text{cs}} R_{\text{cc}} \tilde{m}_T^2   \tilde{p}_T\nonumber\\
    &-\frac{\bpi_B A_{\text{cs}} \tau _B R_0 \left(3 c_s^2-1\right) \tilde{p}_T}{3 \zeta }-A_{\text{cs}} R_0 \tilde{p}_T\nonumber\\
    &+\frac{(j+1) \bnu   \bar{\gamma } n_B p_T \left(\bar{v} A_{\text{cs}} R_{\text{c}} \tilde{m}_T-A_{\text{ccs}} R_0 \tilde{p}_T\right)}{\kappa  (e+p)}\nonumber\\
    &+\sum_{n=0}^\infty \sum_{k=0}^n \sum_{l=0}^{n-k}  (-1)^{n+k+l} \frac{n!(1+n)}{ k! l! (n-k-l)!} \nonumber\\
   &\times(\frac{m_T}{m})^{n-k-l} ( \bar{v} \frac{p_T}{m})^l\nonumber\\
   &\times \left[ \frac{Q_B}{m^2 \kappa }\left((1+j)\bnu (1+\bv^2) m_T^2 \Tilde{p}_T A_{((l+2)*\text{c})s} R_{(n-k-l+1)*\text{c}}\right.\right.\nonumber\\
   &\left.\left. -(1+j)\bnu \bv p_T^2 \Tilde{p}_T  A_{((l+3)*\text{c})s} R_{(n-k-l)*\text{c}} \right.\right.\nonumber\\
   &\left.\left. -(1+j)\bnu \bv m_T^2 \Tilde{p}_T  A_{((l+1)*\text{c})s} R_{(n-k-l+2)*\text{c}} \right.\right.\nonumber\\
    &\left.\left. -\bnu \bv m_T p_T \bgamma^{-1}  A_{((l+1)*\text{c})s} R_{(n-k-l+1)*\text{c}} \right.\right.\nonumber\\
    &\left.\left. +\bnu p_T^2 \bgamma^{-1}  A_{((l+2)*\text{c})s} R_{(n-k-l)*\text{c}} \right)\right.\nonumber\\
    &+\frac{\bpi_B \tau_B }{3\bgamma\zeta} \left.\left((1+j)\tilde{p}_T \tilde{m}_T A_{(((l+1)*\text{c})s} R_{(n-k-l+1)*\text{c}}\right.\right.\nonumber\\
    &\left.\left. -(1+j)\tilde{p}_T^2 \bv A_{(((l+2)*\text{c})s} R_{(n-k-l)*\text{c}}\right.\right.\nonumber\\
    &\left.\left. +\tilde{p}_T \bgamma^{-1} A_{(((l+1)*\text{c})s} R_{(n-k-l)*\text{c}}  \right)\right]
\end{align}
%v3 a ker
\begin{align}
    Y_4^a &= -\frac{1}{2} (j+1) A_0 R_{\text{csss}} \tilde{m}_T^3 \bar{\pi }^{33} -A_0 R_{\text{cs}} \tilde{m}_T \nonumber\\
    &+\frac{1}{2} (j+1) \bpi^t \bar{v}^2 \bar{\gamma }^2 A_0 R_{\text{cccs}} \tilde{m}_T^3-\bar{\gamma } R_{\text{ccs}} A_0 \tilde{m}_T^2 \bar{\pi }^{33}\nonumber\\
    &+(j+1)\bar{\gamma } R_{\text{ccs}} A_c \tilde{m}_T^2 \bpi^t \bar{v} \bar{\gamma }
   \tilde{p}_T+ R_{\text{cs}} \tilde{m}_T \tilde{p}_T  \bar{\gamma } A_c \bar{\pi }^{33}\nonumber\\
   &-\frac{1}{2} R_{\text{cs}} \tilde{m}_T \tilde{p}_T^2 (j+1) A_{\text{ss}} \bar{\pi }^{22}\nonumber\\
   &+\frac{1}{2} R_{\text{cs}} \tilde{m}_T \tilde{p}_T^2 (j+1) \bpi^t \bar{\gamma }^2 A_{\text{cc}} \nonumber
   \end{align}
\begin{align}
   &+\frac{(j+1) \bnu  \bar{\gamma } n_B m_T \left(\bar{v} A_0 R_{\text{ccs}} \tilde{m}_T-A_c R_{\text{cs}} \tilde{p}_T\right)}{\kappa  (e+p)}\nonumber\\
   &-\frac{(j+1) \bpi_B \bar{\gamma } \bar{v} A_c \tau _B R_{\text{cs}} \left(3 c_s^2-1\right)\tilde{m}_T \tilde{p}_T}{3 \zeta }\nonumber\\
   &+\frac{(j+1)   \bpi_B \bar{\gamma } \tau _B \left(3 c_s^2-1\right) A_0 R_{\text{ccs}}  \tilde{m}_T^2}{3 \zeta }\nonumber\\
   &-\frac{\bpi_B \tau _B A_0 R_{\text{cs}} \left(3   c_s^2-1\right) \tilde{m}_T}{3 \zeta }\nonumber\\
   &+\sum_{n=0}^\infty \sum_{k=0}^n \sum_{l=0}^{n-k}  (-1)^{n+k+l} \frac{n!(1+n)}{ k! l! (n-k-l)!} \nonumber\\
   &\times(\frac{m_T}{m})^{n-k-l} ( \bar{v} \frac{p_T}{m})^l\nonumber\\
   &\times \left[ \frac{Q_B}{m^2 \kappa }\left((1+j)\bnu (1+\bv^2) m_T^2 \Tilde{p}_T A_{(l+1)*\text{c}} R_{((n-k-l+2)*\text{c})s}\right.\right.\nonumber\\
   &\left.\left.-\bnu\bv m_T^2 \bgamma^{-1} A_{l*\text{c}} R_{((n-k-l+2)*\text{c})s} \right.\right.\nonumber\\
   &\left.\left. +\bnu m_T p_T \bgamma^{-1} A_{(l+1)*\text{c}} R_{((n-k-l+1)*\text{c})s}\right.\right.\nonumber\\
   &\left.\left.-(1+j) \bnu\bv \Tilde{m}_T p_T^2  A_{(l+2)*\text{c}} R_{((n-k-l+1)*\text{c})s} \right.\right.\nonumber\\
   &\left.\left.-(1+j) \bnu\bv \Tilde{m}_T m_T^2  A_{l*\text{c}} R_{((n-k-l+3)*\text{c})s} \right) \right.\nonumber\\
   &+\frac{\bpi_B \tau_B }{3\bgamma\zeta}\left.\left((1+j)\Tilde{m}_T^2 A_{l*\text{c}} R_{((n-k-l+2)*\text{c})s}\right.\right.\nonumber\\
   &\left.\left. -(1+j)\Tilde{m}_T \Tilde{p}_T \bv A_{(l+1)*\text{c}} R_{((n-k-l+1)*\text{c})s}\right.\right.\nonumber\\
   &\left.\left. + \bgamma^{-1}\Tilde{m}_T \bv A_{l*\text{c}} R_{((n-k-l+1)*\text{c})s} \right)\right]
   \end{align}
   %v3 b ker R_c goea A_c
   \begin{align}
    Y_4^b &=-\frac{1}{2} (j+1) A_c R_{\text{sss}} \tilde{m}_T^3 \bar{\pi }^{33} -A_c R_{\text{s}} \tilde{m}_T \nonumber\\
    &+\frac{1}{2} (j+1) \bpi^t \bar{v}^2 \bar{\gamma }^2 A_c R_{\text{ccs}} \tilde{m}_T^3-\bar{\gamma } R_{\text{cs}} A_c \tilde{m}_T^2 \bar{\pi }^{33}\nonumber\\
    &+(j+1)\bar{\gamma } R_{\text{cs}} A_{\text{cc}} \tilde{m}_T^2 \bpi^t \bar{v} \bar{\gamma }   \tilde{p}_T+ R_{\text{s}} \tilde{m}_T \tilde{p}_T  \bar{\gamma } A_{\text{cc}} \bar{\pi }^{33}\nonumber\\
   &-\frac{1}{2} R_{\text{s}} \tilde{m}_T \tilde{p}_T^2 (j+1) A_{\text{css}} \bar{\pi }^{22}\nonumber\\
   &+\frac{1}{2} R_{\text{s}} \tilde{m}_T \tilde{p}_T^2 (j+1) \bpi^t \bar{\gamma }^2 A_{\text{ccc}} \nonumber\\
   &+\frac{(j+1) \bnu  \bar{\gamma } n_B m_T \left(\bar{v} A_c R_{\text{cs}} \tilde{m}_T-A_{\text{cc}} R_{\text{s}} \tilde{p}_T\right)}{\kappa  (e+p)}\nonumber\\
   &-\frac{(j+1) \bpi_B \bar{\gamma } \bar{v} A_{\text{cc}} \tau _B R_{\text{s}} \left(3 c_s^2-1\right)\tilde{m}_T \tilde{p}_T}{3 \zeta }\nonumber\\
   &+\frac{(j+1)   \bpi_B \bar{\gamma } \tau _B \left(3 c_s^2-1\right) A_c R_{\text{cs}}  \tilde{m}_T^2}{3 \zeta }\nonumber\\
   &-\frac{\bpi_B \tau _B A_c R_{\text{s}} \left(3   c_s^2-1\right) \tilde{m}_T}{3 \zeta }\nonumber\\
   &+\sum_{n=0}^\infty \sum_{k=0}^n \sum_{l=0}^{n-k}  (-1)^{n+k+l} \frac{n!(1+n)}{ k! l! (n-k-l)!} \nonumber\\
   &\times(\frac{m_T}{m})^{n-k-l} ( \bar{v} \frac{p_T}{m})^l\nonumber\\
   &\times \left[ \frac{Q_B}{m^2 \kappa }\left((1+j)\bnu (1+\bv^2) m_T^2 \Tilde{p}_T A_{(l+2)*\text{c}} R_{((n-k-l+1)*\text{c})s}\right.\right.\nonumber\\
   &\left.\left.-\bnu\bv m_T^2 \bgamma^{-1} A_{(l+1)*\text{c}} R_{((n-k-l+1)*\text{c})s} \right.\right.\nonumber
   \end{align}
\begin{align}
   &\left.\left. +\bnu m_T p_T \bgamma^{-1} A_{(l+2)*\text{c}} R_{((n-k-l)*\text{c})s}\right.\right.\nonumber\\
   &\left.\left.-(1+j) \bnu\bv \Tilde{m}_T p_T^2  A_{(l+3)*\text{c}} R_{((n-k-l)*\text{c})s} \right.\right.\nonumber\\
   &\left.\left.-(1+j) \bnu\bv \Tilde{m}_T m_T^2  A_{(l+1)*\text{c}} R_{((n-k-l+2)*\text{c})s} \right) \right.\nonumber\\
   &+\frac{\bpi_B \tau_B }{3\bgamma\zeta}\left.\left((1+j)\Tilde{m}_T^2 A_{(l+1)*\text{c}} R_{((n-k-l+1)*\text{c})s}\right.\right.\nonumber\\
   &\left.\left. -(1+j)\Tilde{m}_T \Tilde{p}_T \bv A_{(l+2)*\text{c}} R_{((n-k-l)*\text{c})s}\right.\right.\nonumber\\
   &\left.\left. + \bgamma^{-1}\Tilde{m}_T \bv A_{(l+1)*\text{c}} R_{((n-k-l)*\text{c})s} \right)\right]
    \end{align}
    %delta nu kernel
    \begin{align}
    Y_5^a &=\frac{\bar{\gamma } A_c n_B R_c \tilde{p}_T}{\kappa  (e+p)}-\frac{\bar{\gamma } \bar{v} n_B R_{\text{cc}}
   \tilde{m}_T}{\kappa  (e+p)}\nonumber\\
   &+\frac{Q_B}{m\kappa}\sum_{n=0}^\infty \sum_{k=0}^n \sum_{l=0}^{n-k}  (-1)^{n+k+l} \frac{n!}{k! l! (n-k-l)!} \nonumber\\
   &\times(\bar{\gamma }\frac{m_T}{m})^{n-k-l} (\bar{\gamma } \bar{v} \frac{p_T}{m})^l\left[\bgamma \tilde{p}_T R_{(n-k-l+1)*\text{c}} A_{(l+1)*\text{c}}\right.\nonumber\\
   &- \bgamma \bv \tilde{m}_T\left.  R_{(n-k-l+2)*\text{c}} A_{l*\text{c}}\right]\\
    Y_5^b &=\frac{\bar{\gamma } A_{\text{cc}} n_B
   \tilde{p}_T}{\kappa  (e+p)}-\frac{\bar{\gamma } \bar{v} A_c n_B R_c \tilde{m}_T}{\kappa  (e+p)}\nonumber\\
    &+\frac{Q_B}{m\kappa}\sum_{n=0}^\infty \sum_{k=0}^n \sum_{l=0}^{n-k}  (-1)^{n+k+l} \frac{n!}{k! l! (n-k-l)!} \nonumber\\
   &\times(\bar{\gamma }\frac{m_T}{m})^{n-k-l} (\bar{\gamma } \bar{v} \frac{p_T}{m})^l\left[\bgamma \tilde{p}_T R_{(n-k-l)*\text{c}} A_{(l+2)*\text{c}}\right.\nonumber\\
   &- \bgamma \bv \tilde{m}_T\left.  R_{(n-k-l+1)*\text{c}} A_{(l+1)*\text{c}}\right]\\
    Y_6^a &=-\frac{A_s n_B R_c \tilde{p}_T}{\kappa  (e+p)}  -\frac{Q_B}{m\kappa}\sum_{n=0}^\infty \sum_{k=0}^n \sum_{l=0}^{n-k}  (-1)^{n+k+l} \nonumber\\
   &\times \frac{n!}{k! l! (n-k-l)!} \tilde{p}_T (\bar{\gamma }\frac{m_T}{m})^{n-k-l} (\bar{\gamma } \bar{v} \frac{p_T}{m})^l \nonumber\\
   &\times R_{(n-k-l+1)*\text{c}} A_{(l*\text{c})s}\\
    Y_6^b &= -\frac{A_{\text{cs}} R_0 n_B  \tilde{p}_T}{\kappa  (e+p)}-\frac{Q_B}{m\kappa}\sum_{n=0}^\infty \sum_{k=0}^n \sum_{l=0}^{n-k}  (-1)^{n+k+l} \nonumber\\
   &\times \frac{n!}{k! l! (n-k-l)!} \tilde{p}_T (\bar{\gamma }\frac{m_T}{m})^{n-k-l} (\bar{\gamma } \bar{v} \frac{p_T}{m})^l \nonumber\\
   &\times R_{(n-k-l)*\text{c}} A_{((l+1)*\text{c})s}\\
    Y_7^a &= -\frac{n_B A_0 R_{\text{cs}}
   \tilde{m}_T}{\kappa  (e+p)} -\frac{Q_B}{m\kappa}\sum_{n=0}^\infty \sum_{k=0}^n \sum_{l=0}^{n-k}  (-1)^{n+k+l} \nonumber\\
   &\times \frac{n!}{k! l! (n-k-l)!} \tilde{m}_T (\bar{\gamma }\frac{m_T}{m})^{n-k-l} (\bar{\gamma } \bar{v} \frac{p_T}{m})^l \nonumber\\
   &\times R_{((n-k-l+1)*\text{c})s} A_{l*\text{c}}\\
    Y_7^b &= -\frac{A_c n_B R_s \tilde{m}_T}{\kappa  (e+p)}  -\frac{Q_B}{m\kappa}\sum_{n=0}^\infty \sum_{k=0}^n \sum_{l=0}^{n-k}  (-1)^{n+k+l} \nonumber\\
   &\times \frac{n!}{k! l! (n-k-l)!} \tilde{m}_T (\bar{\gamma }\frac{m_T}{m})^{n-k-l} (\bar{\gamma } \bar{v} \frac{p_T}{m})^l \nonumber\\
   &\times R_{((n-k-l)*\text{c})s} A_{(l+1)*\text{c}}
    \end{align}
    %shear kernels
    \begin{align}
    Y_8^a &= \bar{\gamma } \bar{v} A_c R_{\text{cc}} \tilde{m}_T \tilde{p}_T-\frac{1}{2} \bar{\gamma }^2 \bar{v}^2 A_0 R_{\text{ccc}} \tilde{m}_T^2\nonumber\\
    &-\frac{1}{2} \bar{\gamma }^2 A_{\text{cc}} R_c \tilde{p}_T^2  +\frac{1}{2} A_{\text{ss}} R_c \tilde{p}_T^2 \\
    Y_8^b &= \bar{\gamma } \bar{v} A_{\text{cc}} R_c \tilde{m}_T \tilde{p}_T-\frac{1}{2} \bar{\gamma }^2 \bar{v}^2 A_c R_{\text{cc}} \tilde{m}_T^2\nonumber\\
    &-\frac{1}{2}  \bar{\gamma }^2 A_{\text{ccc}} R_0 \tilde{p}_T^2+\frac{1}{2}  A_{\text{css}} R_0 \tilde{p}_T^2\\
    Y_9^a &=\bar{\gamma } \bar{v} A_c R_{\text{cc}} \tilde{m}_T \tilde{p}_T-\frac{1}{2} \bar{\gamma }^2 A_{\text{cc}} R_c \tilde{p}_T^2\nonumber\\
    &-\frac{1}{2} \bar{\gamma}^2 \bar{v}^2 A_0 R_{\text{ccc}}\tilde{m}_T^2+\frac{1}{2}  A_0 R_{\text{css}} \tilde{m}_T^2 \\
    Y_9^b &= \bar{\gamma } \bar{v} A_{\text{cc}} R_c \tilde{m}_T \tilde{p}_T-\frac{1}{2} \bar{\gamma }^2 \bar{v}^2 A_c R_{\text{cc}} \tilde{m}_T^2\nonumber\\
    &-\frac{1}{2}  \bar{\gamma }^2  A_{\text{ccc}} R_0 \tilde{p}_T^2+\frac{1}{2} A_c R_{\text{ss}} \tilde{m}_T^2\\
    Y_{10}^a &= \bar{\gamma } \bar{v} A_s R_{\text{cc}} \tilde{m}_T \tilde{p}_T-\bar{\gamma } A_{\text{cs}} R_c \tilde{p}_T^2\\
    Y_{10}^b &= \bar{\gamma } \bar{v} A_{\text{cs}} R_c \tilde{m}_T \tilde{p}_T-\bar{\gamma } A_{\text{ccs}} R_0 \tilde{p}_T^2 \\
    Y_{11}^a &= \bar{\gamma } \bar{v} R_{\text{ccs}} A_0 \tilde{m}_T^2-\bar{\gamma } A_c R_{\text{cs}} \tilde{m}_T \tilde{p}_T\\
    Y_{11}^b &=\bar{\gamma } \bar{v} A_c R_{\text{cs}} \tilde{m}_T^2-\bar{\gamma } A_{\text{cc}} R_s \tilde{m}_T \tilde{p}_T \\
    Y_{12}^a &= A_s R_{\text{cs}} \tilde{m}_T \tilde{p}_T \\
    Y_{12}^b &= A_{\text{cs}} R_s \tilde{m}_T \tilde{p}_T 
    \end{align}
    %bulk kernels
   \begin{align}
    Y^a_{13} &=\frac{(1+j)\tau_B\left(\frac{1}{3}-c_s^2\right)}{\zeta}\left(- \bar{\gamma } \bar{v} A_c R_c  \tilde{p}_T+ \bar{\gamma } A_0  R_{\text{cc}} \tilde{m}_T\right) \nonumber\\
   &-\frac{(j+1) \bbeta m \tau_B}{3 \zeta}\sum_{n=0}^\infty \sum_{k=0}^n \sum_{l=0}^{n-k}  (-1)^{n+k+l} \frac{n!}{k! l! (n-k-l)!} \nonumber\\
   &\times (\bar{\gamma } m_T/m)^{n-k-l} (\bar{\gamma } \bar{v} p_T/m)^l R_{(n-k-l+1)*\text{c}} A_{l*\text{c}} \\
    Y^b_{13} &=\frac{(1+j)\tau_B\left(\frac{1}{3}-c_s^2\right)}{\zeta}\left(- \bar{\gamma } \bar{v} A_{\text{cc}} R_0 \tilde{p}_T+ \bar{\gamma } A_c R_c\tilde{m}_T\right)\nonumber\\
   &-\frac{(j+1) \bbeta m \tau_B}{3 \zeta}\sum_{n=0}^\infty \sum_{k=0}^n \sum_{l=0}^{n-k}  (-1)^{n+k+l} \frac{n!}{k! l! (n-k-l)!} \nonumber\\
   &\times (\bar{\gamma } m_T/m)^{n-k-l} (\bar{\gamma } \bar{v} p_T/m)^l R_{(n-k-l)*\text{c}} A_{(l+1)*\text{c}}
   \end{align}
   %delta mu
   \begin{align}
    Y_{14}^a &=-\frac{\bar{v}^2 \bar{\gamma }^2 A_0 R_{\text{ccc}} \tilde{m}_T^2 \left((j+1) (e+p)- e^+- p^+\right) \bar{\pi }^t}{2 (e+p)}\nonumber\\
   &+\frac{\bar{v} \bar{\gamma }^2 A_c R_{\text{cc}} \tilde{m}_T \tilde{p}_T \left((j+1) (e+p)- e^+- p^+\right)\bar{\pi }^t}{e+p}\nonumber\\
   &+\frac{A_0 R_{\text{css}} \tilde{m}_T^2 \left((j+1) (e+p)- e^+ - p^+\right) \bar{\pi }^{33}}{2   (e+p)} +A_0 R_c\nonumber\\
   &-\frac{\left(\bar{\pi }^t \bar{\gamma }^2 A_{\text{cc}}-\bar{\pi }^{22} A_{\text{ss}}\right) R_c \tilde{p}_T^2 \left((j+1) (e+p)- e^+ - p^+ \right)}{2 (e+p)}\nonumber\\
   &+\bar{\pi }_B \bar{\gamma } \bar{v} A_c R_c \tilde{p}_T \left(3 \zeta  \tau _B \left(c_s^2\right)^+ +\left(3 c_s^2-1\right)\right. \nonumber\\
   &\times\left. \left(\zeta  \left( \tau _B^+ +(1+j) \tau _B\right) - \tau _B \zeta1+ \right)\right)/(3 \zeta^2)\nonumber\\
   &-\bar{\pi }_B \bar{\gamma } A_0 R_{\text{cc}} \tilde{m}_T \left(3 \zeta  \tau _B \left(c_s^2\right)^+ +\left(3 c_s^2-1\right) \right. \nonumber
   \end{align}
\begin{align}
   &\times\left.\left(\zeta  \left( \tau _B^+ +(1+j) \tau _B\right) -\tau _B \zeta^+ \right)\right)/(3 \zeta^2)\nonumber\\
   &+\sum_{n=0}^\infty \sum_{k=0}^n \sum_{l=0}^{n-k}  (-1)^{n+k+l} \frac{n!}{k! l! (n-k-l)!} \nonumber\\
   &\times(\frac{m_T}{m})^{n-k-l} (\bar{v} \frac{p_T}{m})^l\left[\frac{\bpi_B \tau_B m \beta}{3 \zeta}\right.\nonumber\\
   &\times\left(\left.-((1+j)+\tau_B^+/\tau_B)\bgamma^{-1} A_{l*\text{c}}   R_{(n-k-l+1)*\text{c}}\right)\right.\nonumber\\
   &\left.\left.+\zeta^+/\zeta \bgamma^{-1} A_{l*\text{c}}   R_{(n-k-l+1)*\text{c}}  \right)\right.\nonumber\\
   &+\frac{Q_B}{\beta m^2 \kappa}\left((1+j) \Tilde{p}_T \bnu A_{(l+1)*\text{c}}   R_{(n-k-l+1)*\text{c}} \right.\nonumber\\
   &\left.+(1+j) \Tilde{m}_T \bv \bnu A_{l+*\text{c}}   R_{(n-k-l+2)*\text{c}}\right. \nonumber\\
   &\left.+ \Tilde{p}_T \bnu \kappa^+/\kappa A_{(l+1)*\text{c}}   R_{(n-k-l+1)*\text{c}}\right.\nonumber\\
   &\left. - \Tilde{m}_T \bv \bnu \kappa^+/\kappa A_{l*\text{c}}   R_{(n-k-l+2)*\text{c}}\right)\nonumber\\
  & -\frac{\bgamma \bnu \left((1+j) n_B (e+p) +(e+p)n_B^+ -(e^+ +p^+)n_B\right)}{\kappa \bbeta^2(e+p)^2}\nonumber\\
  &\times \left( \bv \Tilde{m}_T^2 A_{l*\text{c}}   R_{(n-k-l+3)*\text{c}} +  \bv \Tilde{p}_T^2 A_{(l+2)*\text{c}}   R_{(n-k-l+1)*\text{c}}\right.\nonumber\\
  &\left. -(1+ \bv^2) \Tilde{m}_T \Tilde{p}_T A_{(l+1)*\text{c}}   R_{(n-k-l+1)*\text{c}} \right)\nonumber\\
  &+\frac{n_B \bnu \bgamma \kappa^+}{\beta^2 \kappa^2 (e+p)}\left(   \bv \Tilde{m}_T^2 A_{l*\text{c}}   R_{(n-k-l+3)*\text{c}}\right.\nonumber\\
  &\left. +  \bv \Tilde{p}_T^2 A_{(l+2)*\text{c}}   R_{(n-k-l+1)*\text{c}}\right.\nonumber\\
  &\left.\left. -(1+ \bv^2) \Tilde{m}_T \Tilde{p}_T A_{(l+1)*\text{c}}   R_{(n-k-l+1)*\text{c}}\right)\right]
   \end{align}
\begin{align}
    Y_{14}^b &= -\frac{\bar{v}^2 \bar{\gamma }^2 A_c R_{\text{cc}} \tilde{m}_T^2 \left((j+1) (e+p)- e^+- p^+\right) \bar{\pi }^t}{2 (e+p)}\nonumber\\
   &+\frac{\bar{v} \bar{\gamma }^2 A_{\text{cc}} R_{\text{c}} \tilde{m}_T \tilde{p}_T \left((j+1) (e+p)- e^+- p^+\right)\bar{\pi }^t}{e+p}\nonumber\\
   &+\frac{A_c R_{\text{ss}} \tilde{m}_T^2 \left((j+1) (e+p)- e^+ - p^+\right) \bar{\pi }^{33}}{2   (e+p)} +A_0 R_c\nonumber\\
   &-\frac{\left(\bar{\pi }^t \bar{\gamma }^2 A_{\text{ccc}}-\bar{\pi }^{22} A_{\text{css}}\right) R_0 \tilde{p}_T^2 \left((j+1) (e+p)- e^+ - p^+ \right)}{2 (e+p)}\nonumber\\
   &+\bar{\pi }_B \bar{\gamma } \bar{v} A_{\text{cc}} R_0 \tilde{p}_T \left(3 \zeta  \tau _B \left(c_s^2\right)^+ +\left(3 c_s^2-1\right)\right. \nonumber\\
   &\times\left. \left(\zeta  \left( \tau _B^+ +(1+j) \tau _B\right) - \tau _B \zeta1+ \right)\right)/(3 \zeta^2)\nonumber\\
   &-\bar{\pi }_B \bar{\gamma } A_c R_{\text{c}} \tilde{m}_T \left(3 \zeta  \tau _B \left(c_s^2\right)^+ +\left(3 c_s^2-1\right) \right. \nonumber\\
   &\times\left.\left(\zeta  \left( \tau _B^+ +(1+j) \tau _B\right) -\tau _B \zeta^+ \right)\right)/(3 \zeta^2)\nonumber\\
   &+\sum_{n=0}^\infty \sum_{k=0}^n \sum_{l=0}^{n-k}  (-1)^{n+k+l} \frac{n!}{k! l! (n-k-l)!} \nonumber\\
   &\times(\frac{m_T}{m})^{n-k-l} (\bar{v} \frac{p_T}{m})^l\left[\frac{\bpi_B \tau_B m \beta}{3 \zeta}\right.\nonumber\\
   &\times\left(\left.-((1+j)+\tau_B^+/\tau_B)\bgamma^{-1} A_{(l+1)*\text{c}}   R_{(n-k-l)*\text{c}}\right)\right.\nonumber\\
   &\left.\left.+\zeta^+/\zeta \bgamma^{-1} A_{(l+1)*\text{c}}   R_{(n-k-l)*\text{c}}  \right)\right.\nonumber
   \end{align}
\begin{align}
   &+\frac{Q_B}{\beta m^2 \kappa}\left((1+j) \Tilde{p}_T \bnu A_{(l+2)*\text{c}}   R_{(n-k-l)*\text{c}} \right.\nonumber\\
   &\left.+(1+j) \Tilde{m}_T \bv \bnu A_{(l+1)+*\text{c}}   R_{(n-k-l+1)*\text{c}}\right. \nonumber\\
   &\left.+ \Tilde{p}_T \bnu \kappa^+/\kappa A_{(l+2)*\text{c}}   R_{(n-k-l)*\text{c}}\right.\nonumber\\
   &\left. - \Tilde{m}_T \bv \bnu \kappa^+/\kappa A_{(l+1)*\text{c}}   R_{(n-k-l+1)*\text{c}}\right)\nonumber\\
  & -\frac{\bgamma \bnu \left((1+j) n_B (e+p) +(e+p)n_B^+ -(e^+ +p^+)n_B\right)}{\kappa \bbeta^2(e+p)^2}\nonumber\\
  &\times \left( \bv \Tilde{m}_T^2 A_{(l+1)*\text{c}}   R_{(n-k-l+2)*\text{c}} +  \bv \Tilde{p}_T^2 A_{(l+3)*\text{c}}   R_{(n-k-l)*\text{c}}\right.\nonumber\\
  &\left. -(1+ \bv^2) \Tilde{m}_T \Tilde{p}_T A_{(l+2)*\text{c}}   R_{(n-k-l)*\text{c}} \right)\nonumber\\
  &+\frac{n_B \bnu \bgamma \kappa^+}{\beta^2 \kappa^2 (e+p)}\left(   \bv \Tilde{m}_T^2 A_{(l+1)*\text{c}}   R_{(n-k-l+2)*\text{c}}\right.\nonumber\\
  &\left. +  \bv \Tilde{p}_T^2 A_{(l+3)*\text{c}}   R_{(n-k-l)*\text{c}}\right.\nonumber\\
  &\left.\left. -(1+ \bv^2) \Tilde{m}_T \Tilde{p}_T A_{(l+2)*\text{c}}   R_{(n-k-l)*\text{c}}\right)\right]
\end{align}
The corresponding field vector reads as $\delta \Phi =(-\delta \beta /\bbeta, \delta v_1, \delta v_2, \delta v_3,\delta \nu_1, \delta \nu_2, \delta \nu_3 ,\delta \pi^{22}, \delta \pi^{33}, \delta \pi^{12}, \delta \pi^{13},$ $ \delta \pi^{23}, \delta \pi_B, \delta \alpha - \alpha \delta \beta /\bbeta )$.
\section{Analytical Rapidity and azimuthal integrals}  \label{sec_integrations}
The rapidity and azimuthal integrations in \autoref{sec_thermal_bg_ker} and \autoref{sec_thermal_pert_ker} can be factorized into terms only depending on rapidity and azimuthal angle, respectively. The integrations in rapidity all have the form 
\begin{equation}
    R_*(k,z) = \frac{1}{2} \int_{-\infty}^{\infty} \ud \eta e^{-z \cosh(\eta)} e^{i k \eta} f_* (\eta),
\end{equation}
where $f_*(\eta)$ is some power of $\cosh(\eta)$ and $\sinh(\eta)$. We will use a shorthand, where the respective power is being denoted by indices $'c'$ and $'s'$, e.g. $R_{ccs}$ is the integrand with $f_{ccs}=\cosh^2(\eta) \sinh(\eta)$. Almost all the appearing integrations up to fourth order in hyperbolic functions (see discussion above) can be expressed using Bessel functions $K_n(z)$ of the second kind,

\begin{align}
	R_0(k,z)&=K_{ik}(z)\\
	R_c(k,z)&= \frac{1}{2}(K_{ik-1}(z)+K_{ik+1}(z))\\
	R_s(k,z)&=\frac{1}{2}(K_{ik-1}(z)+K_{ik+1}(z))\\
	R_{cc}(k,z)&=\frac{1}{4}(K_{ik-2}(z)+2K_{ik}(z)+K_{ik+2}(z))\\
	R_{cs}(k,z)&=\frac{1}{4}(-K_{ik-2}(z)+K_{ik+2}(z))\\
	R_{ss}(k,z)&=\frac{1}{4}(K_{ik-2}(z)-2K_{ik}(z)+K_{ik+2}(z))\\
	R_{ccc}(k,z)&=\frac{1}{8}(K_{ik-3}(z)+3K_{ik-1}(z)\nonumber \\
    &+3K_{ik+1}(z)+K_{ik+3}(z))\\
	R_{ccs}(k,z)&=\frac{1}{8}(-K_{ik-3}(z)-K_{ik-1}(z)\nonumber\\
    &+K_{ik+1}(z)+K_{ik+3}(z))\\
	R_{css}(k,z)&=\frac{1}{8}(-K_{ik-3}(z)-K_{ik-1}(z)\nonumber\\
    &-K_{ik+1}(1)+K_{ik+3}(z))\\
	R_{sss}(k,z)&=\frac{1}{8}(-K_{ik-3}(z)+3K_{ik-1}(z)\nonumber\\
    &-3K_{ik+1}(z)+K_{ik+3}(z))\\
	R_{cccc}(k,z)&=\frac{1}{16}(K_{ik-4}(z)+4K_{ik-2}(z)+6K_{ik}(z)\nonumber\\
    &+4K_{ik+2}(z)+K_{ik+4}(z))\\
	R_{cccs}(k,z)&=\frac{1}{16}(-K_{ik-4}(z)-2K_{ik-2}(z)+2K_{ik+2}(z)\nonumber\\
    &+K_{ik+4}(z))\\
	R_{ccss}(k,z)&=\frac{1}{16}(K_{ik-4}(z)-2K_{ik}(z)+K_{ik+4}(z))\\
	R_{csss}(k,z)&=\frac{1}{16}(-K_{ik-4}(z)+2K_{ik-2}(z)-2K_{ik+2}(z)\nonumber\\
    &+K_{ik+4}(z))\\
	R_{ssss}(k,z)&=\frac{1}{16}(K_{ik-4}(z)-4K_{ik-2}(z)+6K_{ik}(z)\nonumber\\
    &-4K_{ik+2}+K_{ik+4}(z)).
\end{align}

Similarly to the rapidity integrations, the azimuthal integration can be factorized out and written as
\begin{equation}
    A_* (m,z)= \frac{1}{2\pi} \int_0^{2\pi} \ud \phi e^{z \cos(\phi)} e^{im\phi} g_* (\phi).
\end{equation}
As before we employ a similar notation where $g_*(\phi)$ is a power of $\sin(\phi)$ and $\cos(\phi)$, where the integration $A_{ccss}$ is given by the integrand $g_{ccss}=\cos^2(\phi) \sin^2(\phi)$. All the relevant integrations up to fourth power in trigonometric functions can be expressed in terms of modified Bessel functions $I_m(z)$ of the first kind,

\begin{align}
	A_0(m,z)&=I_m(z)\\
	A_c(m,z)&=\frac{1}{2}(I_{m-1}(z)+I_{m+1}(z))\\
	A_s(m,z)&=\frac{i}{2}(I_{m-1}(z)-I_{m+1}(z))\\
	A_{cc}(m,z)&=\frac{1}{4}(I_{m-2}(z)+2I_m(z) +I_{m+2}(z))\\
	A_{cs}(m,z)&=\frac{i}{4}(I_{m-2}(z)-I_{m+2}(z))\\
	A_{ss}(m,z)&=\frac{1}{4}(-I_{m-2}(z)+2I_m(z)-I_{m+2}(z))\\
	A_{ccc}(m,z)&=\frac{1}{8}(I_{m-3}(z)+3I_{m-1}(z)\nonumber\\
	&+3I_{m+1}(z)+I_{m+3}(z))\\
	A_{ccs}(m,z)&=\frac{i}{8}(I_{m-3}(z)+I_{m-1}(z)\nonumber\\
	&-I_{m+1}(z)-I_{m+3}(z))\\
	A_{css}(m,z)&=\frac{1}{8}(-I_{m-3}(z)+I_{m-1}(z)\nonumber\\
	&+I_{m+1}(z)-I_{m-3}(z))\\
	A_{sss}(m,z)&=\frac{i}{8}(-I_{m-3}(z)+3I_{m-1}(z)\nonumber\\
	&-3I_{m+1}(z)+I_{m+3}(z))\\
	A_{cccc}(m,z)&=\frac{1}{16}(I_{m-4}(z)+4I_{m-2}(z)\nonumber\\
	&+6I_{m}(z)+4I_{m+2}(z)+I_{m+4}(z))\\
	A_{cccs}(m,z)&=\frac{i}{16}(I_{m-4}(z)+2I_{m-2}(z)\nonumber \\
	&-2I_{m+2}(z)-I_{m+4}(z))\\
	A_{ccss}(m,z)&=\frac{1}{16}(-I_{m-4}(z)+2I_{m}(z)\nonumber\\
	&-I_{m+4}(z))\\
	A_{csss}(m,z)&=\frac{i}{16}(-I_{m-4}(z)+2I_{m-2}(z)\nonumber\\
	&-2I_{m+2}(z)+I_{m+4}(z))\\
	A_{ssss}(m,z)&=\frac{1}{16}(i_{m-4}(z)-4I_{m-2}(z)+6I_{m}(z)\nonumber\\
	&-4I_{m+2}(z)+I_{m+4}(z)).
\end{align}

\clearpage
\bibliography{references}

\end{document}